\documentclass[10pt,final,journal,twocolumn,singlespaced]{IEEEtran}
\usepackage{algorithm}
\usepackage{algpseudocode}
\usepackage{amsmath}
\usepackage{amssymb}
\usepackage{bm}
\usepackage{booktabs}
\usepackage{color}
\usepackage{cite}
\usepackage{epsfig}
\usepackage{epstopdf}
\usepackage{flafter}
\usepackage{float}
\usepackage{ifthen}
\usepackage{multirow}
\usepackage{amsfonts}
\usepackage{makecell} 
\usepackage{pifont}
\usepackage{subfigure}
\usepackage{times}
\usepackage{threeparttable}
\usepackage{url}
 
\usepackage{subfigure,graphicx}
\usepackage{float}
\usepackage{color,multirow}
\usepackage{makecell}
\usepackage{setspace}
\usepackage{cite} 
\usepackage{url} 
\usepackage{ifthen} 
\usepackage[T1]{fontenc}
\usepackage{graphicx}
\usepackage{changepage}
\usepackage{rotating}  
\usepackage[colorlinks,linkcolor=red,anchorcolor=gray,citecolor=green,urlcolor=blue]{hyperref}
\usepackage{graphics}
\usepackage{booktabs}

\newcommand{\tabincell}[2]{\begin{tabular}{@{}#1@{}}#2\end{tabular}}

\newtheorem{theorem}{Theorem}
\newtheorem{definition}{Definition}
\def\BibTeX{{\rm B\kern-.05em{\sc i\kern-.025em b}\kern-.08em
    T\kern-.1667em\lower.7ex\hbox{E}\kern-.125emX}}
\begin{document}
\title{Self-Supervised Nonlinear Transform-Based Tensor Nuclear Norm for Multi-Dimensional Image Recovery}
\author{Yi-Si Luo, Xi-Le Zhao, \IEEEmembership{Member, IEEE}, Tai-Xiang Jiang, \IEEEmembership{Member, IEEE}, Yi Chang, \IEEEmembership{Member, IEEE},\\ Michael K. Ng, \IEEEmembership{Senior Member, IEEE}, and Chao Li
\thanks{This research is supported by NSFC (No. 61876203, 61772003), the Applied Basic 
Research Project of Sichuan Province (No. 2021YJ0107), the Key Project of Applied Basic Research in Sichuan Province (No. 2020YJ0216), and National Key Research and Development Program of China (No. 2020YFA0714001).}
\thanks{Y.-S. Luo and X.-L. Zhao are with the School of Mathematical Sciences, University of Electronic Science and Technology of China, Chengdu, P.R.China (e-mail: yisiluo1221@foxmail.com; xlzhao122003@163.com).}
\thanks{T.-X. Jiang is with the Financial Intelligence and Financial Engineering Research Key Laboratory of Sichuan province, School of Economic Information Engineering, Southwestern University of Finance and Economics, Chengdu, P.R.China (e-mail: taixiangjiang@gmail.com).}
\thanks{Y. Chang is with the Artificial Intelligence Research Center, Peng Cheng Laboratory, Shenzhen, P.R.China (e-mail: yichang@hust.edu.cn).}
\thanks{Michael K. Ng is with the Department of Mathematics, The University of Hong Kong, Hong Kong (e-mail: mng@maths.hku.hk).}
\thanks{C. Li is with RIKEN Center on Advanced Intelligence Project, Tokyo 103-0027, Japan (e-mail: chao.li@riken.jp).}}
\maketitle
\begin{abstract}
In this paper, we study multi-dimensional image recovery. Recently, transform-based tensor nuclear norm minimization methods are considered to capture low-rank tensor structures to recover third-order tensors in multi-dimensional image processing applications. The main characteristic of such methods is to perform the linear transform along the third mode of third-order tensors, and then compute tensor nuclear norm minimization on the transformed tensor so that the underlying low-rank tensors can be recovered. The main aim of this paper is to propose a nonlinear multilayer neural network to learn a nonlinear transform via the observed tensor data under self-supervision. The proposed network makes use of low-rank representation of transformed tensors and data-fitting between the observed tensor and the reconstructed tensor to construct the nonlinear transformation. Extensive experimental results on tensor completion, background subtraction, robust tensor completion, and snapshot compressive imaging are presented to demonstrate that the performance of the proposed method is better than that of state-of-the-art methods.
\end{abstract}
\section{Introduction}\label{sec:introduction}
\begin{figure}[t]
\tiny
\setlength{\tabcolsep}{0.9pt}
\begin{center}
\begin{tabular}{ccccc}
\rotatebox{90}{\tabincell{c}{\scriptsize\;\;\;\;\;\;\;Tensor\\\;\;\;\;\;\;\;\;\;\scriptsize completion}}&
\includegraphics[width=0.11\textwidth]{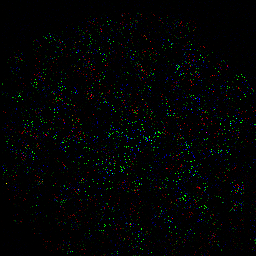}&
\includegraphics[width=0.11\textwidth]{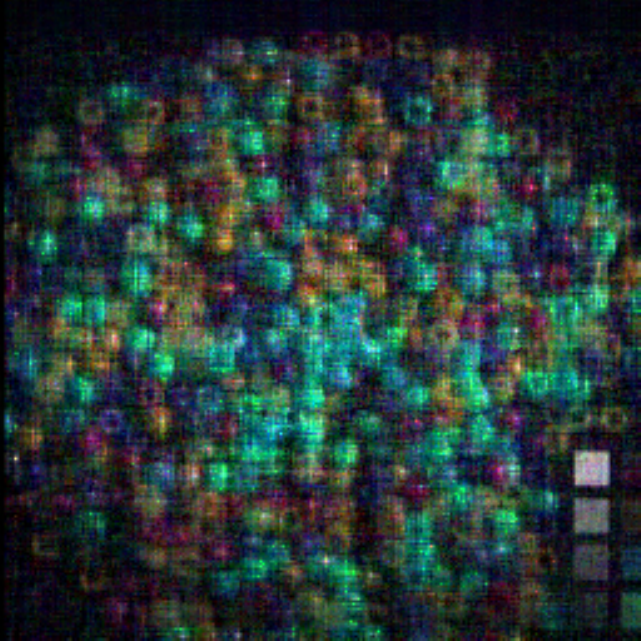}&
\includegraphics[width=0.11\textwidth]{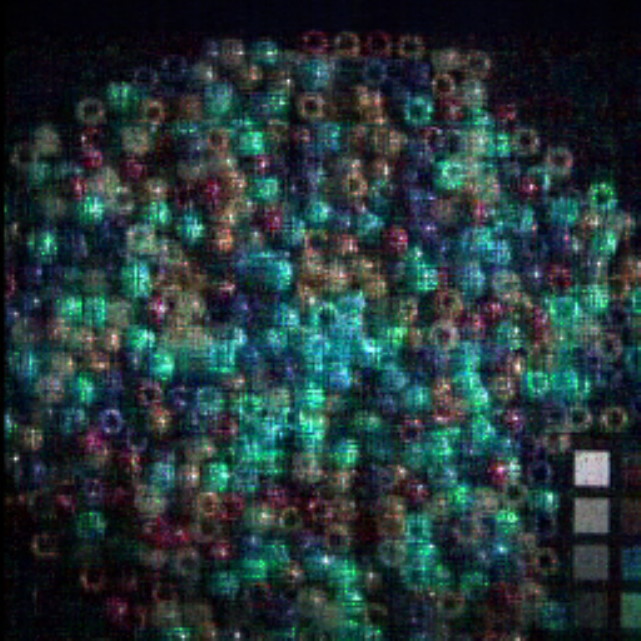}&
\includegraphics[width=0.11\textwidth]{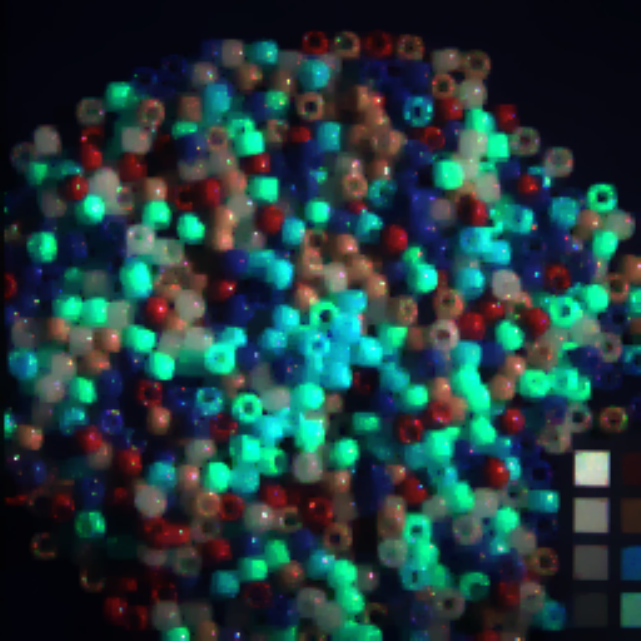}\\
\vspace{0.08cm}
~&Observed &TNN \cite{TNN_LRTC} 20.0dB&FTNN\cite{FTNN} 21.0dB&SSNT-TV 24.6dB\\
\rotatebox{90}{\tabincell{c}{\;\;\;\;\scriptsize \;\;Robust tensor\\\;\;\;\;\;\;\scriptsize completion}}&
\includegraphics[width=0.11\textwidth]{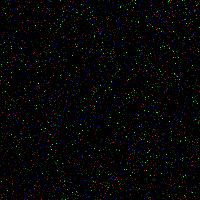}&
\includegraphics[width=0.11\textwidth]{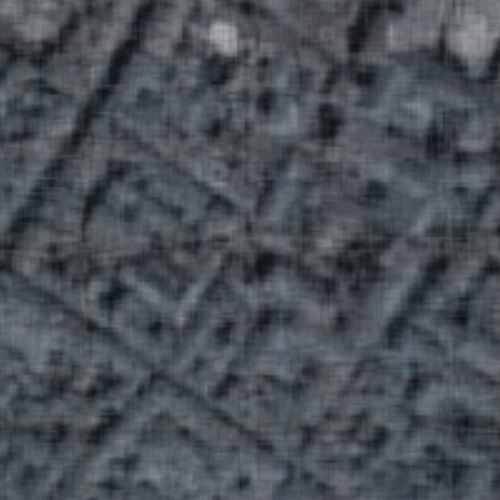}&
\includegraphics[width=0.11\textwidth]{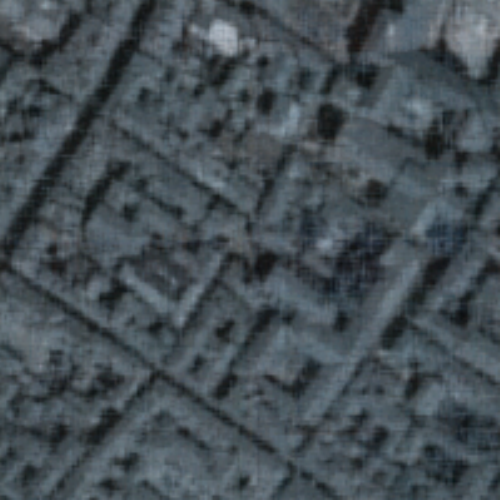}&
\includegraphics[width=0.11\textwidth]{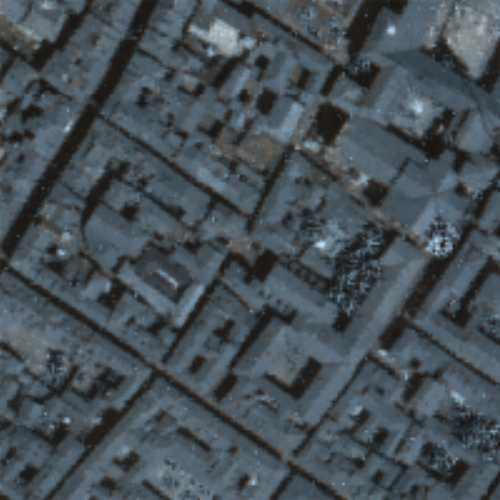}\\
~&Observed &TNN\cite{TNN_LRTC} 23.7dB& UTNN\cite{Haar} 25.9dB&SSNT-TV 28.1dB\\
\rotatebox{90}{\tabincell{c}{\scriptsize Snapshot\\\scriptsize compressive imaging}}&
\includegraphics[width=0.11\textwidth]{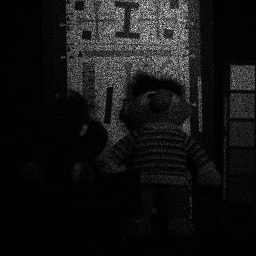}&
\includegraphics[width=0.11\textwidth]{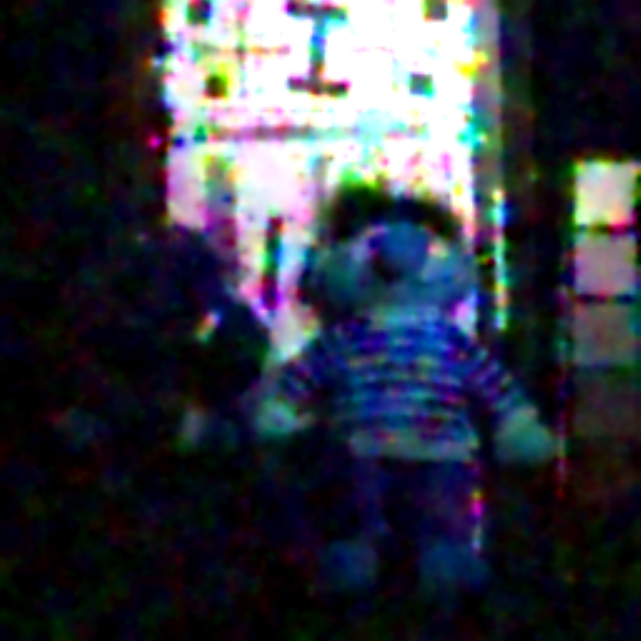}&
\includegraphics[width=0.11\textwidth]{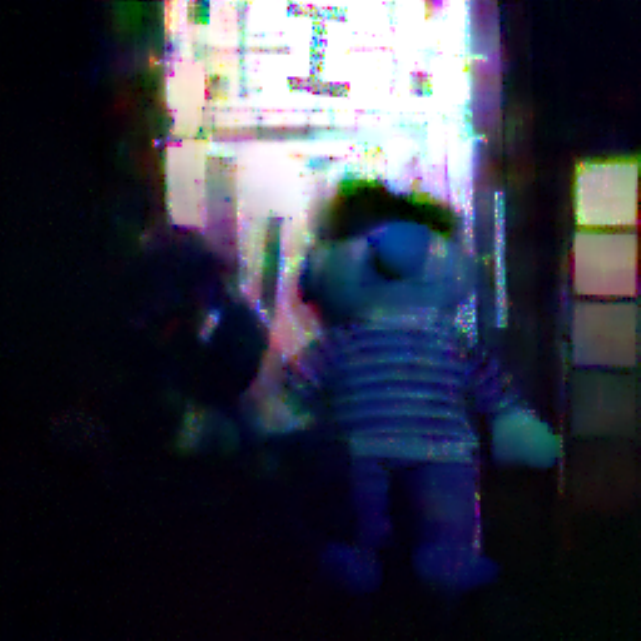}&
\includegraphics[width=0.11\textwidth]{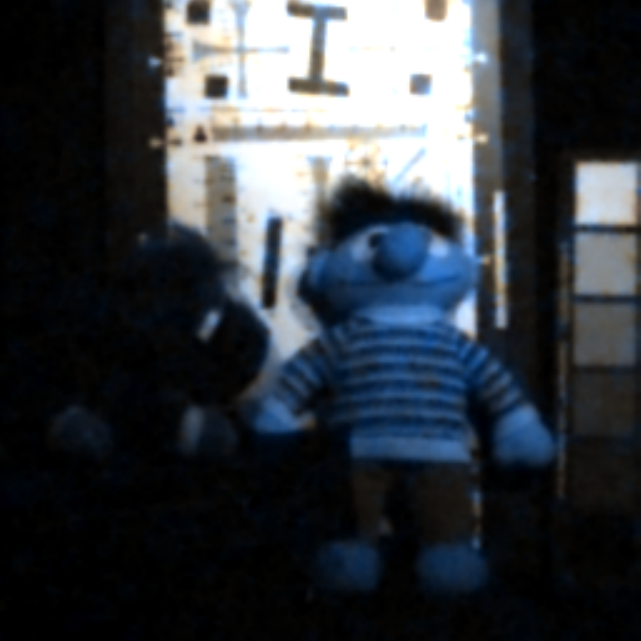}\\
~&Observed &GAP-TV\cite{GAP-TV} 21.2dB&DeSCI\cite{DeSCI} 23.2dB&SSNT-TV 27.1dB\\
\vspace{-0.5cm}
\end{tabular}
\end{center}
\caption{The recovered results and PSNR values by different methods on different inverse problems. Three rows respectively list the recovered results for tensor completion on {\it Beads} with SR=0.05, the recovered results for robust tensor completion on {\it Pavia} with SR=0.05, and the recovered results for snapshot compressive imaging on {\it Toys} with SR=0.25. The proposed SSNT-TV obtains the best PSNR values and qualitative results compared with state-of-the-art methods.\label{First_show}}\vspace{-0.6cm}
\end{figure}
\IEEEPARstart{M}{ANY} real-world images are multi-dimensional, such as multispectral images (MSIs), videos, and magnetic resonance images (MRIs). However, in many applications, multi-dimensional images are incomplete or essentially degraded \cite{MSIs_CVPR16} due to irresistible factors such as low light or failure of sensors. Thus, it is of the tremendous need to recover/restore the high-quality underlying images from the observed images, which is one of the important imaging problems \cite{TIP_2017}. \par
Mathematically, a multi-dimensional image can be represented by a third-order tensor \cite{Tensor_app,TCI_1,TCI_MRI}, which preserves the multi-direction structure. Since most real-world images have low-rank structures \cite{tTNN,TPAMI_TC,TNN_TRPCA,exact_TC,ChaoLi,TCI_2,TCI_TC}, the restoration of the observed image is usually formulated as the following low-rank tensor recovery problem:
\begin{equation}\label{eq_1}
\min_{\mathcal X}\;\lambda\;rank({\mathcal X}) + L({\mathcal X},{\mathcal O}),
\end{equation}
where $\mathcal O$ denotes the observed tensor, $\mathcal X$ denotes the underlying low-rank tensor, $L({\mathcal X},{\mathcal O})$ is the fidelity loss function, 
and $\lambda$ is the trade-off parameter.\par
\begin{table*}[!t]
\caption{The milestone of the transforms induced TNN. The {\bf DATA-DEPENDENT} transforms are highlighted by {\bf BOLDFACE}.\label{milestones}}\vspace{-0.4cm}
\footnotesize
\begin{center}
\setlength{\tabcolsep}{12pt}
\begin{spacing}{0.95}
\begin{tabular}{ccccc}
\toprule
{\sl DFT/DCT}&{\sl Orthogonal}&{\sl Invertible}&{\sl Non-invertible}&{\sl Nonlinear}\\
\midrule
Zhang et al., 2014\cite{TNN_LRTC}&Song et al., 2020\cite{Haar}&Kernfeld et al., 2015\cite{T_prod}&{Jiang et al., 2020}\cite{FTNN}&{\bf Ours}\\
Madathil and George, 2018\cite{Mada_2018}&{\bf Ng et al., 2020}\cite{Patch_tube}&Lu et al., 2019\cite{DCTNN}&{\bf Jiang et al., 2020}\cite{DTNN}&~\\
Xu et al., 2019\cite{DCT_Zhao}&~&~&{\bf Kong and Lin, 2021}\cite{Q_rank}&~\\
\bottomrule
\end{tabular}
\end{spacing}
\end{center}
\vspace{-0.85cm}
\end{table*}
Different from matrices, the definition of tensor rank is not unique. Several definitions of tensor ranks are proposed. The CP rank (see for example \cite{CP_Tucker}) is defined as the smallest number of rank one tensor decomposition. However, computing the CP rank is an NP-hard problem and its convex surrogate is not clear. The Tucker rank was studied for tensors by considering the ranks of unfolding matrices from tensors, see for example \cite{CP_Tucker}. However, the sum of the nuclear norm of unfolding matrices is not the convex envelope of the sum of the rank of frontal slices \cite{NIPS_2013}. In this paper, we focus on the tensor tubal-rank, which is based on the tensor singular value decomposition (t-SVD) \cite{TSVD}. It allows new extensions of familiar matrix analysis to the multilinear setting while avoiding the loss of information inherent in flattening of the tensor \cite{Comp}. 
The minimization of the tubal-rank is an NP-hard problem. Zhang {et al.} \cite{TNN_LRTC} built a convex surrogate of the tensor tubal-rank, named the tensor nuclear norm (TNN). Thus, model (\ref{eq_1}) is re-formulated as follows:
\begin{equation}
\min_{\mathcal X}\;\lambda\;\lVert{\mathcal X}\rVert_{\rm TNN} + L({\mathcal X},{\mathcal O}).\label{opt}
\end{equation}\par
Note that the TNN of a tensor is computed by summing the nuclear norm of each transformed frontal slice where a transform is applied along the third mode of the tensor \cite{TNN_LRTC}. Thus, model (\ref{opt}) can be re-formulated as follows:
\begin{equation}
\min_{\mathcal X}\;\lambda\;\sum_k 
\lVert l({\mathcal X})^{(k)}\rVert_* + L({\mathcal X},{\mathcal O}),
\end{equation}
where $l({\mathcal X})$ is the transformed tensor under the transform $l$ and the superscript refers to the $k$-th frontal slice of the transformed tensor. More precisely, the discrete Fourier transform (DFT) is used, see \cite{TNN_LRTC,TSVD,T_prod}. Since TNN is convex, model (\ref{opt}) can be addressed by many convex optimization algorithms.\par
\begin{figure*}[!t]
\centering
\setlength{\abovecaptionskip}{0.1cm}
\includegraphics [width = 0.99\linewidth]{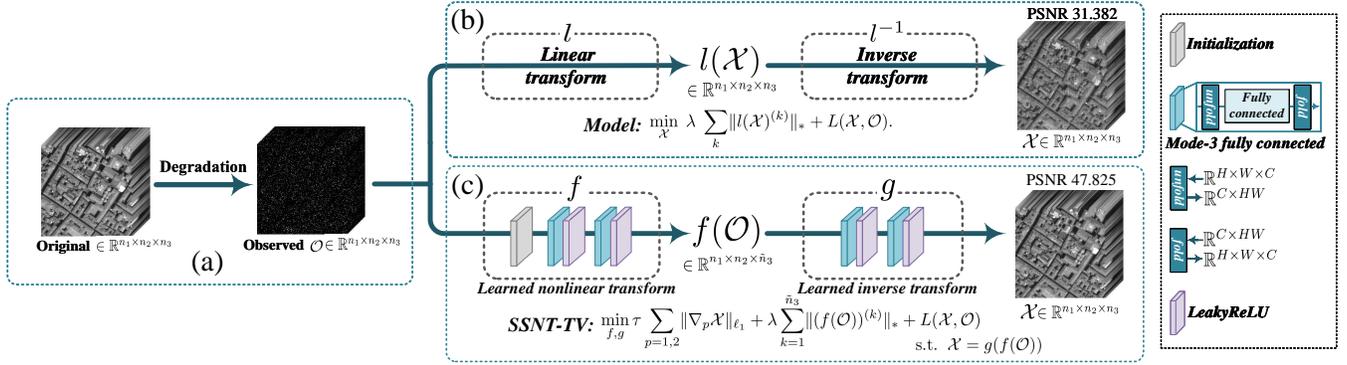}
\vspace{-0.1cm} 
\caption{The pipeline of linear transform-based TNN and the proposed SSNT for low-rank tensor recovery. (a) The degradation process. (b) The classical linear transform-based TNN for low-rank tensor recovery. (c) The proposed SSNT-TV for low-rank tensor recovery.\label{dongji}}\vspace{-0.3cm} 
\end{figure*}

 
In the literature, other transforms were considered and studied, for instance, the use of discrete cosine transform (DCT) \cite{DCT_Zhao,Mada_2018} for real arithmetic computation and other unitary transforms \cite{Haar}. The motivation is that when a suitable transform is applied to the third-mode of a tensor, a smaller low-rank representation of the transformed tensor can be obtained, and therefore the underlying low-rank tensor can be more easily recovered, see \cite{Haar,DCTNN}. In Figure \ref{dongji}(b), we describe a tensor completion process by using a linear transform in the TNN.\par
To explore a better low-rank representation of the transformed tensor, Jiang {et al.} \cite{FTNN} suggested to use the non-invertible framelet transform (a redundant basis) to represent low-rank transformed tensors. Along this research direction, data-adaptive transforms were proposed and studied. Kong {et al.} \cite{Q_rank} proposed the data-dependent transform to capture the low Q-rank tensor structure. Jiang {et al.} \cite{DTNN} proposed to learn low-rank coding coefficients using dictionary approach. Ng {et al.} \cite{Patch_tube} used the left singular vectors of the unfolding matrix to establish the patched-tube unitary transform.\par
Nevertheless, all of the aforementioned transforms are linear which limits their capability to model the complex and nonlinear nature of real-world data. In this paper, we employ a nonlinear multilayer neural network to represent the transform. The neural network consists of the composition of weighted linear functions with nonlinear activation functions. By optimization of low-rank representation of transformed tensors and data-fitting between the observed tensor and the reconstructed tensor, both the nonlinear neural network and the underlying low-rank ${\mathcal X}$ can be learned simultaneously under self-supervision. We call such transform to be Self-Supervised Nonlinear Transform (SSNT). Based on the universal approximation theorem of neural networks \cite{Universial}, the proposed SSNT could approximate to any functions and thus it can obtain a better and lower-rank representation than linear transforms.\par 
\begin{table}[!t]
\caption{Notations used in this paper.\label{tab_notation}}
\begin{center}
\setlength{\tabcolsep}{2.8pt}
\begin{spacing}{1.00}
\begin{tabular}{c|c}
\Xhline{1pt}
Notations & Interpretations\\
\hline
${\bf X},{\mathcal X}$ & matrix, tensor\\
${\mathcal X}_{ijk}$ & the $i,j,k$-th element of $\mathcal X$\\
${\mathcal X}(:,:,k)$ or ${\mathcal X}^{(k)}$ & the $k$-th frontal slice of $\mathcal X$\\
${\mathcal X}(i,j,:)$ & the $i,j$-th tube of $\mathcal X$\\
$\nabla_p$ & \tabincell{c}{the difference operator along the  \vspace{-0.1cm}\\$p$-th dimension}\\
$\lVert{\bf X}\rVert_*$ & the nuclear norm of $\bf X$\\
$\lVert{\mathcal X}\rVert_F$ & \tabincell{c}{the tensor Frobenius norm  \vspace{-0.1cm} \\$\lVert{\mathcal X}\rVert_F=\sqrt{\langle {\mathcal X},{\mathcal X}\rangle}=\sqrt{\sum_{ijk}{\mathcal X}_{ijk}^2}$}\\
$\lVert {\mathcal X}\rVert_{\ell_1}$ & \tabincell{c}{the tensor $\ell_1$-norm  \vspace{-0.1cm} \\$\lVert {\mathcal X}\rVert_{\ell_1}=\sum_{ijk}|{\mathcal X}_{ijk}|$}\\
${\tt unfold}_3(\cdot)$ &  \tabincell{c}{the mode-3 unfolding operator \vspace{-0.1cm} \\ ${\tt unfold}_3(\cdot):{\mathbb R}^{n_1\times n_2\times n_3}\rightarrow{\mathbb R}^{n_3\times n_1n_2}$}\\ 
${\tt fold}_3(\cdot)$ &  \tabincell{c}{the mode-3 folding operator \vspace{-0.1cm} \\ ${\tt fold}_3(\cdot):{\mathbb R}^{n_3\times n_1n_2}\rightarrow{\mathbb R}^{n_1\times n_2 \times n_3}$}\\ 
$\times_3$ &  \tabincell{c}{the mode-3 tensor-matrix product \vspace{-0.1cm} \\ ${\mathcal X}\times_3{\bf A}={\tt fold}_3({\bf A}{\tt unfold}_3({\mathcal X}))$}\\
\hline
\Xhline{1pt}
\end{tabular}
\end{spacing}
\end{center}
\vspace{-0.7cm}
\end{table}
Generally, only considering tensor low-rankness is not sufficient to recover the multi-dimensional images with complex image details. Thus, other hand-crafted priors are needed. The proposed SSNT can be easily combined with many hand-crafted regularizers. In this paper, we consider the simple and efficient total variation (TV) \cite{TV_1} regularization. The TV term could explore the spatial local smoothness of the tensor, which improves the recovery quality. In Figure \ref{dongji}(c), we describe a tensor completion process by using TV-regularized SSNT.\par
We summarize the contributions of this paper as follows:
\begin{itemize}
\item 
We propose the self-supervised nonlinear transform-based TNN to exploit the complex and nonlinear nature inside multi-dimensional images. A nonlinear multilayer neural network is utilized to represent the transform, which is self-supervisedly learned by only using the observed data. The nonlinear modeling capability of SSNT is believed to faithfully capture the implicit low-rankness of the data, which helps to obtain a better low-rank representation. To explore the spatial local correlation, we further employ the spatial TV regularization to deliver the local smooth prior for better performance.
\item 
We directly use the gradient descent algorithm to minimize the SSNT-based TNN and develop ADMM-like algorithm to tackle the proposed TV-regularized optimization problem. Extensive experiments are conducted on different multi-dimensional inverse problems including tensor completion, background subtraction, robust tensor completion (RTC), and snapshot compressive imaging (SCI) with various types of multi-dimensional images. The superiority of the proposed method is demonstrated as compared with state-of-the-art methods.
\end{itemize}\par
The outline of this paper is given as follows. In Sec. \ref{notation}, we give preliminaries of tensors. In Sec. \ref{Sec:Pro}, we present the proposed method. In Sec. \ref{Sec:Exp}, experimental results are reported to demonstrate the performance of the proposed network is better than that of state-of-the-art methods. Finally, some concluding remarks are given in Sec. \ref{Sec:Con}.
\section{Preliminaries}\label{notation}
The primary notations used in this paper are introduced in Table \ref{tab_notation}. In addition, we introduce the following definitions and theorem, which are derived from \cite{TSVD}. \par
\begin{definition}{\rm (t-product)}
The tensor-tensor product ${\mathcal C}={\mathcal A}*{\mathcal B}$ is defined by ${\mathcal C}(i,j,:)=\sum_{k=1}^{n_2}{\mathcal A}(i,k,:)*{\mathcal B}(k,j,:)$, where $*$ denotes the circular convolution between two vectors. 
\end{definition}
\begin{definition}{\rm (Conjugate transpose)}
The conjugate transpose of ${\mathcal A}\in{\mathbb R}^{n_1\times n_2\times n_3}$, denoted as ${\mathcal A}^H$, is defined by $({\mathcal A}^H)^{(1)}=({\mathcal A}^{(1)})^H$ and $({\mathcal A}^H)^{(i)}=({\mathcal A}^{(n_3+2-i)})^H(i=2,\cdots,n_3)$. 
\end{definition}
\begin{definition}{\rm (Identity tensor)}
${\mathcal I}\in{\mathbb R}^{n_1\times n_1\times n_3}$ is called an identity tensor if ${\mathcal I}^{(1)}$ is an identity matrix and ${\mathcal I}^{(k)}(k=2,\cdots,n_3)$ are zero matrices. 
\end{definition}
\begin{definition}{\rm (Orthogonal tensor)}
The tensor ${\mathcal Q}$ is orthogonal if ${\mathcal Q}*{\mathcal Q}^H={\mathcal Q}^H*{\mathcal Q}={\mathcal I}$. ${\mathcal A}\in{\mathbb R}^{n_1\times n_2\times n_3}$ is $f$-diagonal if ${\mathcal A}^{(i)}(i=1,\cdots,n_3)$ are diagonal matrices. 
\end{definition}
\begin{theorem}
{\rm (t-SVD)} Any ${\mathcal A}\in{\mathbb R}^{n_1\times n_2\times n_3}$ can be decomposed as ${\mathcal A}={\mathcal U}*{\mathcal S}*{\mathcal V}^H$, where ${\mathcal U}\in{\mathbb R}^{n_1\times n_1\times n_3}$ and ${\mathcal V}\in{\mathbb R}^{n_2\times n_2\times n_3}$ are orthogonal and ${\mathcal S}\in{\mathbb b}^{n_1\times n_2\times n_3}$ is $f$-diagonal. 
\end{theorem}
\begin{definition}
{\rm (Tensor tubal-rank)} Given the t-SVD: ${\mathcal A}={\mathcal U}*{\mathcal S}*{\mathcal V}^H$, where ${\mathcal A}\in{\mathbb R}^{n_1\times n_2\times n_3}$, the tubal-rank $rank_t({\mathcal A})$ is defined as the number of nonzero singular tubes of $\mathcal S$. 
\end{definition}
\begin{definition}
{\rm (TNN)} The tensor nuclear norm of ${\mathcal A}\in{\mathbb R}^{n_1\times n_2\times n_3}$ is defined as $\lVert{\mathcal A}\rVert_{\rm TNN} = \sum_{k=1}^{n_3}\lVert({{\mathcal A}\times_3{\bf F}})^{(k)}\rVert_*$, where ${\bf F}\in{\mathbb R}^{n_3\times n_3}$ denotes the DFT matrix. 
\end{definition}
\section{Proposed SSNT}\label{Sec:Pro}
In this section, we introduce the structure of the proposed nonlinear transform. Using the proposed transform, we build the optimization model and the corresponding algorithm for low-rank tensor recovery. The TV regularization is suggested to characterize the spatial local smoothness of the data, and the ADMM-like algorithm is presented to tackle the TV-based low-rank tensor recovery model. 
\subsection{Network Architechture of SSNT} 
Classical linear transforms in the t-SVD framework are generally represented by matrices, {e.g.}, the DFT matrix \cite{TNN_LRTC}, the DCT matrix \cite{DCT_Zhao}, or the data-dependent matrix \cite{Q_rank}. \par
Under the motivation of building a more expressive nonlinear transform, we propose to use a multilayer neural network to represent the transform. We call such transform the SSNT. The proposed SSNT has a hierarchical structure containing linear weights and nonlinear activation function. Thus, SSNT can be interpreted as a nonlinear multilayer neural network. \par
Formally, we suggest the nonlinear mode-3 fully connected (NoFC$_3$) layer as the unit of SSNT. A single NoFC$_3$ layer is formulated as
\begin{equation}
w_i({\mathcal X}) = \sigma({\mathcal X}\times_3{\bf W}_i),
\end{equation}
where $\sigma(\cdot)$ denotes the nonlinear activation function and ${\bf W}_i$ is the learnable matrix. In this paper, we use the LeakyReLU \cite{PReLU} as the nonlinear activation function $\sigma$. Consistent with the classical TNN, we employ the neural network along mode-3 to explore the interaction of frontal slices.\par
The proposed SSNT is formulated as
\begin{equation}
f({\mathcal X}) = w_p\circ w_{p-1}\circ\cdots\circ w_1({\mathcal X}_0),\label{f}
\end{equation}
where ${\mathcal X}_0={\rm Init}({\mathcal X})$, and ${\rm Init(\cdot)}$ is the initialization function, which can be chosen based on different problems. Here, $\circ$ denotes the composition of functions and $p$ denotes the number of NoFC$_3$ layers in $f$. Similarly, the inverse transform is
\begin{equation}
g({\mathcal X}) = w_{p+q}\circ w_{p+q-1}\circ\cdots\circ w_{p+1}({\mathcal X}),\label{g}
\end{equation}where $q$ denotes the number of NoFC$_3$ layers in $g$. 
If we omit the nonlinear function $\sigma$, $f$ and $g$ degrade to multilayer linear transforms. If we further let $p=q=1$, then $f$ and $g$ degrade to linear transforms, which include DFT \cite{TNN_LRTC}, DCT \cite{DCT_Zhao,Mada_2018}, orthogonal transforms \cite{Haar,Patch_tube}, invertible linear transforms \cite{DCTNN,T_prod}, and non-invertible transforms \cite{Q_rank,FTNN,DTNN}, {i.e.}, the linear transform is just a special case of our transform.\par
Our SSNT $f$ and the inverse transform $g$ are self-supervisedly learned by minimizing the transformed TNN and the fidelity loss. The detailed optimization strategy is illustrated as follows. 
\subsection{SSNT for Low-Rank Tensor Recovery}
\subsubsection{Optimization Model}
Basically, our optimization model for low-rank tensor recovery could be the same as the TNN model (\ref{opt}). The only difference is the transform. However, in our framework, the desired tensor ${\mathcal X}$ is difficult to be directly optimized via ADMM, as SSNT is nonlinear and is not pre-defined. Alternatively, we suggest training $f$ and $g$ by minimizing the SSNT-based TNN and obtain the desired tensor ${\mathcal X}$ via the learned transforms.\par
Given the observed data ${\mathcal O}\in{\mathbb R}^{n_1\times n_2\times n_3}$, the proposed optimizing model for low-rank tensor recovery is
\begin{equation}
\begin{split}
&\min_{f,g}\;\lambda\sum_{k=1}^{{\tilde n}_3}\lVert (f({\mathcal O}))^{(k)}\rVert_*+L({\mathcal X},{\mathcal O})\\
&\;{\rm s.t.}\;\;{\mathcal X} = g(f({\mathcal O})),\label{opt_rewr}
\end{split}
\end{equation}
where $f:{\mathbb R}^{n_1\times n_2\times n_3}\rightarrow{\mathbb R}^{n_1\times n_2\times {\tilde n}_3}$ and $g:{\mathbb R}^{n_1\times n_2\times {\tilde n}_3}\rightarrow{\mathbb R}^{n_1\times n_2\times n_3}$ are the learnable transforms defined by Eq. (\ref{f}) and Eq. (\ref{g}). $L({\mathcal X},{\mathcal O})$ is the fidelity term, which has different forms in different inverse problems. The desired tensor ${\mathcal X}$ is obtained by the learned transforms, i.e., ${\mathcal X} = g(f({\mathcal O}))$.\par
In model (\ref{opt_rewr}), the SSNT $f$ transforms $\mathcal O$ into a low-rank representation, and the inverse transform $g$ transforms the low-rank representation to the desired tensor. By minimizing the SSNT-based TNN $\sum_{k=1}^{{\tilde n}_3}\lVert (f({\mathcal O}))^{(k)}\rVert_*$, the transforms $f,g$ can be self-supervisedly learned and the reconstruction ${\mathcal X}$ can be obtained via the learned $f$ and $g$. With the nonlinear modeling ability of SSTN, minimizing the SSTN-based TNN could obtain a lower-rank transformed tensor $f({\mathcal O})$, which results in better low-rank representation. \par
\subsubsection{Algorithm}
Let ${\mathcal L}_1= \lambda\sum_{k}\lVert (f({\mathcal O}))^{(k)}\rVert_*$ and ${\mathcal L}_2 = L(g(f({\mathcal O})),{\mathcal O})$, the loss function corresponding to (\ref{opt_rewr}) is
\begin{equation}
{\mathcal L} = {\mathcal L}_1 + {\mathcal L}_2.
\end{equation}
It is expected to minimize ${\mathcal L}$ via updating $f$ and $g$. This is equivalent to update learnable matrices ${\bf W}_1,\cdots,{\bf W}_{p+q}$. The gradient of ${\mathcal L}_1$ on the $i,j$-th element of ${\bf W}_m$ is
\begin{equation}
\begin{split}
\frac{\partial {\mathcal L}_1}{\partial ({\bf W}_m)_{ij}} &=  \lambda\sum_k \frac{\partial\lVert (f({\mathcal O}))^{(k)}\rVert_*}{\partial ({\bf W}_m)_{ij}} \\&=\lambda\sum_k \sum_{s,t}\frac{\partial\lVert (f({\mathcal O}))^{(k)}\rVert_*}{\partial ((f({\mathcal O}))^{(k)})_{st}}\frac{\partial ((f({\mathcal O}))^{(k)})_{st}}{\partial ({\bf W}_m)_{ij}}.\label{9}
\end{split}
\end{equation}
The subgradient of the nuclear norm \cite{nuc_grad} is
\begin{equation}
\frac{\partial \lVert (f({\mathcal O}))^{(k)}\rVert_*}{\partial (f({\mathcal O}))^{(k)}} \ni {\bf \widetilde U}_k{\bf \widetilde V}_k^{T},\label{10}
\end{equation}
where $(f({\mathcal O}))^{(k)}={\bf U}_k{\bf S}_k{\bf V}_k^T$ is the matrix singular value decomposition, and ${\bf \widetilde U}_k,{\bf \widetilde V}_k$ are ${\bf U}_k,{\bf V}_k$ truncated to the first $s_k$ columns and rows. Here, $s_k$ denotes the number of non-zero elements in ${\bf S}_k$. Integrating (\ref{9}) and (\ref{10}), we have 
\begin{equation}
\frac{\partial {\mathcal L}_1}{\partial ({\bf W}_m)_{ij}} \ni \lambda\sum_k \sum_{s,t}({\bf \widetilde U}_k{\bf \widetilde V}_k^{T})_{st}\frac{\partial ((f({\mathcal O}))^{(k)})_{st}}{\partial ({\bf W}_m)_{ij}}.\label{grad_1}
\end{equation}
The gradient of ${\mathcal L}_2$ on ${\bf W}_m$ is 
\begin{equation}
\frac{\partial {\mathcal L}_2}{\partial ({\bf W}_m)_{ij}} = \sum_{r,s,t} \frac{\partial {\mathcal L}_2}{\partial (g(f({\mathcal O})))_{rst}}\frac{\partial (g(f({\mathcal O})))_{rst}}{\partial ({\bf W}_m)_{ij}}.\label{grad_2}
\end{equation}
With gradients (\ref{grad_1}) and (\ref{grad_2}) and the maximum iteration $t_{max}$, model (\ref{opt_rewr}) can be addressed by most gradient descent-based algorithms. In this paper, we adopt the adaptive moment estimation (Adam) \cite{ADAM}. 
\subsection{TV-Regularized SSNT for Tensor Recovery}
In model (\ref{opt_rewr}), we only consider tensor low-rankness, which would be sometimes not sufficient to explore the spatial local correlation of data. Thus, we propose the TV regularized SSNT (termed as SSNT-TV) for tensor recovery. The spatial local smoothness can be faithfully exploited by TV regularization for better multi-dimensional image recovery performance.
\subsubsection{Optimization Model}
By introducing the TV regularization, the model of SSNT-TV for tensor recovery is
\begin{equation}
\begin{split}
&\min_{f,g}\tau\sum_{p=1,2}\lVert\nabla_p{\mathcal X}\rVert_{\ell_1}+\lambda\sum_{k=1}^{{\tilde n}_3}\lVert (f({\mathcal O}))^{(k)}\rVert_*+L({\mathcal X},{\mathcal O})\\
&\;{\rm s.t.}\;\;{\mathcal X} = g(f({\mathcal O})),\label{tv}
\end{split}
\end{equation}
where $\sum_{p=1,2}\lVert\nabla_p{\mathcal X}\rVert_{\ell_1}$ is the spatial TV term and $\tau$ is the weight parameter of the TV term. 
\subsubsection{Algorithm}
To address the model (\ref{tv}), we introduce an ADMM-like algorithm \cite{admmdiptv}. By introducing auxiliary variables ${\mathcal V}_p\;(p=1,2)$, we re-formulate model (\ref{tv}) as  
\begin{equation}
\begin{split}
&\min_{f,g,{\mathcal V}_p}\tau\sum_{p}\lVert{\mathcal V}_p\rVert_{\ell_1}+\lambda\sum_{k=1}^{{\tilde n}_3}\lVert (f({\mathcal O}))^{(k)}\rVert_*+L(g(f({\mathcal O})),{\mathcal O})\\&\;{\rm s.t.}\;\;{\mathcal V}_p = \nabla_p\big{(}g(f({\mathcal O}))\big{)},\;p=1,2.\label{opt_tv_re}
\end{split}
\end{equation}
The augmented Lagrangian function of (\ref{opt_tv_re}) is
\begin{equation}
\begin{split}
L_{\beta}&(f,g,{\mathcal V}_p,\Lambda_p) =\tau\sum_{p}\lVert{\mathcal V}_p\rVert_{\ell_1}+\lambda\sum_{k=1}^{{\tilde n}_3}\lVert (f({\mathcal O}))^{(k)}\rVert_*
\\&+L(g(f({\mathcal O})),{\mathcal O}) 
+\sum_p\Big{(}\langle\Lambda_p,\nabla_p\big{(}g(f({\mathcal O}))\big{)}-{\mathcal V}_p\rangle
\\&+\frac{\beta}{2}\lVert\nabla_p\big{(}g(f({\mathcal O}))\big{)}-{\mathcal V}_p\rVert_F^2\Big{)},
\end{split}
\end{equation}where $\beta$ is the penalty parameter and $\Lambda_p$ is the Lagrangian multiplier. We alternatively update $f,g$, ${\mathcal V}_p$, and $\Lambda_p$ as follows.\par 
$f,g$ {\bf sub-problem}: The transforms $f$ and $g$ at $t$-th iteration is determined by
\begin{equation}
\begin{split}
f^{t+1},g^{t+1} &\in \arg\min_{f,g}\lambda\sum_{k=1}^{{\tilde n}_3}\lVert (f({\mathcal O}))^{(k)}\rVert_*\\&+L(g(f({\mathcal O})),{\mathcal O})\\
&+\frac{\beta}{2}\sum_p\lVert \nabla_p\big{(}g(f({\mathcal O}))\big{)}-{\mathcal V^t_p}+\frac{\Lambda_p^t}{\beta}\rVert_F^2.\\
\label{fg_sub}
\end{split}
\vspace{-0.1cm}
\end{equation}
Similar to the optimization scheme of (\ref{opt_rewr}), the model (\ref{fg_sub}) can be easily obtained by Adam. \par
${\mathcal V}_p$ {\bf sub-problem}: The variable ${\mathcal V}_p$ at $t$-th iteration is
\begin{equation}
\begin{split}
{\mathcal V}_p^{t+1} &= \arg\min_{{\mathcal V}_p}\;\tau\lVert{\mathcal V}_p\rVert_{\ell_1}\\&+\frac{\beta}{2}\lVert{\mathcal V}_p -\Big{(}\nabla_p(g^t(f^t({\mathcal O})))+\frac{\Lambda_p^t}{\beta}\Big{)}\rVert_F^2.\label{V_sub}
\end{split}
\end{equation}
Model (\ref{V_sub}) has a closed-form solution implies that
\begin{equation}
{\mathcal V}_p^{t+1} = {\tt Soft}_{\frac{\tau}{\beta}}\Big{(}\nabla_p(g^t(f^t({\mathcal O})))+\frac{\Lambda_p^t}{\beta}\Big{)},\label{V_p_sub}
\end{equation}
where ${\tt Soft}_v(\cdot)$ denotes the soft-thresholding operator with threshold value $v$.\par
$\Lambda_p$ {\bf updating}: The multiplier $\Lambda_p$ is updated by 
\begin{equation}
\Lambda_p^{t+1} = \Lambda_p^t +\beta\Big{(}\nabla_p(g^t(f^t({\mathcal O})))-{\mathcal V}_p^t\Big{)}.\label{Lambda_sub}
\end{equation}
\par
Our algorithm for solving model (\ref{tv}) is summarized in Algorithm \ref{alg_1}.
\begin{algorithm}[t]
\renewcommand\arraystretch{1.2}
\caption[Caption for LOF]{SSNT-TV for Tensor Recovery}
\begin{algorithmic}[1]
\renewcommand{\algorithmicrequire}{\textbf{Input:}} 
\Require
The observed tensor ${\mathcal O}$; trade-off parameters $\tau$ and $\lambda$; Lagrange parameter $\beta$; maximum iteration $t_{max}$.
\renewcommand{\algorithmicrequire}{\textbf{Initialization:}} 
\Require ${\mathcal X}_0 = {\rm Init}({\mathcal X})$, ${\mathcal{V}}_p=\nabla_p{\mathcal X}_0$, $\Lambda_p = {\bf 0}$, $t=0$.
\While {$t<t_{max}$}
\State Update $f$ and $g$ via (\ref{fg_sub});
\State Update ${\mathcal V}_p$ via Eq. (\ref{V_p_sub});
\State Update ${\Lambda}_p$ via Eq. (\ref{Lambda_sub});
\State t=t+1;
\EndWhile
\renewcommand{\algorithmicrequire}{\textbf{Output:}}
\Require The recovered tensor ${\mathcal X} = g(f({\mathcal O}))$.
\end{algorithmic}
\label{alg_1}
\end{algorithm}
\section{Experiments}\label{Sec:Exp}
In this section, we introduce four tensor-based inverse problems. Each of these problems can be addressed using SSNT and SSNT-TV, where the only difference is the fidelity term $L({\mathcal X},{\mathcal O})$. We would like to emphasize that the proposed SSNT is self-supervisedly trained by only using the observed data. Thus, no training data and training/testing data splitting are required. Experiments and comparisons with state-of-the-arts are conducted on each problem to illustrate the effectiveness of our method. All experiments are conducted on the platform of Windows 10 with an Intel(R) Core i5-9400f CPU and RTX 2080 GPU with 24 GB RAM.
\subsection{Tensor Completion}
The tensor completion \cite{TPAMI_TC,TPAMI_pos} aims at recovering the original tensor from the incompleted tensor with random sampling, where the key issue is how to build a low-rank representation of the desired tensor since the definition of tensor rank is not unique. The low-rank tensor completion is formulated as 
\begin{equation}
\min_{\mathcal X}\;rank({\mathcal X}),\;\;{\rm s.t.}\;\;{\mathcal X}_\Omega = {\mathcal O}_\Omega,
\end{equation}
where $\mathcal O$ is the incompleted observation and $\Omega$ is the observed set. To make the entries of the desired tensor in $\Omega$ as close as that of the observation, the fidelity term of the proposed model for tensor completion is 
\begin{equation}
L({\mathcal X},{\mathcal O}) = \lVert{\mathcal P}_\Omega({\mathcal X} - {\mathcal O})\rVert_F^2, 
\end{equation}
where ${\mathcal P}_\Omega(\cdot)$ is the projection function that keeps the elements in $\Omega$ and making others be zero. To ensure the entries in the observed set exactly equal to that of the observation, the final result of the tensor completion is obtained by ${\mathcal X} = {\mathcal P}_{\Omega^C}\big{(}{g(f({\mathcal O}))}\big{)} + {\mathcal P}_{\Omega}\big{(}{\mathcal O}\big{)}$, where $\Omega^C$ denotes the complementary set of $\Omega$.
\begin{table*}[!t]
\begin{center}
\caption{The quantitative results by different methods on different data for tensor completion. The {\bf BEST} values are highlighted by {\bf BOLDFACE}, and the \underline{SECOND-BEST} values are highlighted by \underline{UNDERLINED}.\label{TC_tab}}
\scriptsize
\setlength{\tabcolsep}{2.9pt}
\begin{spacing}{0.95}
\begin{tabular}{clcccccccccccccccc}
\toprule
\multirow{2}*{Data}&SR&\multicolumn{3}{c}{0.05}&\multicolumn{3}{c}{0.1}&\multicolumn{3}{c}{0.15}&\multicolumn{3}{c}{0.2}&\multicolumn{3}{c}{0.25}
&\;\multirow{2}*{\tabincell{c}{Time\\(s)}}\\
\cmidrule{2-17}
~&Metric&\;\;\;PSNR&SSIM&SAM\;\;\;&PSNR&SSIM&SAM\;\;\;&PSNR&SSIM&SAM\;\;\;&PSNR&SSIM&SAM\;\;\;&PSNR&SSIM&SAM&~\\
\midrule
\multirow{6}*{\tabincell{c}{
HSI {\it WDC mall}\\{(256$\times$256$\times$191)}\\}}
~&Observed&\;\;\; 14.567&0.076&1.351 \;\;\;&14.801&0.118&1.253 \;\;\;&15.050&0.158&1.176 \;\;\;&15.312&0.199&1.109 \;\;\;&15.594&0.239&1.049&$-$\\
~&TRLRF\cite{TRLRF}&\;\;\; 27.044&0.854&0.209 \;\;\;&29.463&0.912&0.164 \;\;\;&29.959&0.920&0.160 \;\;\;&29.671&0.918&0.168 \;\;\;&30.589&0.931&0.156&1806\\
~&TNN\cite{TNN_LRTC}&\;\;\; 29.513&0.916&0.197 \;\;\;&33.249&0.962&0.144 \;\;\;&36.109&0.979&0.113 \;\;\;&38.311&0.986&0.093 \;\;\;&40.075&0.990&0.079&668\\
~&FTNN\cite{FTNN}&\;\;\; 32.776&0.955&0.131 \;\;\;&37.752&\underline{0.983}&0.095 \;\;\;&41.311&\underline{0.991}&0.074 \;\;\;&43.874&0.994&0.062 \;\;\;&45.954&\underline{0.996}&0.053&1742\\
~&SSNT&\;\;\; \underline{40.118}&\underline{0.992}&\underline{0.055} \;\;\;&\underline{44.764}&\bf{0.997}&\underline{0.040} \;\;\;&\underline{46.591}&\bf{0.998}&\underline{0.034} \;\;\;&\underline{47.657}&\underline{0.998}&\underline{0.031} \;\;\;&\underline{48.556}&\bf{0.999}&\underline{0.029}&547\\
~&SSNT-TV&\;\;\; \bf{41.155}&\bf{0.994}&\bf{0.050} \;\;\;&\bf{45.387}&\bf{0.997}&\bf{0.037} \;\;\;&\bf{47.291}&\bf{0.998}&\bf{0.032} \;\;\;&\bf{48.990}&\bf{0.999}&\bf{0.028} \;\;\;&\bf{49.917}&\bf{0.999}&\bf{0.026}&721\\
\midrule
\multirow{6}*{\tabincell{c}{
HSI {\it Pavia}\\{(200$\times$200$\times$80)}\\}}
~&Observed&\;\;\; 12.191&0.042&1.355 \;\;\;&12.426&0.071&1.254 \;\;\;&12.674&0.098&1.177 \;\;\;&12.939&0.125&1.110 \;\;\;&13.220&0.150&1.049&$-$\\
~&TRLRF\cite{TRLRF}&\;\;\; 28.232&0.888&0.113 \;\;\;&29.484&0.915&0.102 \;\;\;&30.918&0.936&0.087 \;\;\;&31.572&0.944&0.084 \;\;\;&32.028&0.950&0.082&124\\
~&TNN\cite{TNN_LRTC}&\;\;\; 26.002&0.822&0.174 \;\;\;&31.382&0.938&0.111 \;\;\;&35.429&0.971&0.080 \;\;\;&37.867&0.981&0.066 \;\;\;&40.171&0.987&0.055&68\\
~&FTNN\cite{FTNN}&\;\;\; 32.345&0.954&0.079 \;\;\;&37.821&0.985&0.052 \;\;\;&42.066&\underline{0.992}&\underline{0.039} \;\;\;&45.266&\underline{0.996}&0.030 \;\;\;&48.447&\underline{0.997}&0.024&304\\
~&SSNT&\;\;\; \underline{38.755}&\underline{0.990}&\underline{0.027} \;\;\;&\underline{46.164}&\underline{0.998}&\underline{0.016} \;\;\;&\underline{50.803}&\bf{0.999}&\bf{0.011} \;\;\;&\underline{52.021}&\bf{1.000}&\underline{0.010} \;\;\;&\underline{53.075}&\bf{1.000}&\underline{0.009}&257\\
~&SSNT-TV&\;\;\; \bf{38.837}&\bf{0.993}&\bf{0.026} \;\;\;&\bf{47.825}&\bf{0.999}&\bf{0.013} \;\;\;&\bf{50.994}&\bf{0.999}&\bf{0.011} \;\;\;&\bf{52.741}&\bf{1.000}&\bf{0.009} \;\;\;&\bf{54.381}&\bf{1.000}&\bf{0.008}&306\\
\midrule
\multirow{6}*{\tabincell{c}{
MSI {\it Balloons}\\{(256$\times$256$\times$31)}\\}}
~&Observed&\;\;\; 13.529&0.205&1.389 \;\;\;&13.762&0.248&1.278 \;\;\;&14.010&0.286&1.194 \;\;\;&14.272&0.320&1.123 \;\;\;&14.554&0.350&1.059&$-$\\
~&TRLRF\cite{TRLRF}&\;\;\; 30.062&0.883&0.244 \;\;\;&34.450&0.952&0.167 \;\;\;&38.868&0.982&0.112 \;\;\;&39.907&0.985&0.103 \;\;\;&40.288&0.986&0.101&125\\
~&TNN\cite{TNN_LRTC}&\;\;\; 26.321&0.850&0.267 \;\;\;&34.521&0.961&0.161 \;\;\;&38.822&0.982&0.111 \;\;\;&41.355&0.990&0.087 \;\;\;&43.253&0.993&0.071&57\\
~&FTNN\cite{FTNN}&\;\;\; 35.067&0.974&0.111 \;\;\;&39.640&0.990&0.069 \;\;\;&43.187&\underline{0.995}&0.049 \;\;\;&45.419&0.997&0.040 \;\;\;&47.609&\underline{0.998}&0.033&245\\
~&SSNT&\;\;\; \underline{38.021}&\underline{0.987}&\underline{0.078} \;\;\;&\underline{43.337}&\underline{0.996}&\underline{0.052} \;\;\;&\underline{46.646}&\bf{0.998}&\underline{0.039} \;\;\;&\underline{48.504}&\underline{0.998}&\underline{0.034} \;\;\;&\underline{49.426}&\bf{0.999}&\underline{0.028}&369\\
~&SSNT-TV&\;\;\; \bf{40.662}&\bf{0.994}&\bf{0.047} \;\;\;&\bf{44.622}&\bf{0.997}&\bf{0.036} \;\;\;&\bf{47.164}&\bf{0.998}&\bf{0.030} \;\;\;&\bf{49.183}&\bf{0.999}&\bf{0.025} \;\;\;&\bf{50.066}&\bf{0.999}&\bf{0.024}&381\\
\midrule
\multirow{6}*{\tabincell{c}{
MSI {\it Beads}\\{(256$\times$256$\times$31)}\\}}
~&Observed&\;\;\; 14.414&0.187&1.406 \;\;\;&14.646&0.227&1.295 \;\;\;&14.899&0.267&1.211 \;\;\;&15.165&0.309&1.139 \;\;\;&15.438&0.349&1.073&$-$\\
~&TRLRF\cite{TRLRF}&\;\;\; 18.010&0.449&0.688 \;\;\;&23.255&0.738&0.476 \;\;\;&26.211&0.845&0.356 \;\;\;&31.150&0.948&0.218 \;\;\;&32.259&0.958&0.197&126\\
~&TNN\cite{TNN_LRTC}&\;\;\; 19.976&0.584&0.580 \;\;\;&23.284&0.773&0.434 \;\;\;&26.004&0.866&0.344 \;\;\;&28.283&0.916&0.278 \;\;\;&30.230&0.944&0.230&72\\
~&FTNN\cite{FTNN}&\;\;\; 20.958&0.694&0.404 \;\;\;&25.168&0.860&0.274 \;\;\;&28.468&0.927&0.209 \;\;\;&31.023&\underline{0.957}&0.167 \;\;\;&33.223&\underline{0.973}&0.136&241\\
~&SSNT&\;\;\; \underline{24.218}&\bf{0.846}&\underline{0.261} \;\;\;&\underline{30.815}&\underline{0.963}&\underline{0.127} \;\;\;&\underline{34.798}&\underline{0.983}&\underline{0.093} \;\;\;&\underline{38.080}&\bf{0.991}&\bf{0.072} \;\;\;&\underline{40.276}&\bf{0.994}&\underline{0.061}&310\\
~&SSNT-TV&\;\;\; \bf{24.594}&\underline{0.839}&\bf{0.202} \;\;\;&\bf{31.419}&\bf{0.968}&\bf{0.121} \;\;\;&\bf{35.380}&\bf{0.986}&\bf{0.087} \;\;\;&\bf{38.280}&\bf{0.991}&\underline{0.074} \;\;\;&\bf{40.508}&\bf{0.994}&\bf{0.060}&424\\
\midrule
\multirow{6}*{\tabincell{c}{
MRI {\it Brain}\\{(181$\times$217$\times$181)}\\}}
~&Observed&\;\;\; 5.868&0.107&1.352 \;\;\;&6.103&0.134&1.253 \;\;\;&6.352&0.161&1.176 \;\;\;&6.615&0.188&1.109 \;\;\;&6.895&0.215&1.049&$-$\\
~&TRLRF\cite{TRLRF}&\;\;\; 19.228&0.554&0.379 \;\;\;&20.554&0.635&0.333 \;\;\;&21.109&0.672&0.311 \;\;\;&21.525&0.701&0.294 \;\;\;&21.861&0.721&0.283&1329\\
~&TNN\cite{TNN_LRTC}&\;\;\; 19.122&0.550&0.344 \;\;\;&21.221&0.655&0.296 \;\;\;&22.730&0.720&0.266 \;\;\;&23.977&0.768&0.242 \;\;\;&25.052&0.803&0.223&358\\
~&FTNN\cite{FTNN}&\;\;\; 20.182&0.681&0.271 \;\;\;&22.529&0.769&0.230 \;\;\;&24.064&0.815&0.206 \;\;\;&25.289&0.847&0.189 \;\;\;&26.337&0.870&0.175&1307\\
~&SSNT&\;\;\; \underline{20.680}&\underline{0.672}&\underline{0.246} \;\;\;&\underline{22.830}&\underline{0.753}&\underline{0.211} \;\;\;&\underline{24.766}&\underline{0.821}&\underline{0.186} \;\;\;&\underline{25.952}&\underline{0.851}&\underline{0.173} \;\;\;&\underline{26.891}&\underline{0.872}&\underline{0.160}&628\\
~&SSNT-TV&\;\;\; \bf{22.695}&\bf{0.769}&\bf{0.211} \;\;\;&\bf{23.903}&\bf{0.805}&\bf{0.192} \;\;\;&\bf{24.882}&\bf{0.825}&\bf{0.182} \;\;\;&\bf{26.019}&\bf{0.853}&\bf{0.167} \;\;\;&\bf{26.961}&\bf{0.873}&\bf{0.156}&664\\
\bottomrule
\end{tabular}
\end{spacing}
\end{center}
\vspace{-0.3cm}
\end{table*}
\begin{figure*}[!h]
\tiny
\setlength{\tabcolsep}{0.9pt}
\begin{center}
\begin{tabular}{c}
\vspace{-0.2cm}
\includegraphics[width=0.9\textwidth]{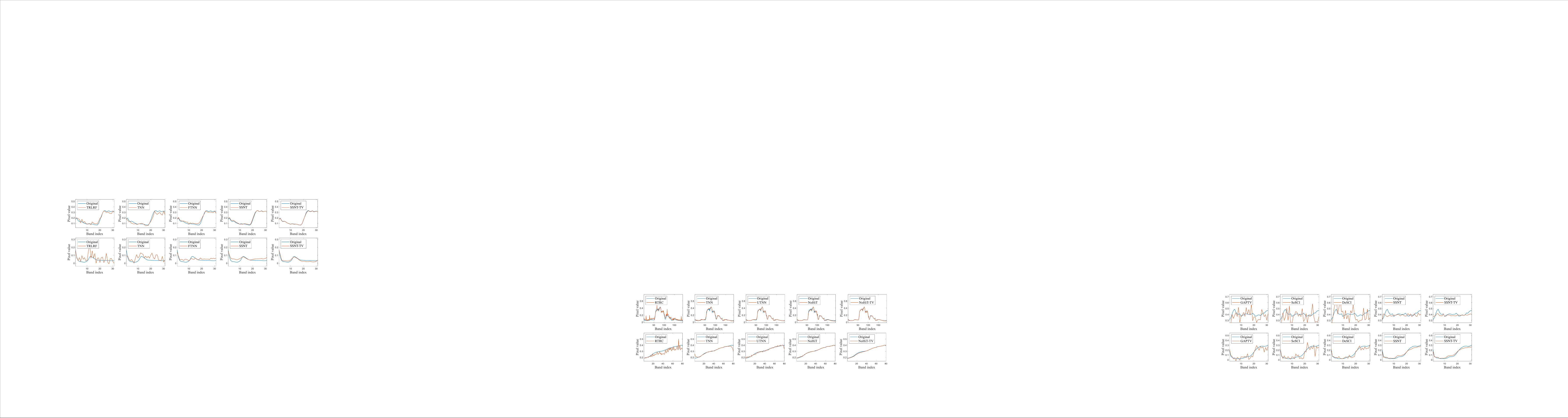}\\
\end{tabular}
\end{center}
\caption{The spectral curves of the tensor completion results by different methods on MSIs {\it Balloons} and {\it Beads} with SR = 0.05.\label{TC_spec}}\vspace{-0.2cm} 
\end{figure*}
\begin{figure*}[!t]
\footnotesize
\setlength{\tabcolsep}{0.9pt}
\begin{center}
\begin{tabular}{ccccccc}
\includegraphics[width=0.139\textwidth]{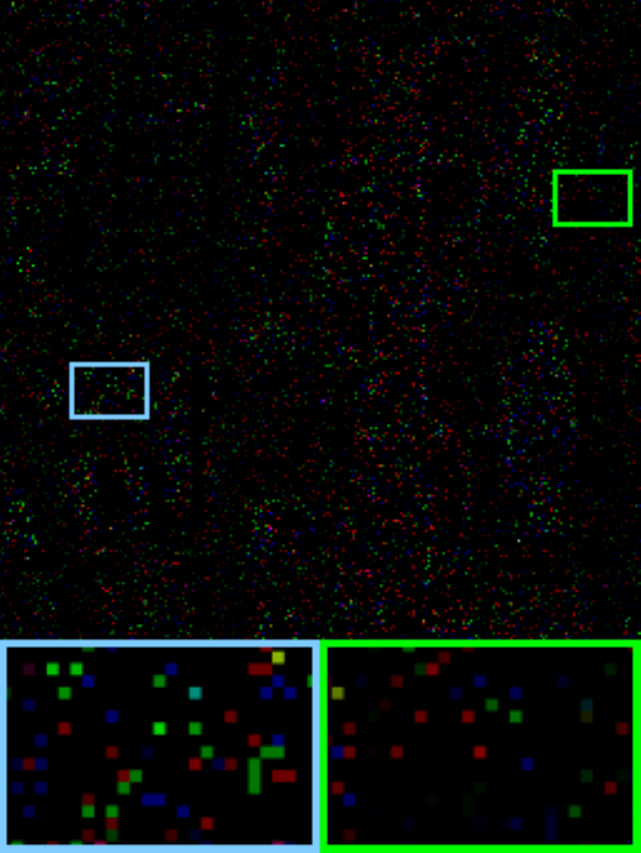}&
\includegraphics[width=0.139\textwidth]{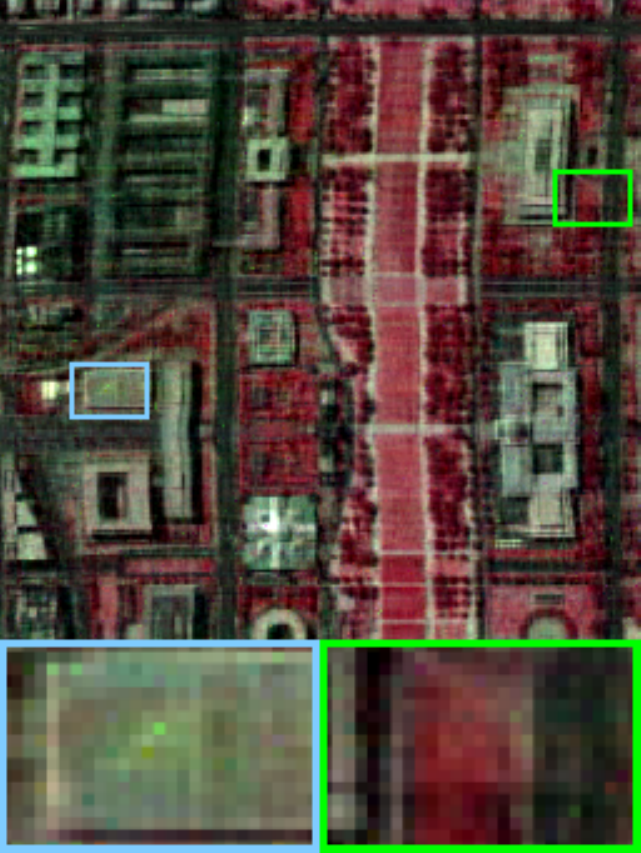}&
\includegraphics[width=0.139\textwidth]{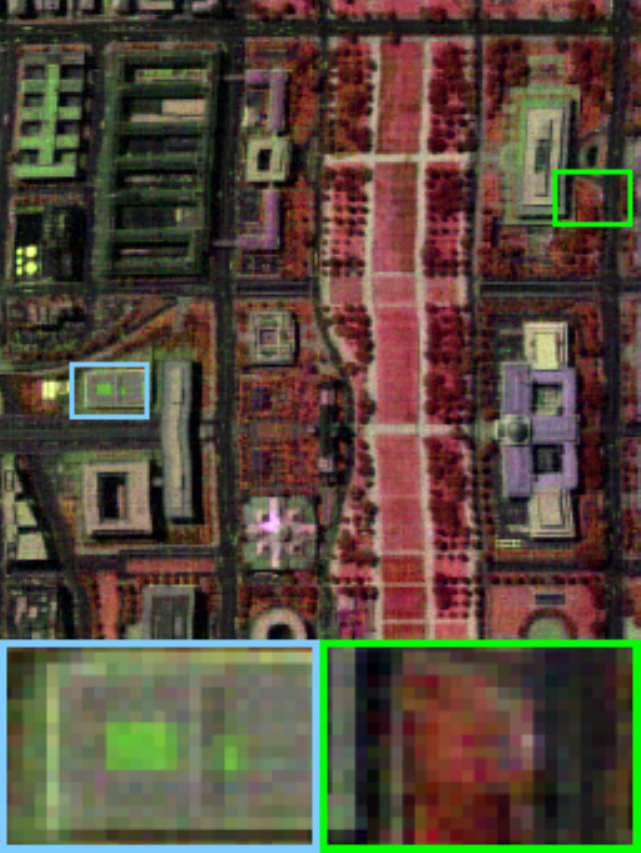}&
\includegraphics[width=0.139\textwidth]{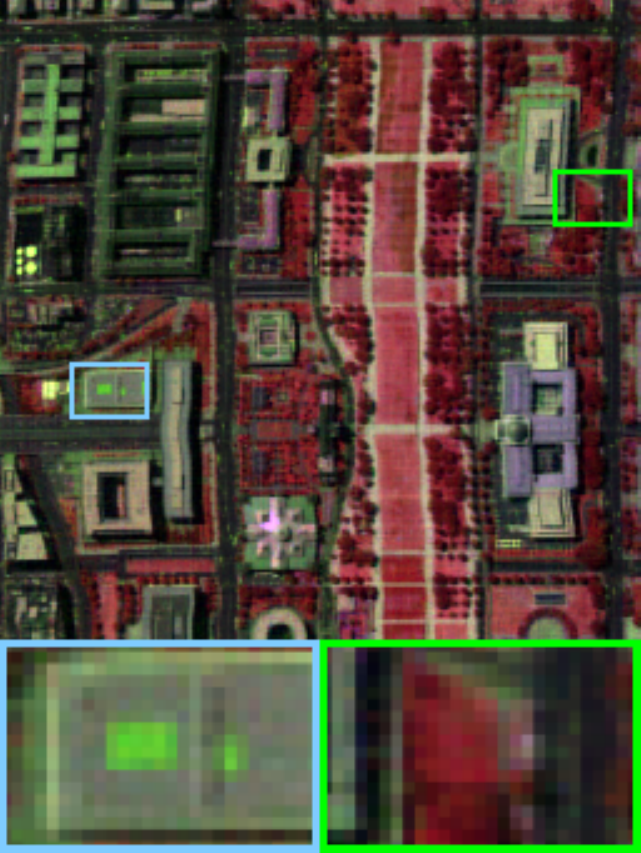}&
\includegraphics[width=0.139\textwidth]{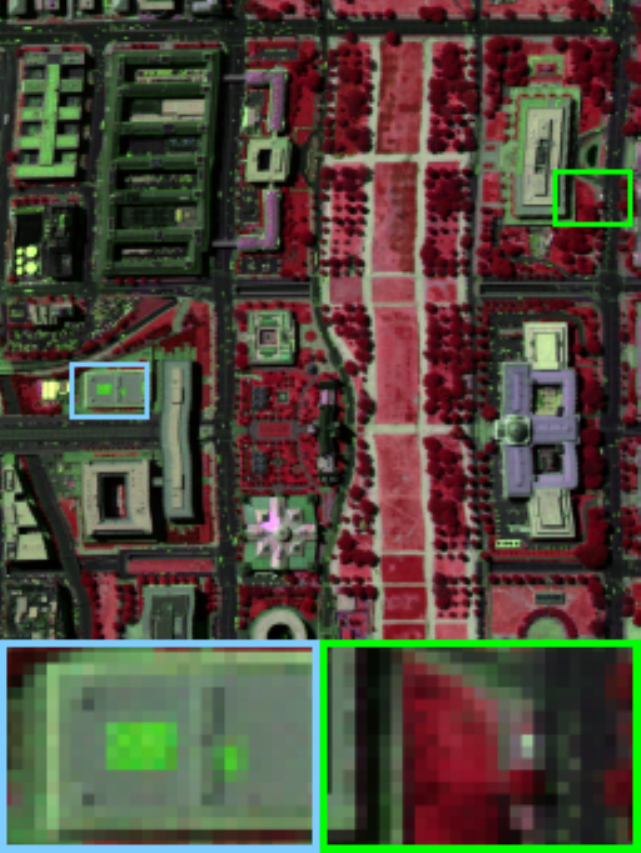}&
\includegraphics[width=0.139\textwidth]{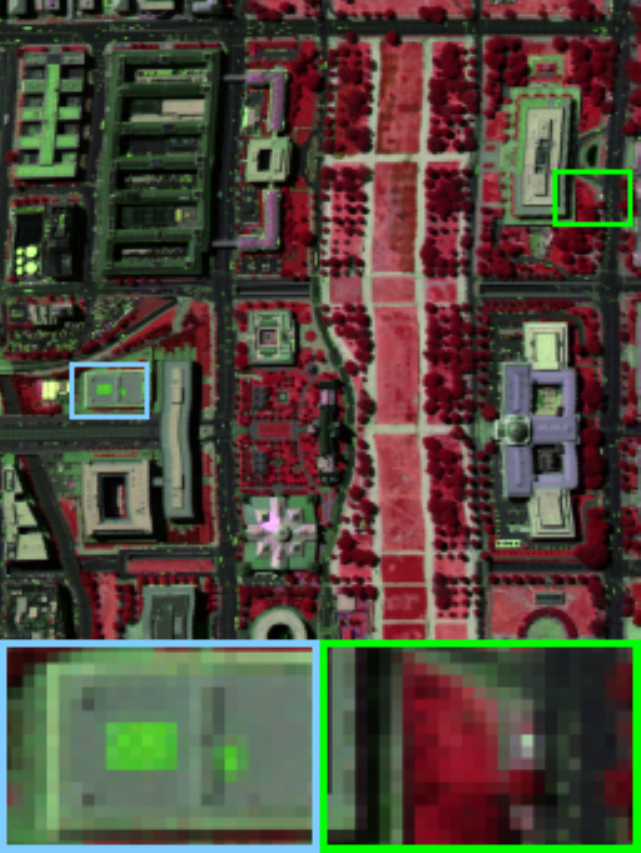}&
\includegraphics[width=0.139\textwidth]{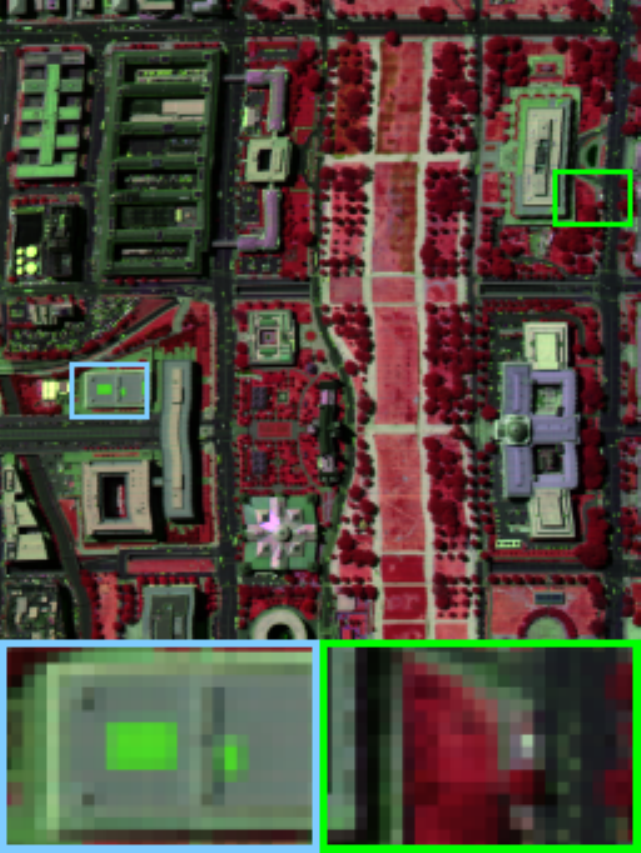}\\
\vspace{0.1cm}
PSNR 14.567 dB & PSNR 27.044 dB&PSNR 29.513 dB&PSNR 32.776 dB&PSNR 40.118 dB&PSNR 41.155 dB&PSNR Inf\\
\includegraphics[width=0.139\textwidth]{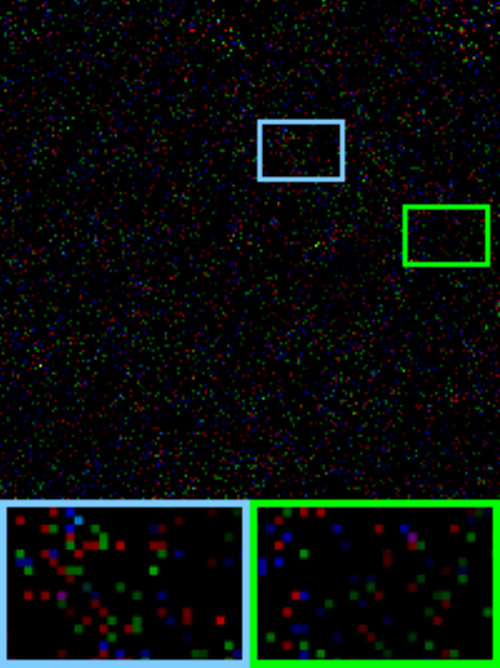}&
\includegraphics[width=0.139\textwidth]{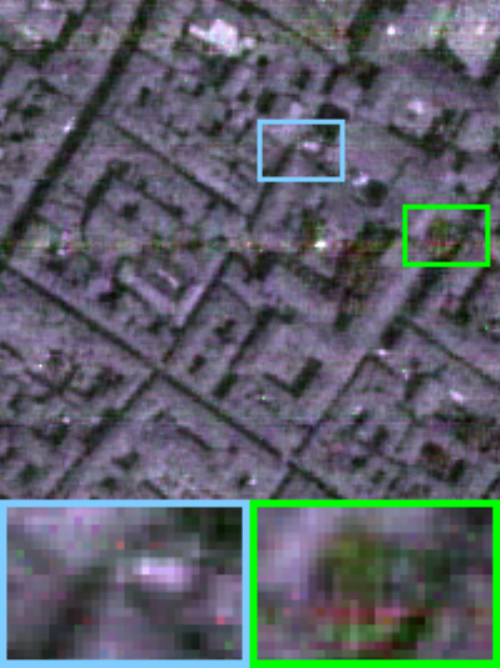}&
\includegraphics[width=0.139\textwidth]{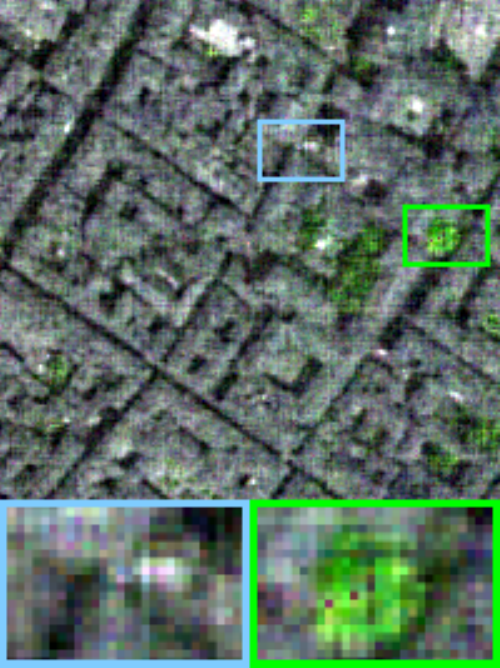}&
\includegraphics[width=0.139\textwidth]{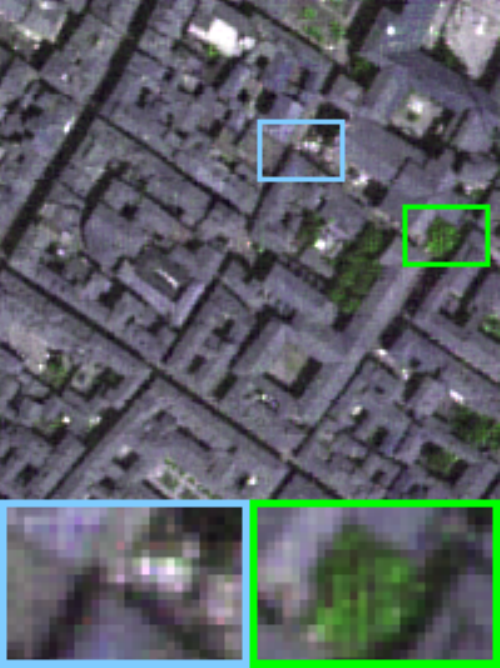}&
\includegraphics[width=0.139\textwidth]{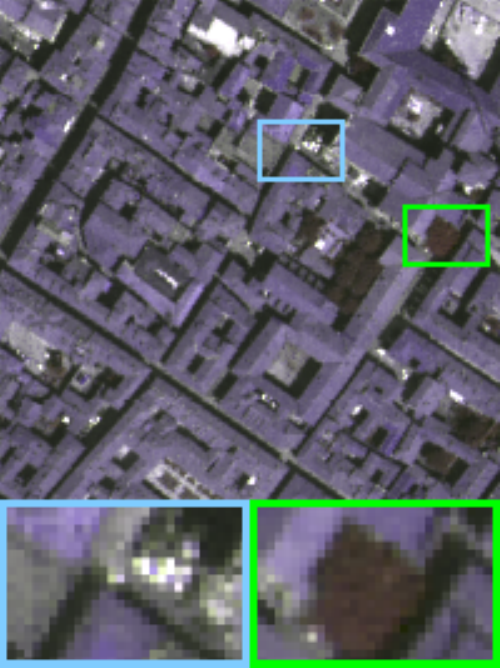}&
\includegraphics[width=0.139\textwidth]{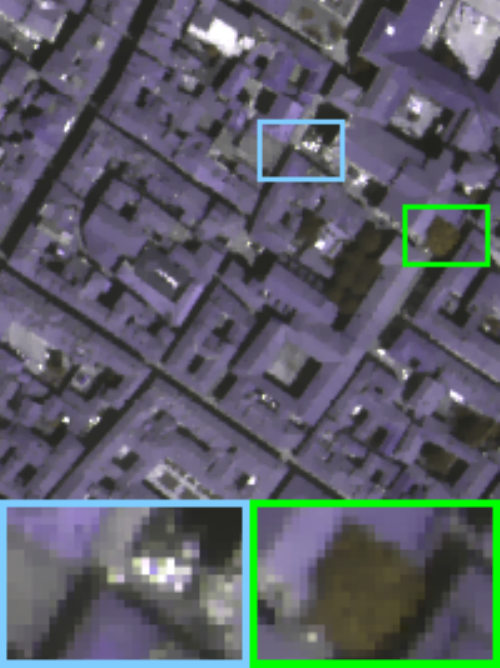}&
\includegraphics[width=0.139\textwidth]{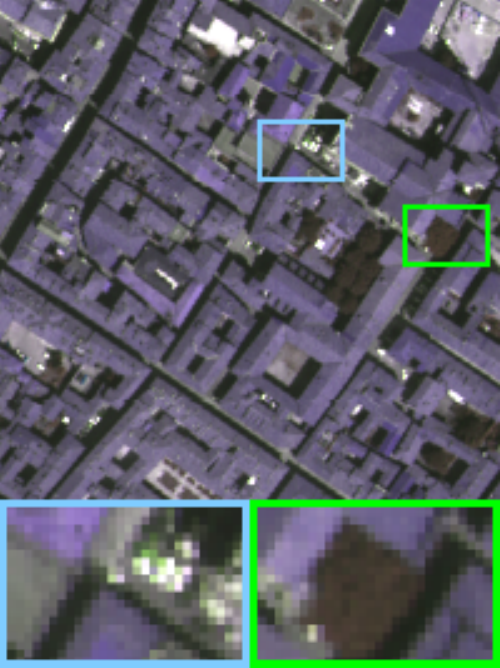}\\
\vspace{0.1cm}
PSNR 12.191 dB & PSNR 28.232 dB&PSNR 26.002 dB&PSNR 32.345 dB&PSNR 38.755 dB&PSNR 38.837 dB&PSNR Inf\\
\includegraphics[width=0.139\textwidth]{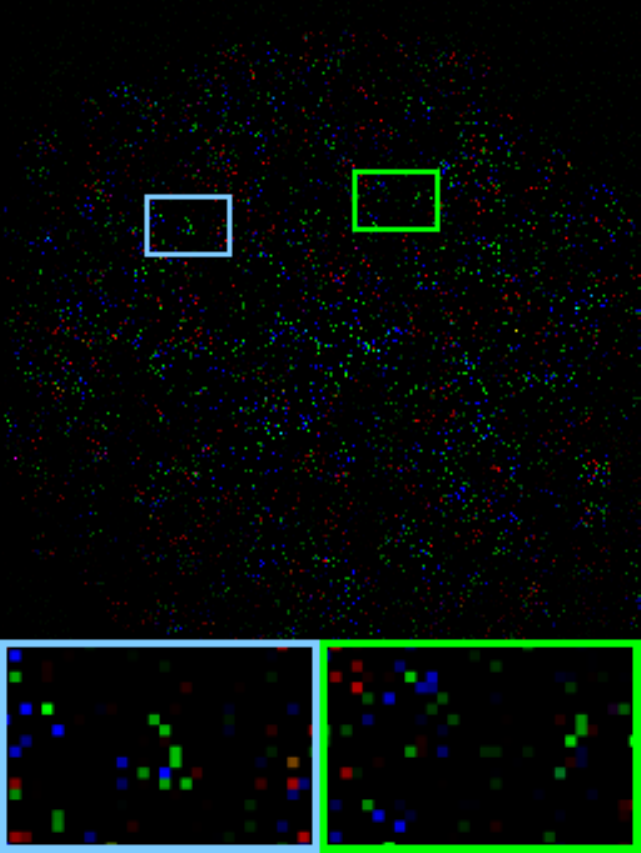}&
\includegraphics[width=0.139\textwidth]{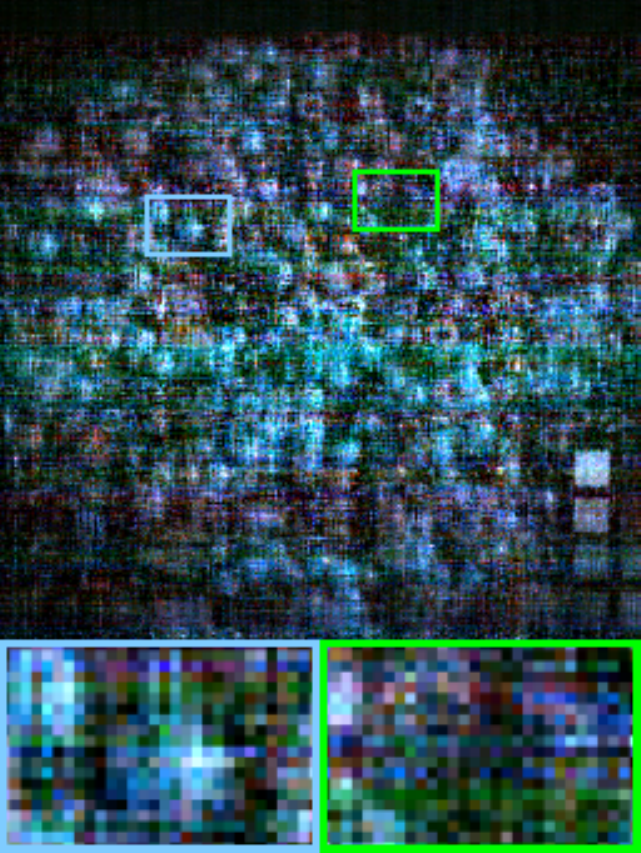}&
\includegraphics[width=0.139\textwidth]{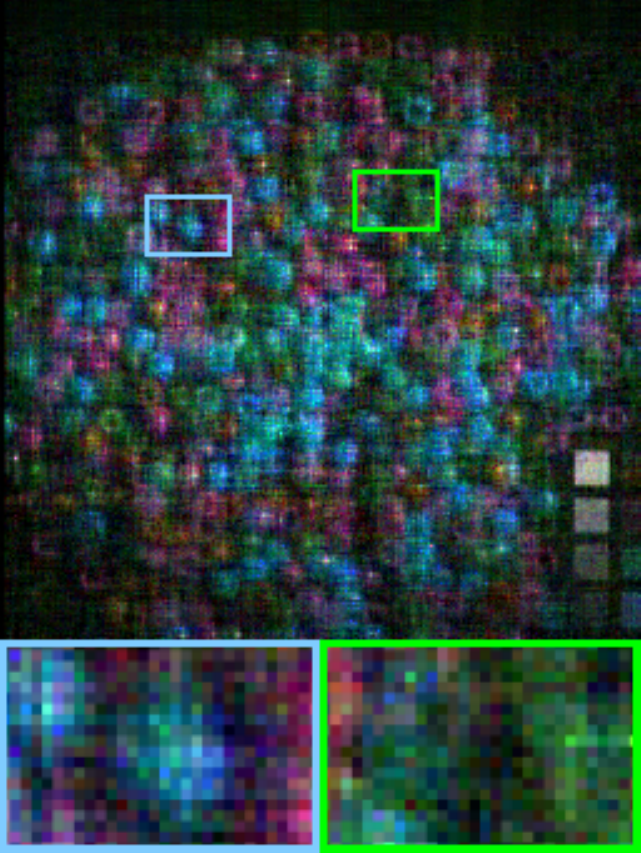}&
\includegraphics[width=0.139\textwidth]{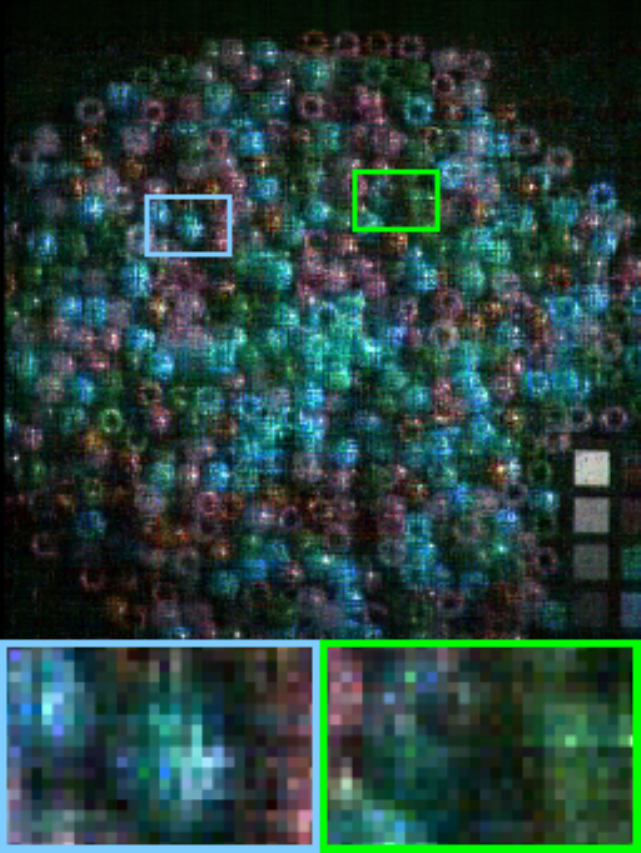}&
\includegraphics[width=0.139\textwidth]{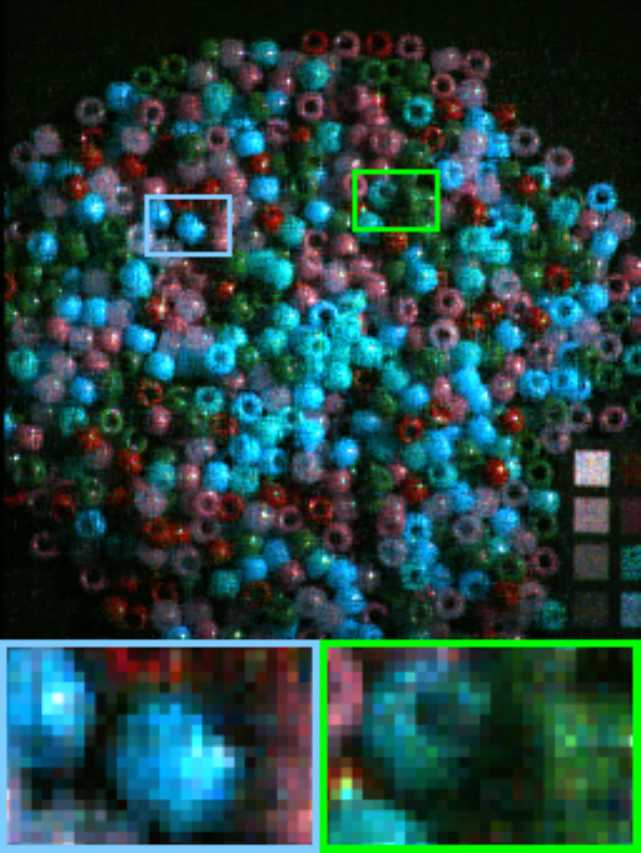}&
\includegraphics[width=0.139\textwidth]{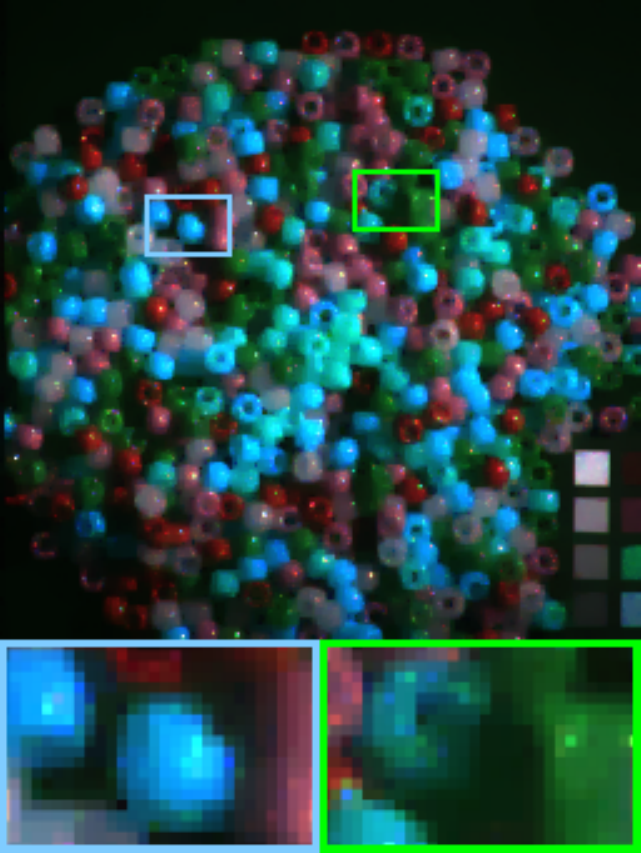}&
\includegraphics[width=0.139\textwidth]{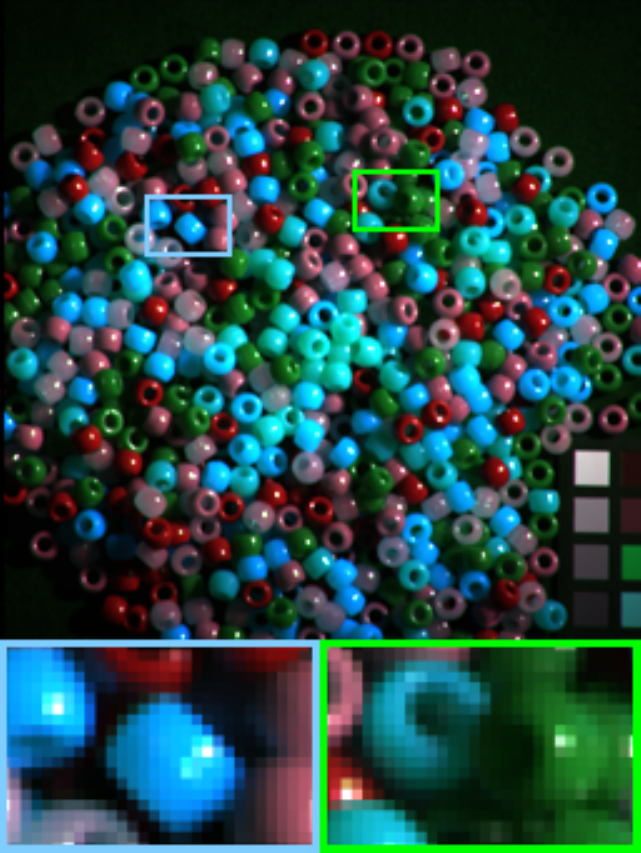}\\
\vspace{0.1cm}
PSNR 14.414 dB & PSNR 18.010 dB&PSNR 19.976 dB&PSNR 20.958 dB&PSNR 24.218 dB&PSNR 24.594 dB&PSNR Inf\\
\includegraphics[width=0.139\textwidth]{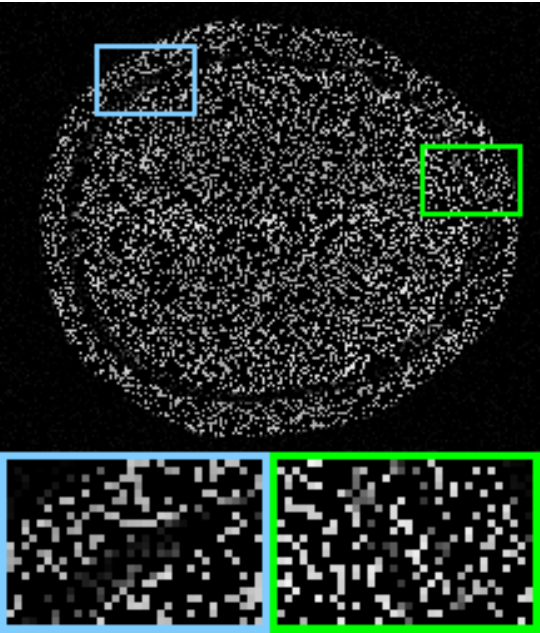}&
\includegraphics[width=0.139\textwidth]{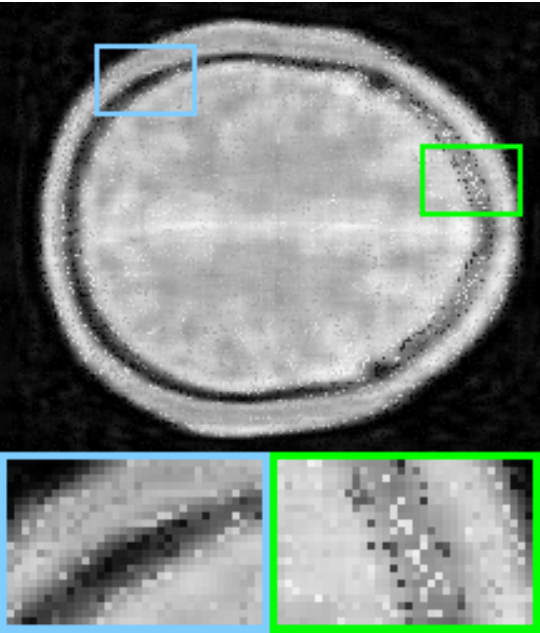}&
\includegraphics[width=0.139\textwidth]{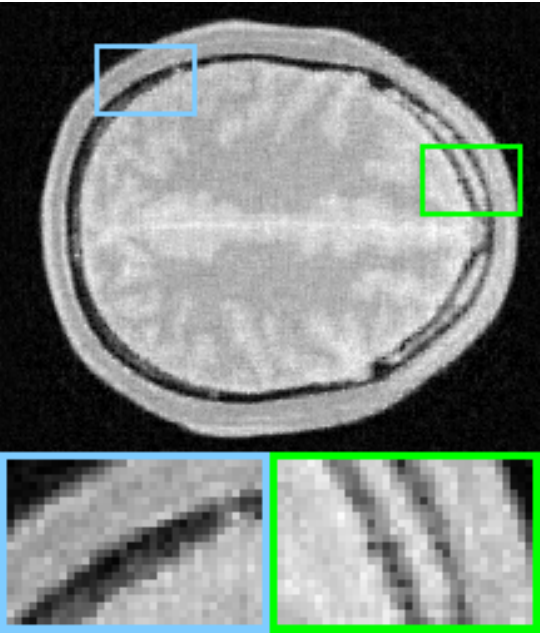}&
\includegraphics[width=0.139\textwidth]{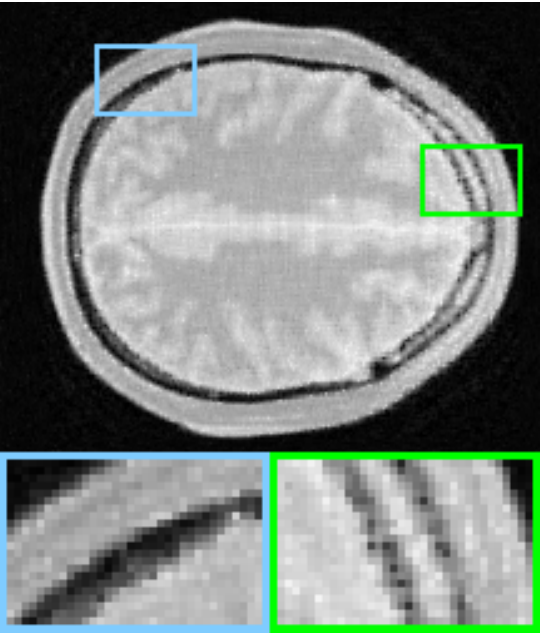}&
\includegraphics[width=0.139\textwidth]{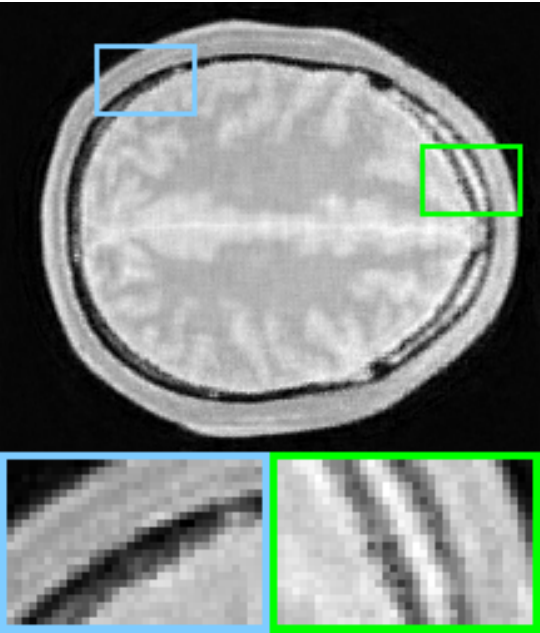}&
\includegraphics[width=0.139\textwidth]{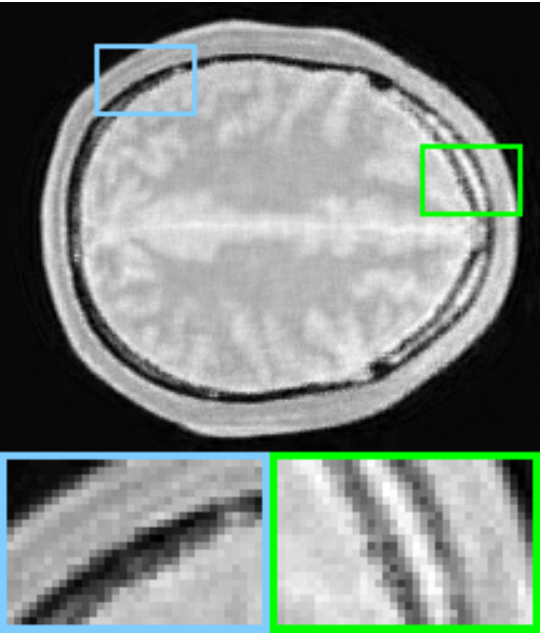}&
\includegraphics[width=0.139\textwidth]{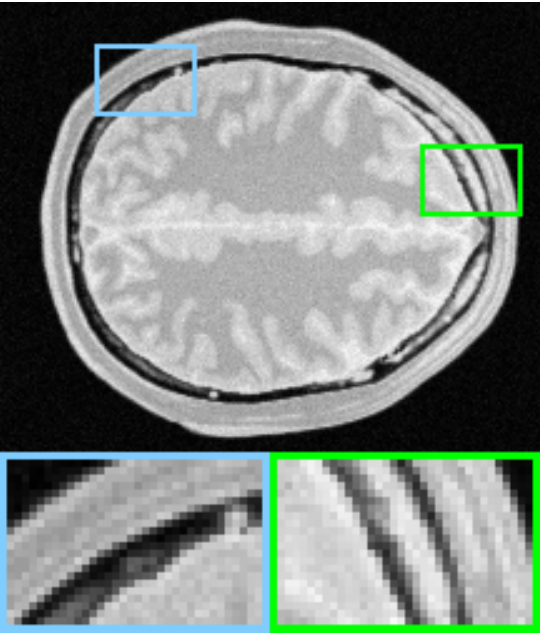}\\
PSNR 6.895 dB & PSNR 21.861 dB&PSNR 25.052 dB&PSNR 26.337 dB&PSNR 26.891 dB&PSNR 26.961 dB&PSNR Inf\\
\vspace{-0.3cm}
Observed & TRLRF \cite{TRLRF} &TNN \cite{TNN_LRTC}&FTNN \cite{FTNN}&SSNT&SSNT-TV&Original\\
\end{tabular}
\end{center}
\caption{The recovered results by different methods for tensor completion on HSI {\it WDC mall} (composed of the 50-th, 100-th, and the 150-th bands) with SR = 0.05, HSI {\it Pavia} (composed of the 1-st,10-th, and the 20-th bands) with SR = 0.05, MSI {\it Beads} (composed of the 10-th, 20-th, and the 30-th bands) with SR = 0.05, and MRI {\it Brain} (the 50-th band) with SR = 0.25. \label{TC_fig}}\vspace{-0.3cm}
\end{figure*}
\subsubsection{Experimental Settings}
To illustrate the effectiveness of our method for tensor completion, we collected three types of multi-dimensional images including MSIs\footnote{\url{https://www.cs.columbia.edu/CAVE/databases/multispectral/}} \cite{CAVE} ({\it Balloons} and {\it Beads}), hyperspectral images (HSIs)\footnote{\url{https://engineering.purdue.edu/biehl/MultiSpec/hyperspectral.html}} ({\it Pavia} and {\it WDC mall}), and MRI\footnote{\url{https://brainweb.bic.mni.mcgill.ca/brainweb}} ({\it Brain}). Five cases with sampling rate (SR) $0.05,0.1,0.15,0.2,0.25$ are established.\par
The competing methods for tensor completion are: The tensor ring decomposition-based method TRLRF \cite{TRLRF}, the linear transform-based methods TNN (induced by DFT) \cite{TNN_LRTC} and FTNN (induced by framelet transform) \cite{FTNN}. The hyperparameters of competing methods are set for their best performance. For the proposed method, we set $\lambda = N\times 10^{-7}$, $\tau = 0.01N$, $\beta=1$, $p=q=2$, ${\tilde n}_3=2n_3$, and $t_{max}=7000$, where $N = n_1n_2n_3$ is the total number of elements of the data. The initialization function ${\rm Init(\cdot)}$ for tensor completion is the linear interpolation that used in \cite{DTNN}, which provides an ideal initialization with less time. \par
We use three numerical evaluation indices: peak signal to noise ratio (PSNR), structural similarity (SSIM), and spectral angle mapper (SAM) \cite{sam}. Higher PSNR and SSIM values correspond to better quality, while lower SAM value represents a smaller spectral angle between the ground truth and the recovered result.\par
\subsubsection{Experimental Results}
The numerical results for tensor completion are illustrated in Table \ref{TC_tab}. We can see that the proposed SSNT could achieve better PSNR and SSIM values than competing methods, which verifies that SSNT recovers the low-rank tensor more precisely. Also, SSNT achieves better SAM values, which shows that SSNT preferably exploits the correlation along the third mode. We can see that SSNT-TV outperforms SSNT from the perspective of PSNR, SSIM, and SAM, which shows the effectiveness of TV regularization to enhance the spatial recovered quality.\par
Some visual examples of the results for tensor completion are shown in Fig. \ref{TC_fig}. We can see that SSNT and SSNT-TV recover the images better than competing methods. SSNT-TV has better recovery in the spatial domain, especially according to the results on {\it Beads}. This is due to the consideration of the spatial local smoothness delivered by the spatial TV regularization.\par
In addition, we plot the spectral curves of the recovered results in Fig. \ref{TC_spec}. We can see that SSNT and SSNT-TV more faithfully capture the nonlinear nature of spectral curves, which shows the nonlinear modeling capability of SSNT.
\subsection{Background Subtraction}
The background subtraction aims at subtracting low-rank background from the original video, which is formulated as
\begin{equation}
\min_{{\mathcal X},{\mathcal S}}\;\lambda\;rank({\mathcal X})+\lVert{\mathcal S}\rVert_{\ell_1},\;\;{\rm s.t.}\;\;{\mathcal X}+{\mathcal S} = {\mathcal O},
\end{equation}
where $\mathcal O$ is the original video, $\mathcal X$ is the desired low-rank component, and $\mathcal S$ is the sparse foreground. The fidelity term of the proposed model for background subtraction is
\begin{equation}
L({\mathcal X},{\mathcal O}) = \lVert{\mathcal X} - {\mathcal O}\rVert_{\ell_1}. 
\end{equation} 
\subsubsection{Experimental Settings}
Five video frames\footnote{\url{http://trace.eas.asu.edu/yuv/} and \url{http://jacarini.dinf.usherbrooke.ca/static/dataset/}} that contain low-rank background and sparse foreground are selected. \par
The competing methods for the background subtraction are: The variational-based matrix robust principal component analysis method FastRPCA \cite{MRPCA}, the linear transform-based methods TNN \cite{TNN_TRPCA} and DCTNN (induced by DCT) \cite{DCTNN}. For our hyperparameters, we set $\lambda = N\times 10^{-3}$. We directly use the original tensor as the input of SSNT and SSNT-TV, i.e., there is no initialization for background subtraction.
\subsubsection{Experimental Results}
\begin{figure*}[!h]
\footnotesize
\setlength{\tabcolsep}{0.9pt}
\begin{center}
\begin{tabular}{cccccc}
\vspace{0.1cm}
\includegraphics[width=0.139\textwidth]{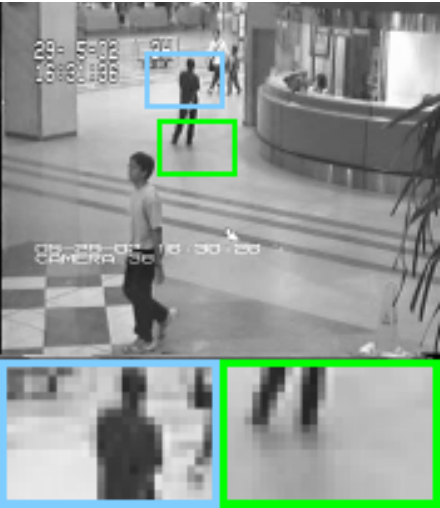}&
\includegraphics[width=0.139\textwidth]{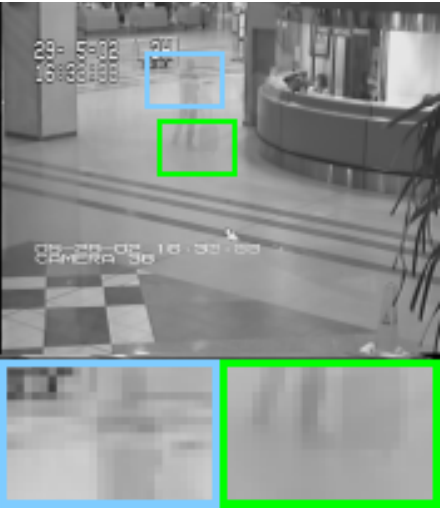}&
\includegraphics[width=0.139\textwidth]{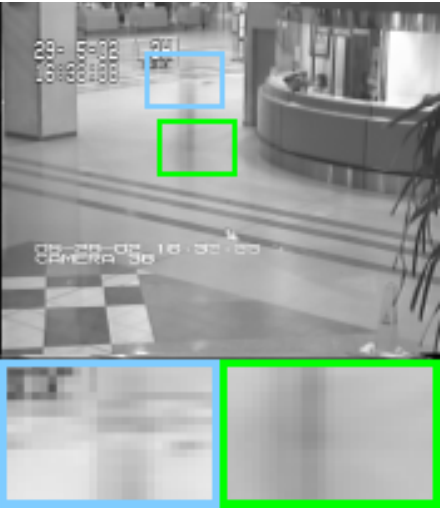}&
\includegraphics[width=0.139\textwidth]{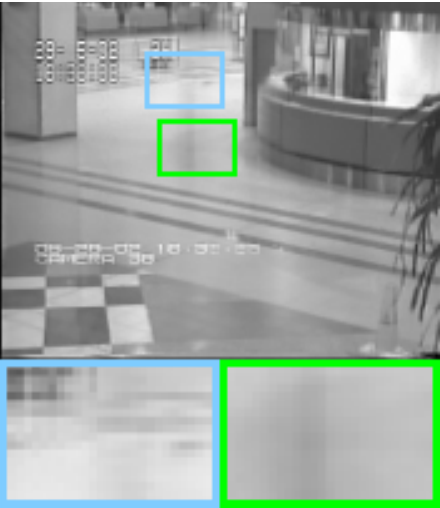}&
\includegraphics[width=0.139\textwidth]{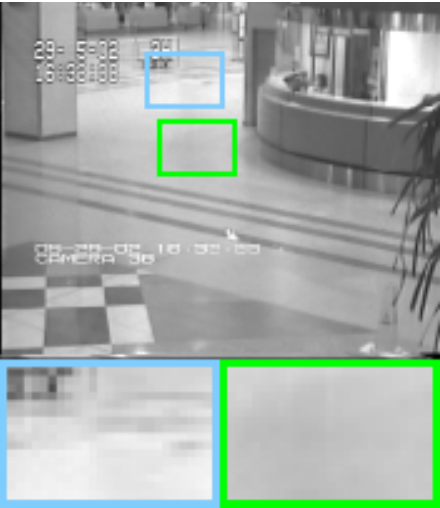}&
\includegraphics[width=0.139\textwidth]{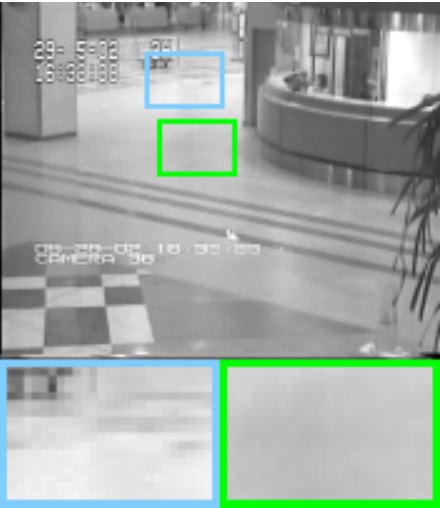}\\
\vspace{0.1cm}
\includegraphics[width=0.139\textwidth]{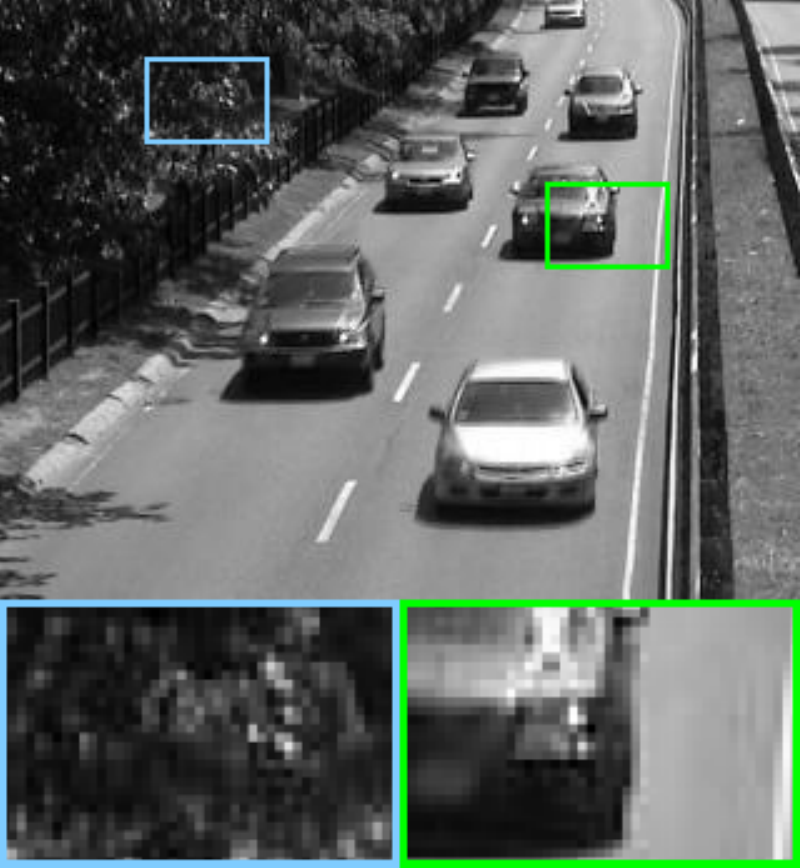}&
\includegraphics[width=0.139\textwidth]{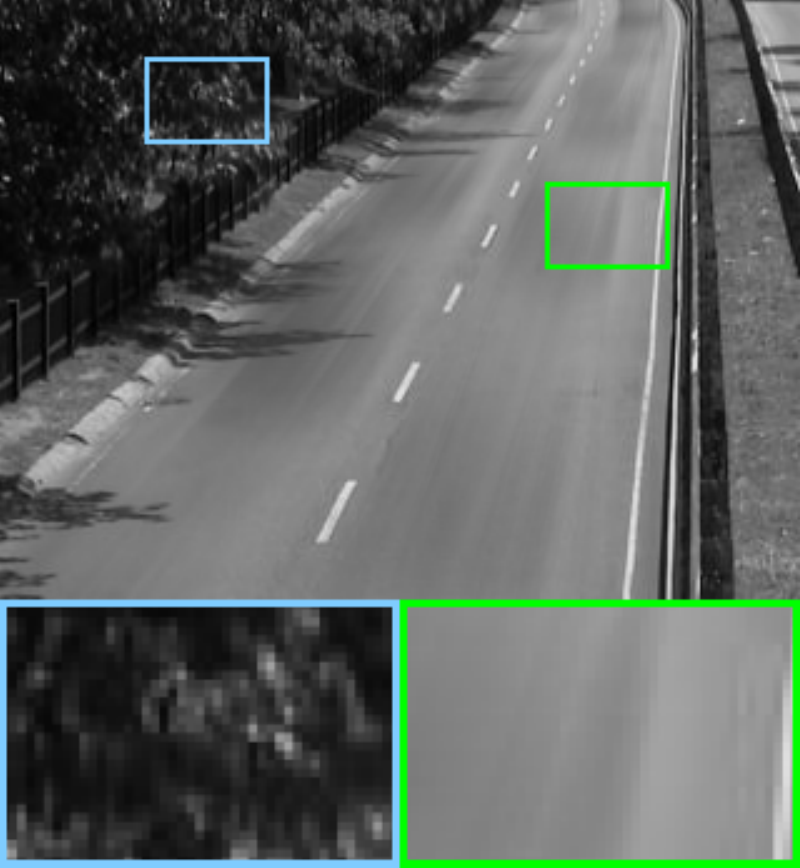}&
\includegraphics[width=0.139\textwidth]{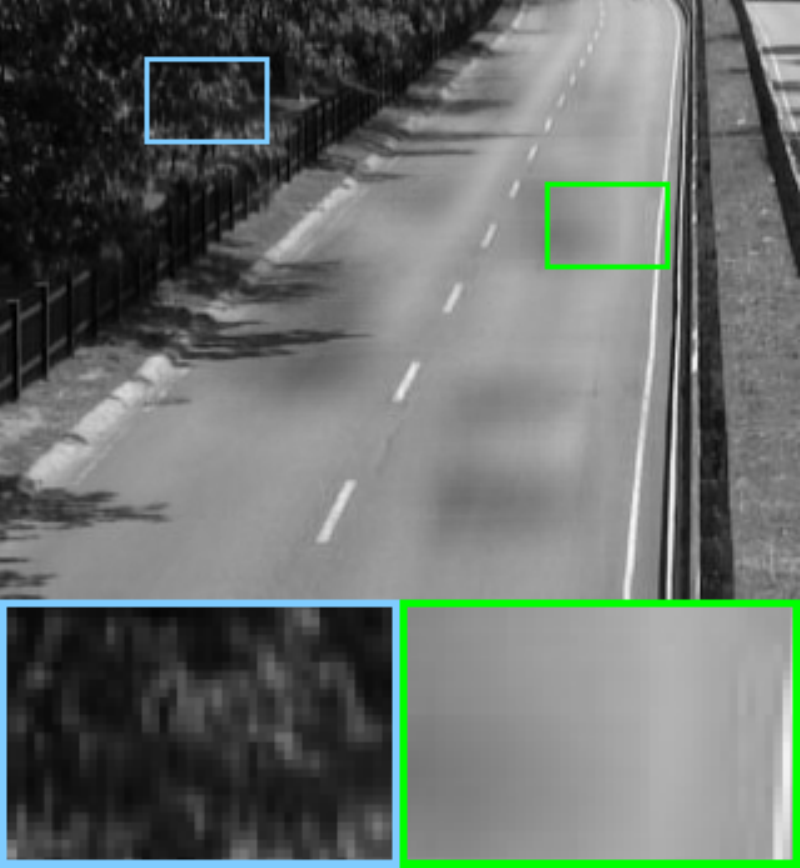}&
\includegraphics[width=0.139\textwidth]{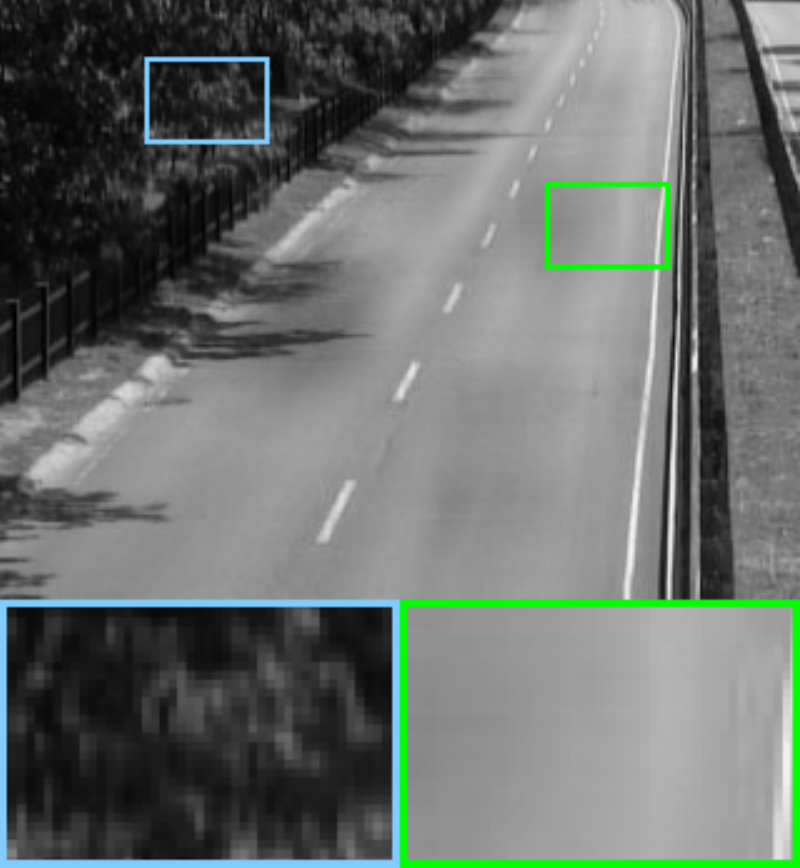}&
\includegraphics[width=0.139\textwidth]{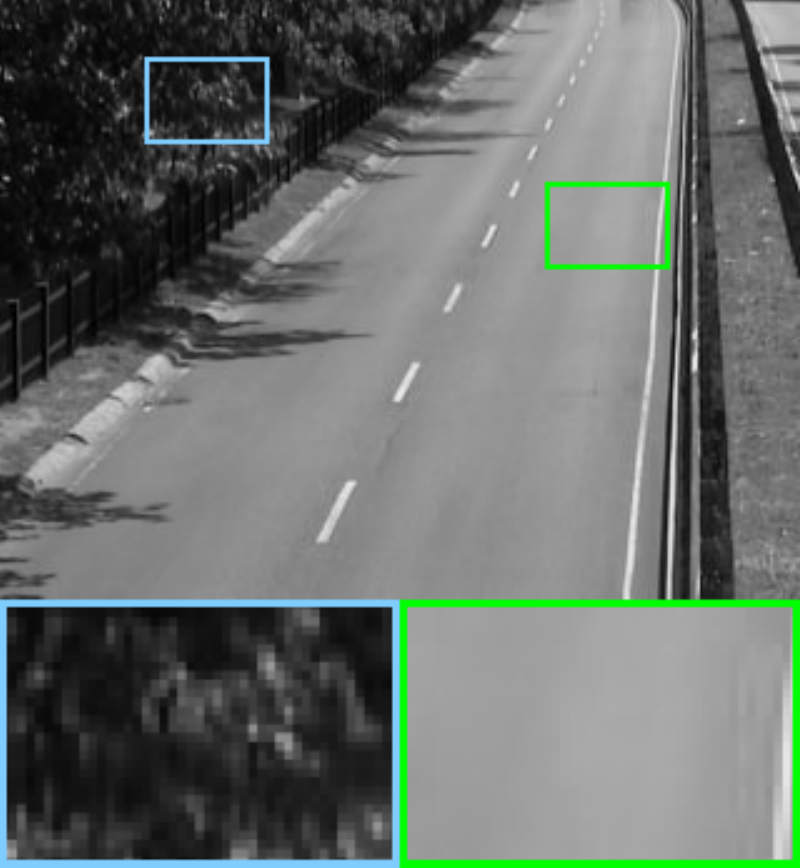}&
\includegraphics[width=0.139\textwidth]{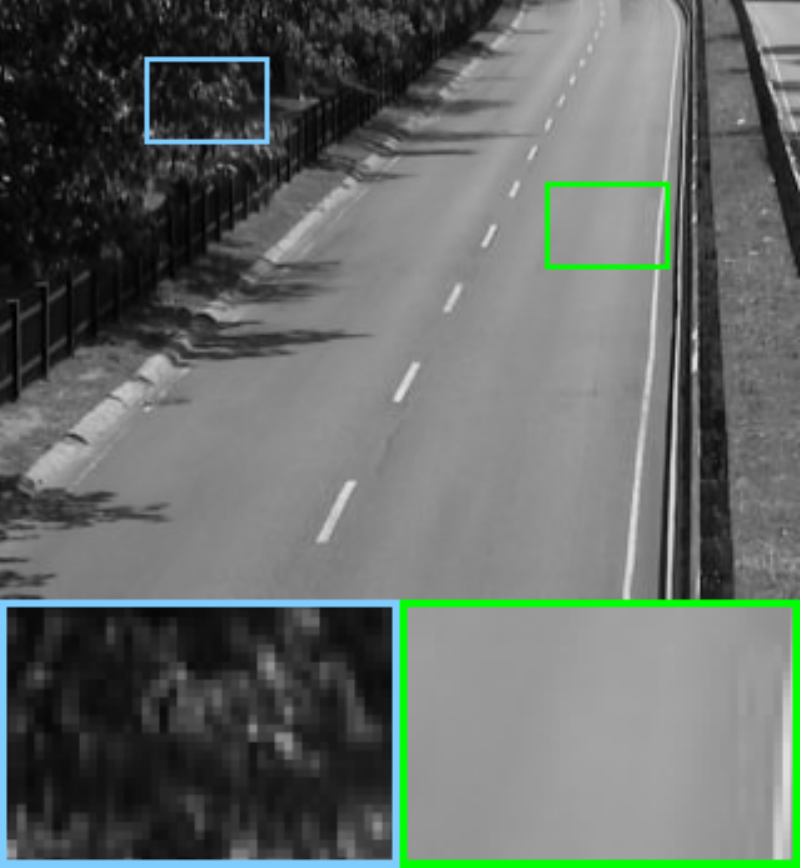}\\
\vspace{0.1cm}
\includegraphics[width=0.139\textwidth]{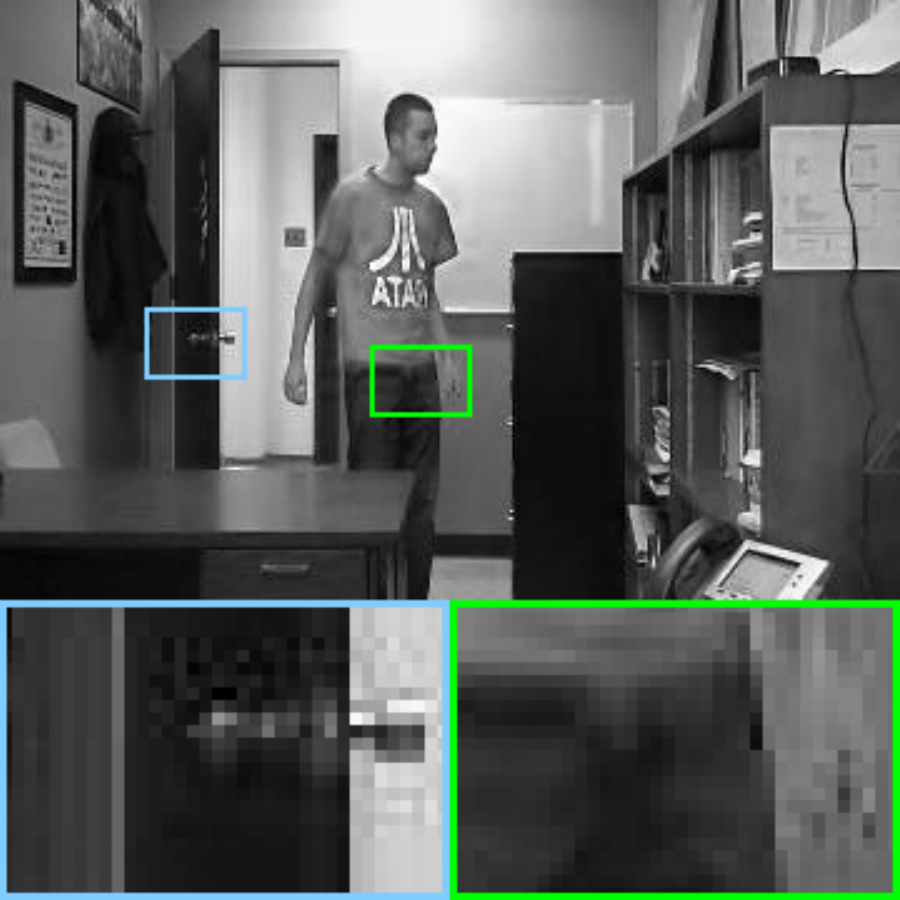}&
\includegraphics[width=0.139\textwidth]{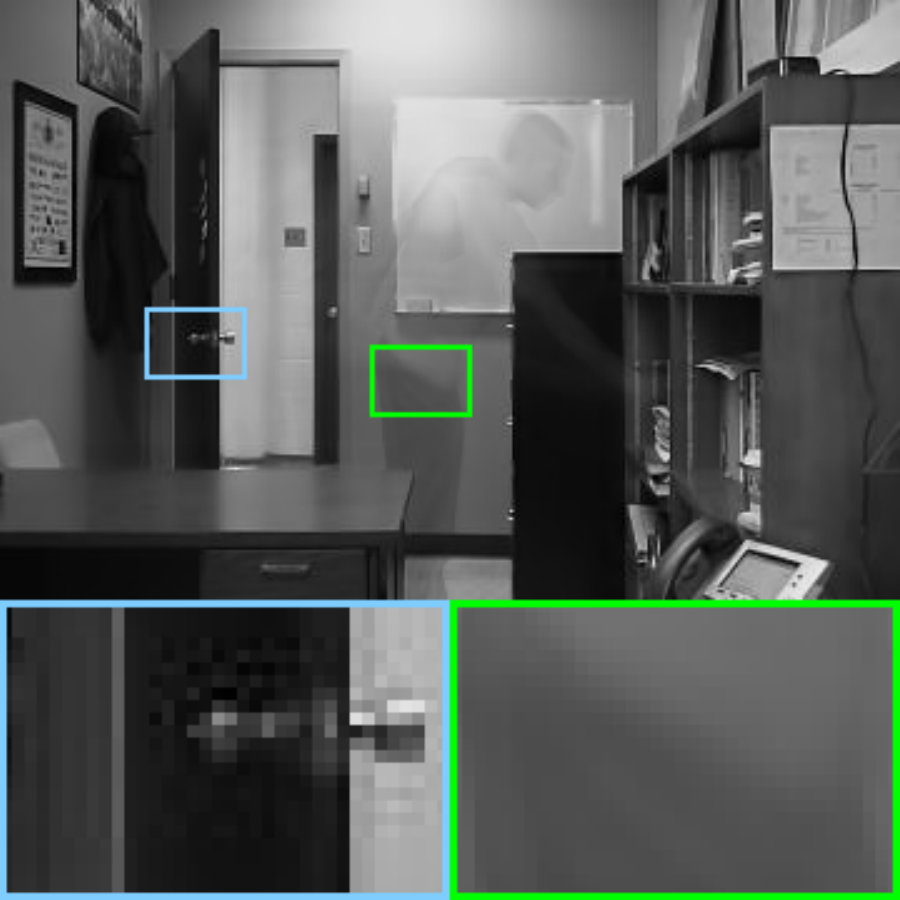}&
\includegraphics[width=0.139\textwidth]{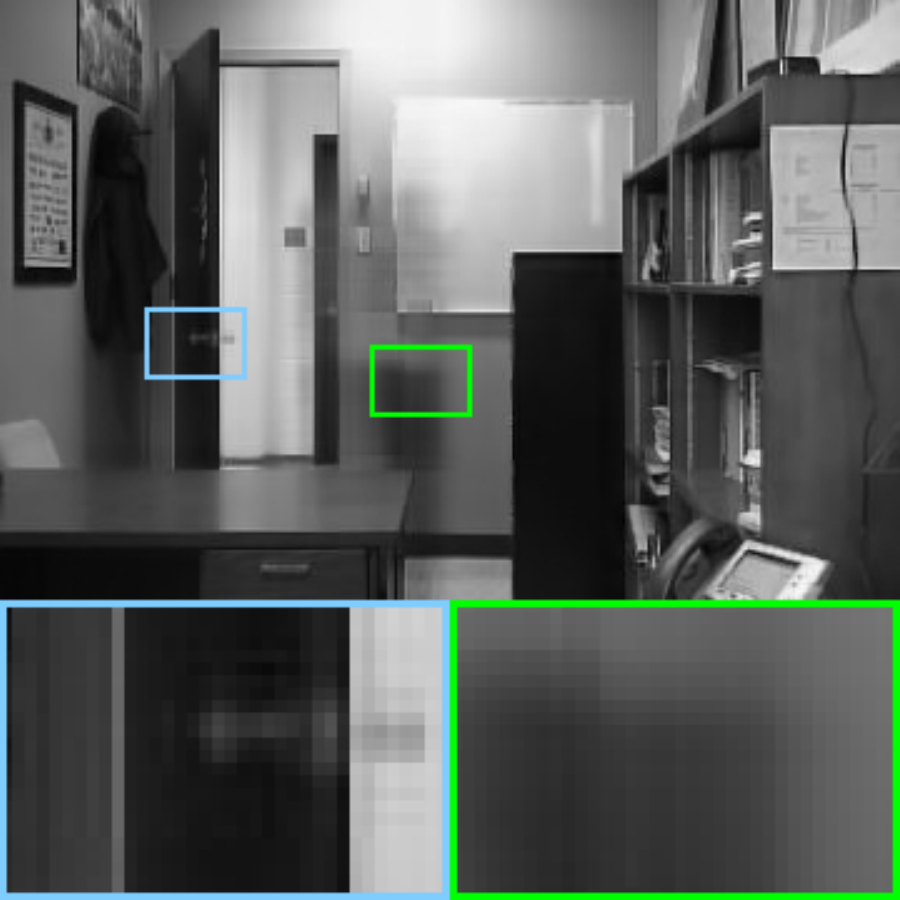}&
\includegraphics[width=0.139\textwidth]{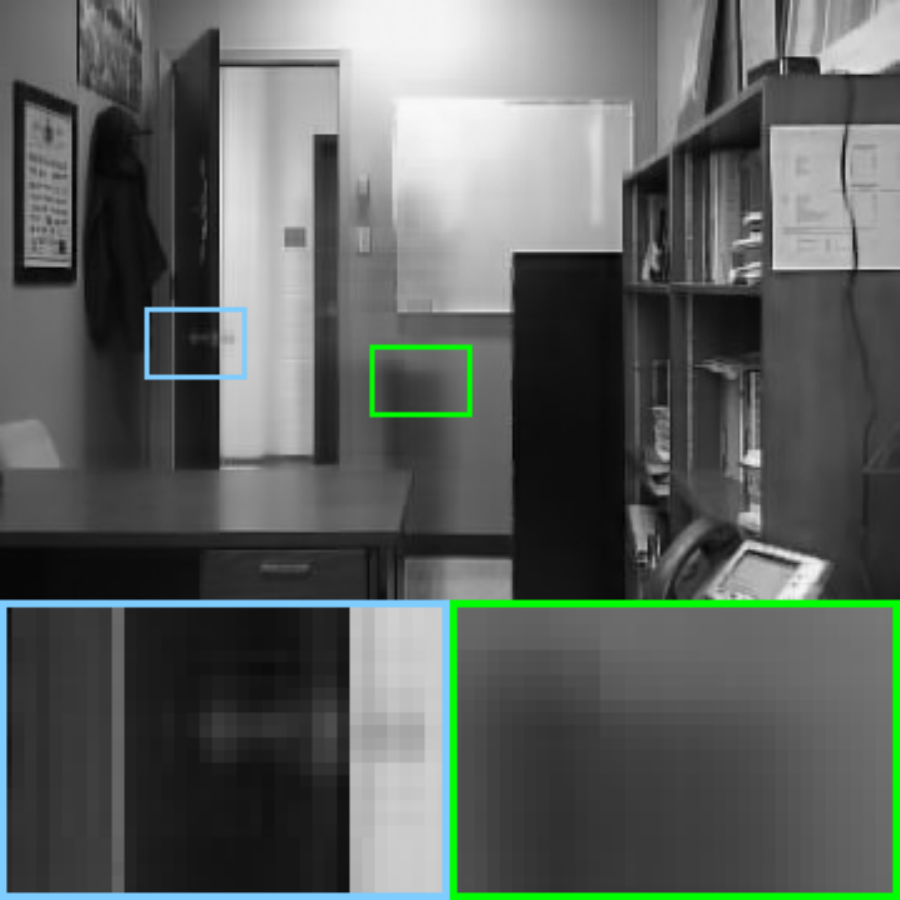}&
\includegraphics[width=0.139\textwidth]{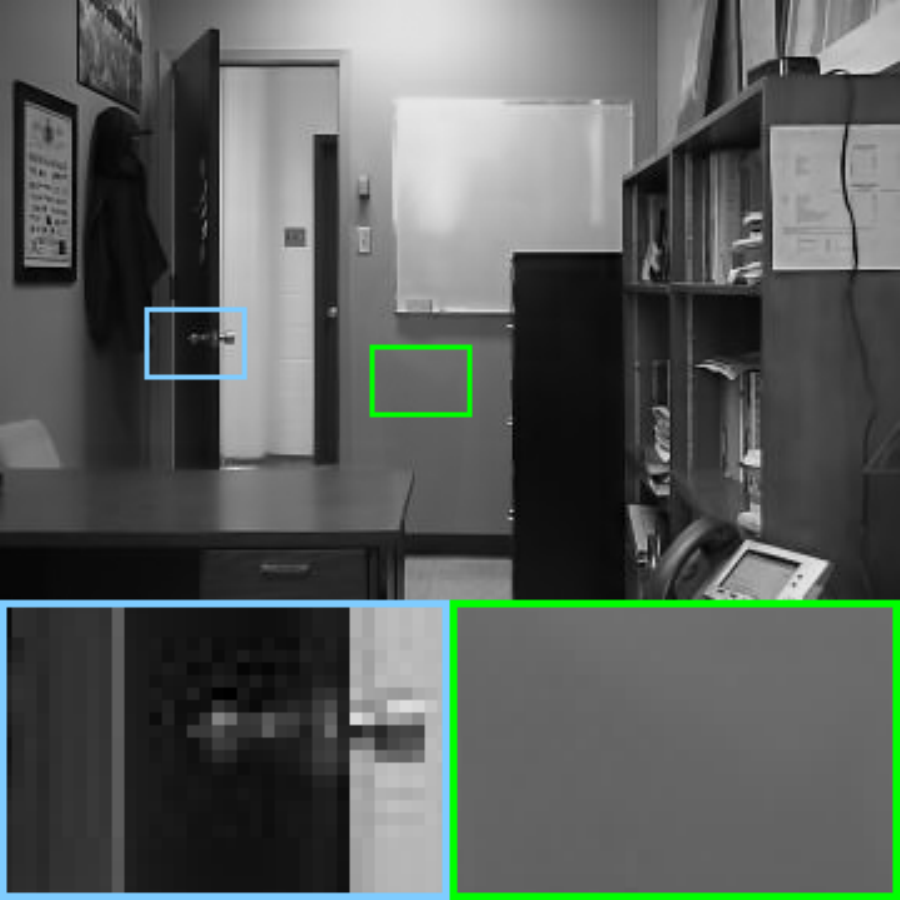}&
\includegraphics[width=0.139\textwidth]{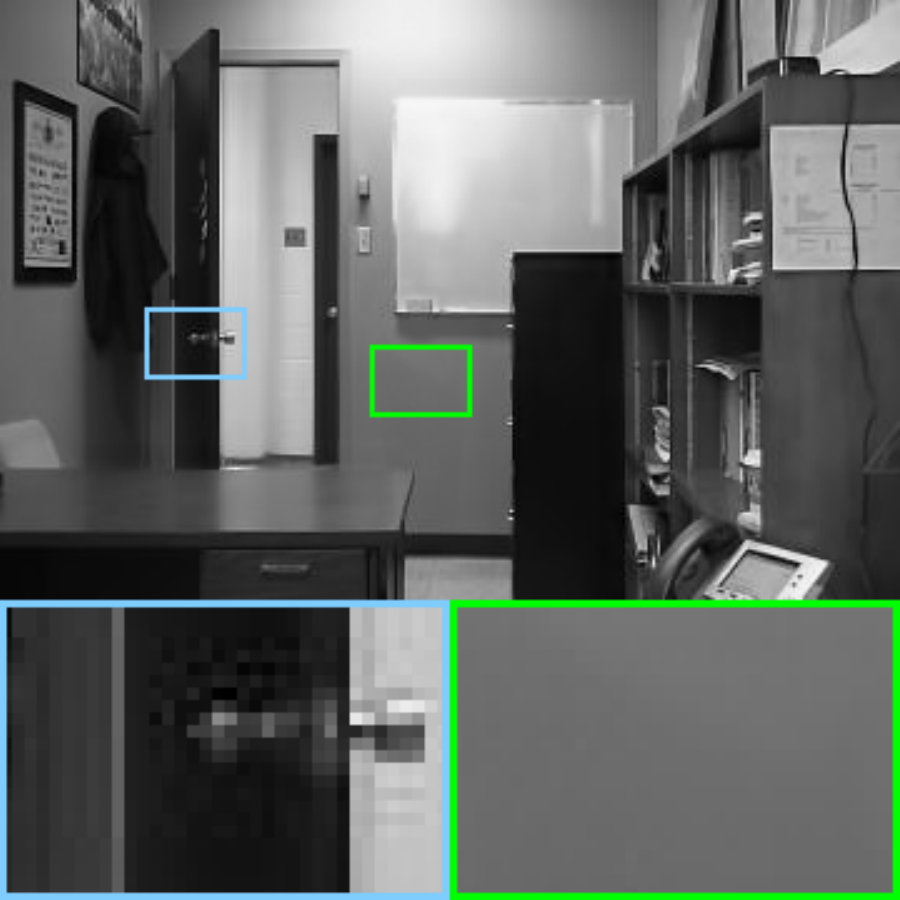}\\
\vspace{0.1cm}
\includegraphics[width=0.139\textwidth]{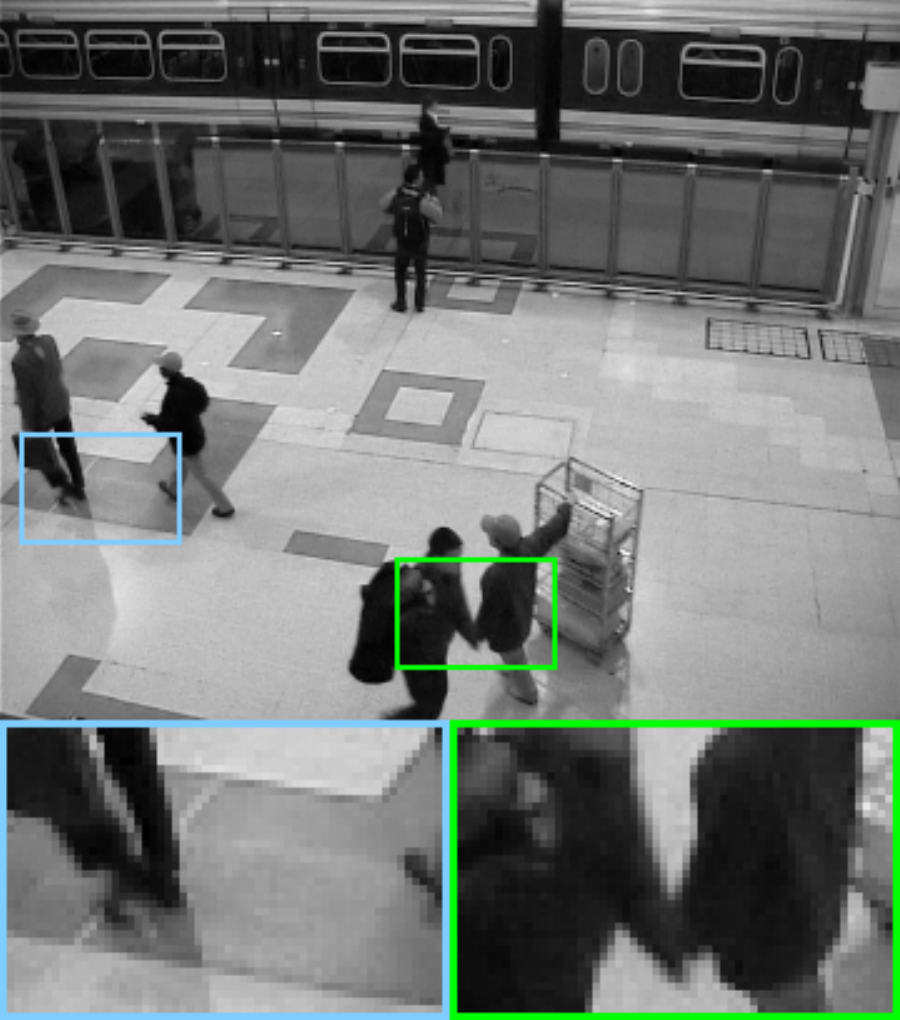}&
\includegraphics[width=0.139\textwidth]{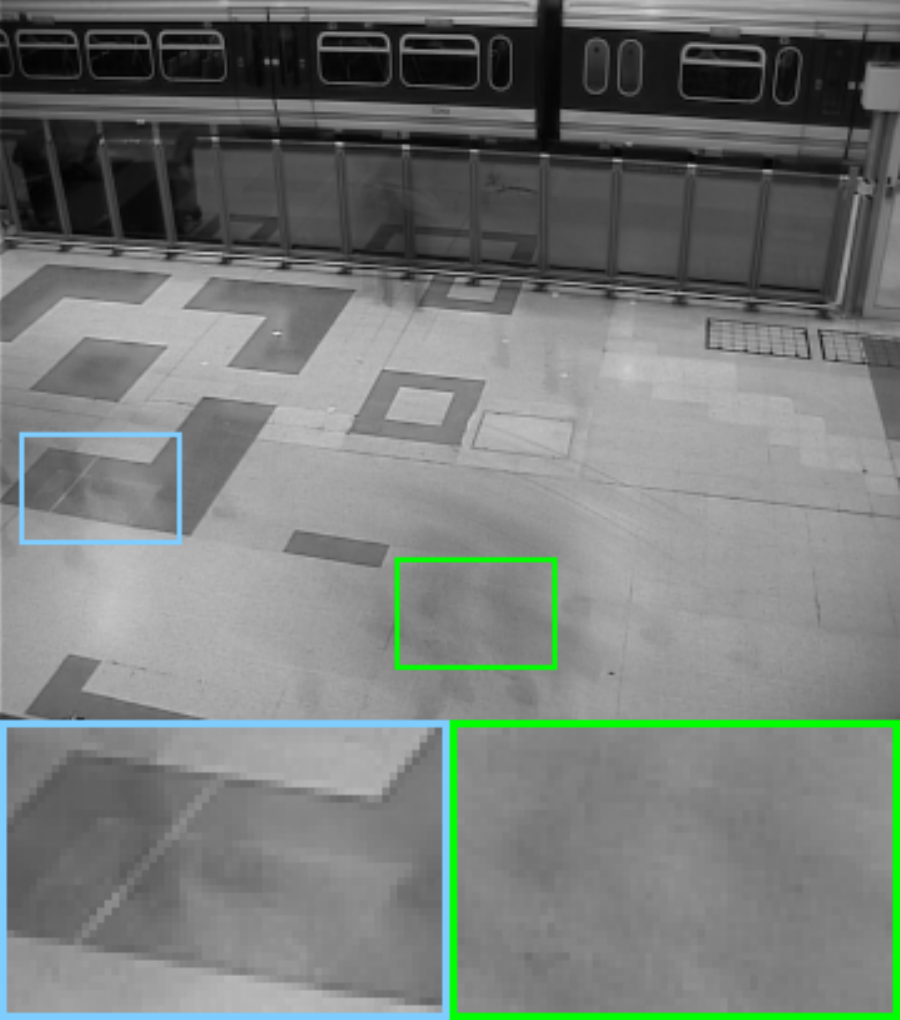}&
\includegraphics[width=0.139\textwidth]{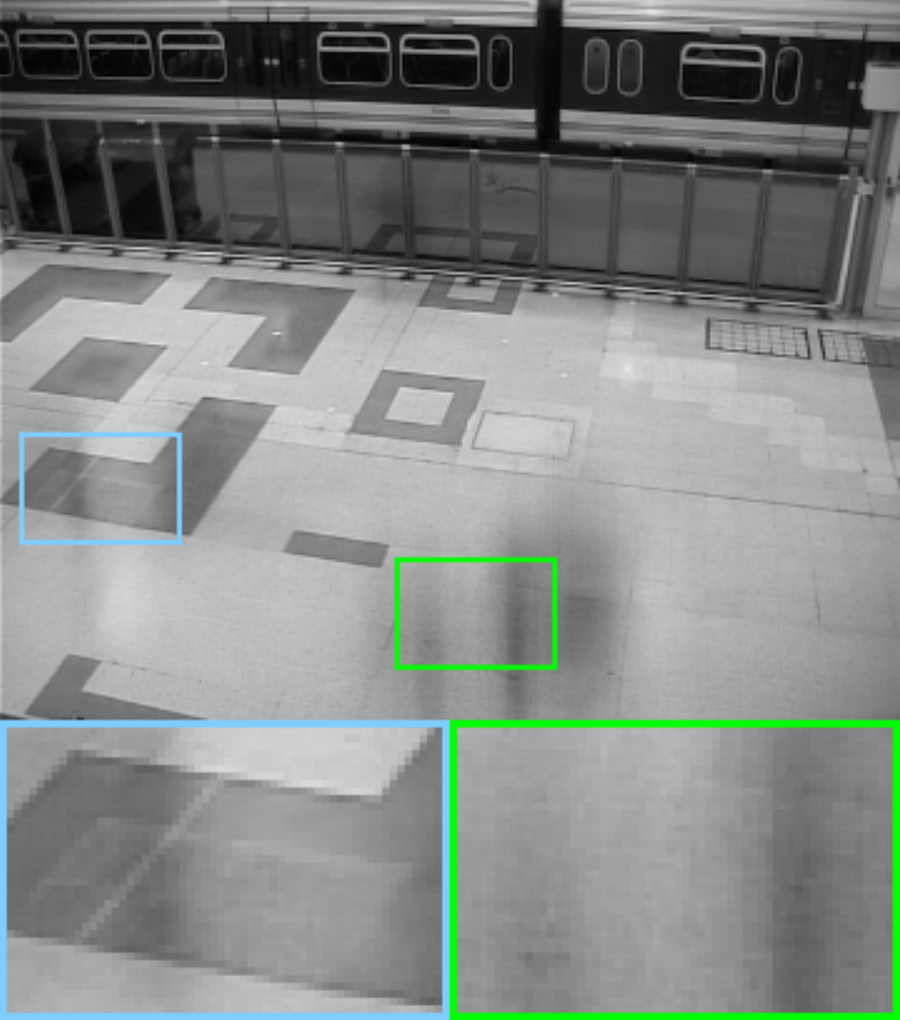}&
\includegraphics[width=0.139\textwidth]{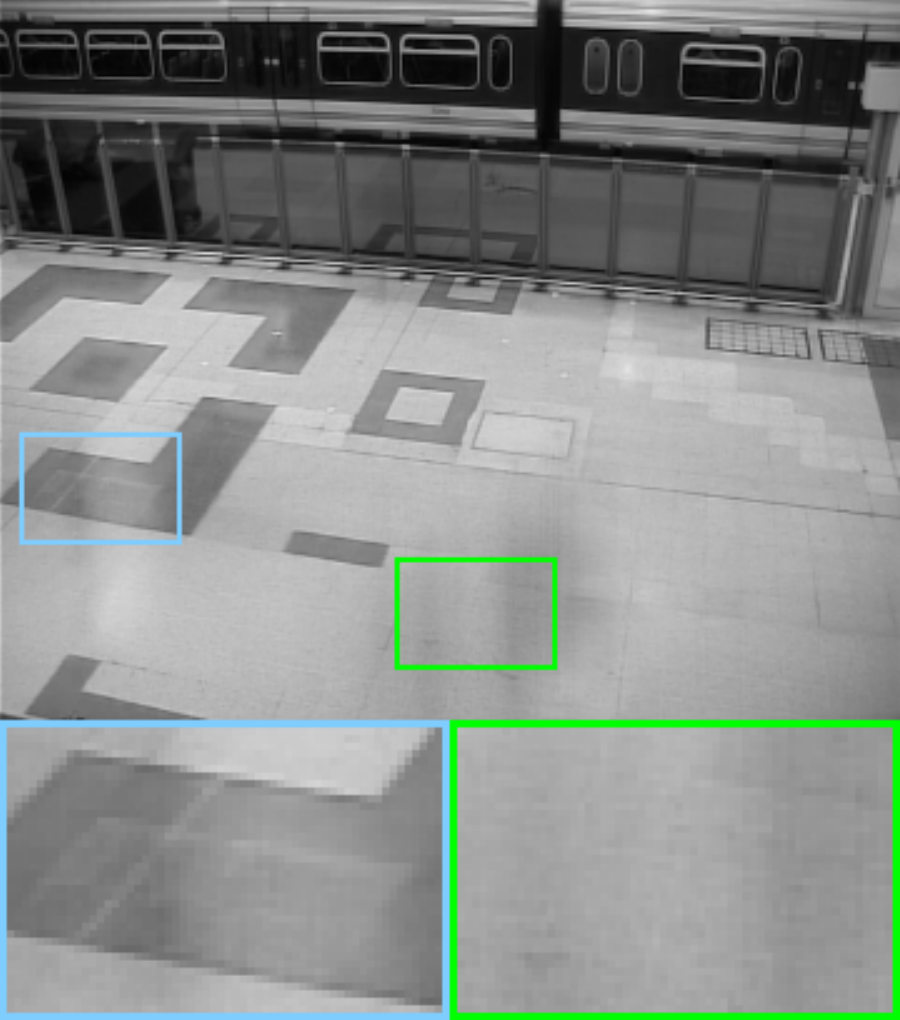}&
\includegraphics[width=0.139\textwidth]{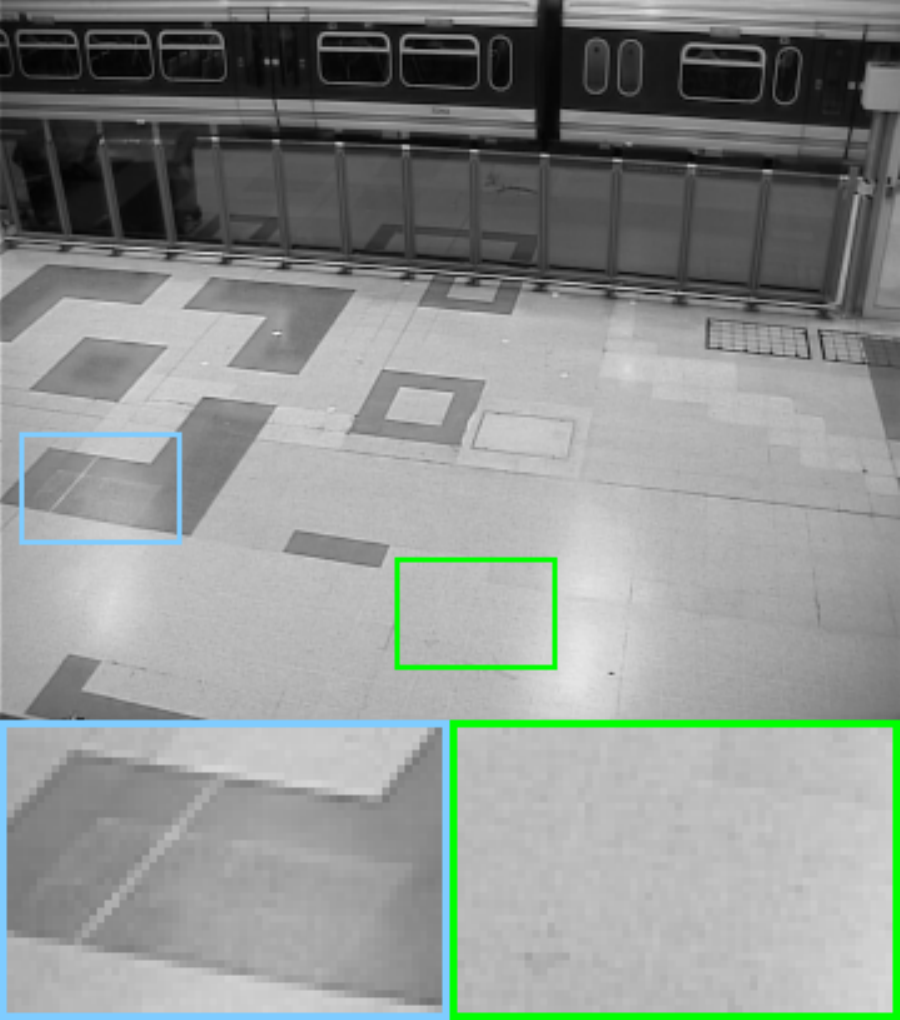}&
\includegraphics[width=0.139\textwidth]{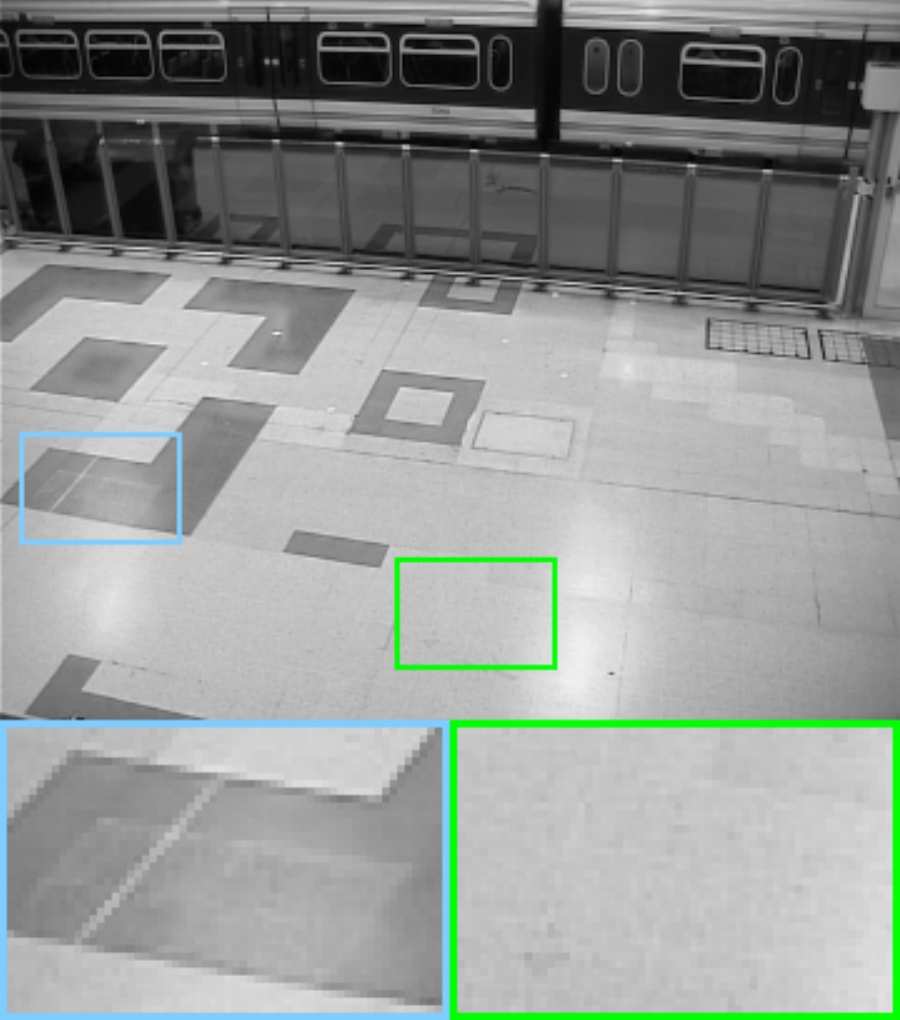}\\
\vspace{0.1cm}
\includegraphics[width=0.139\textwidth]{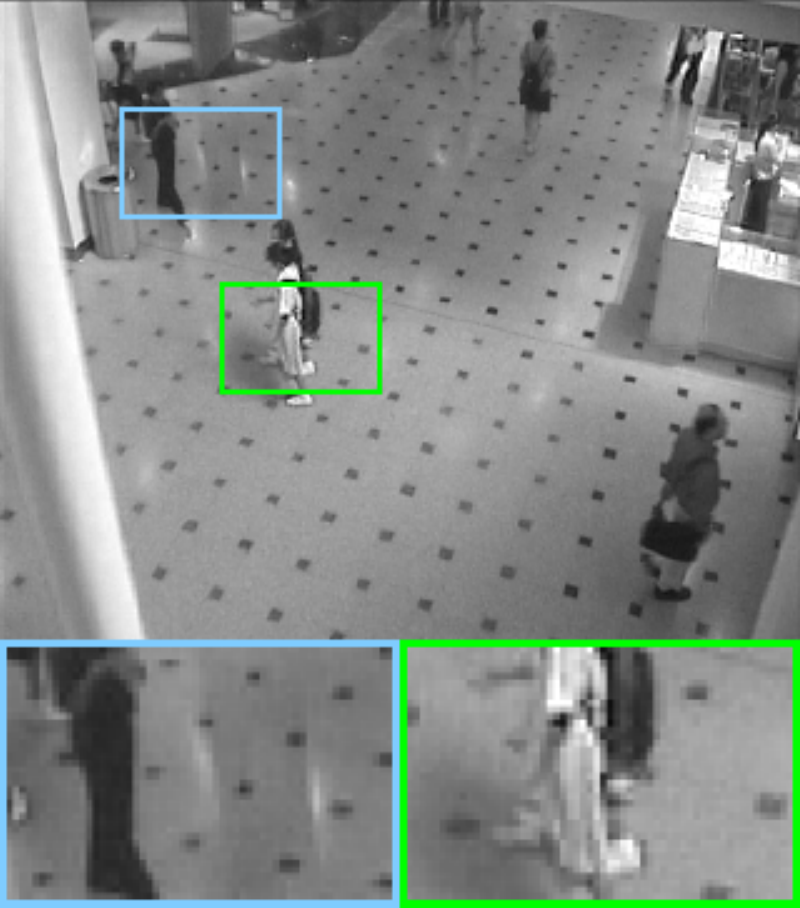}&
\includegraphics[width=0.139\textwidth]{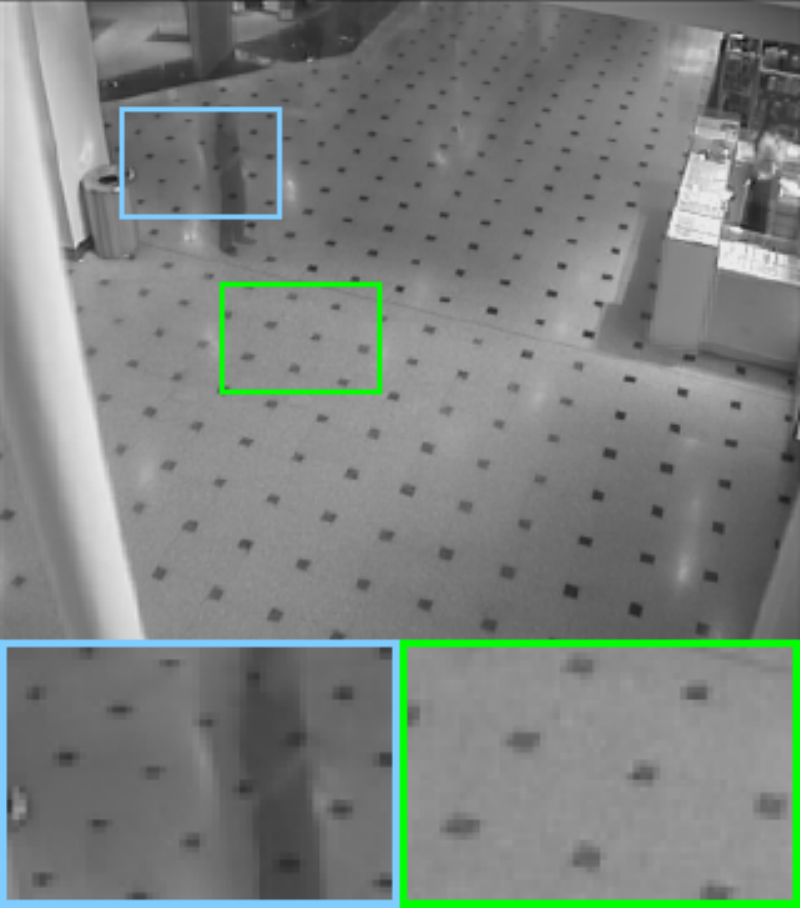}&
\includegraphics[width=0.139\textwidth]{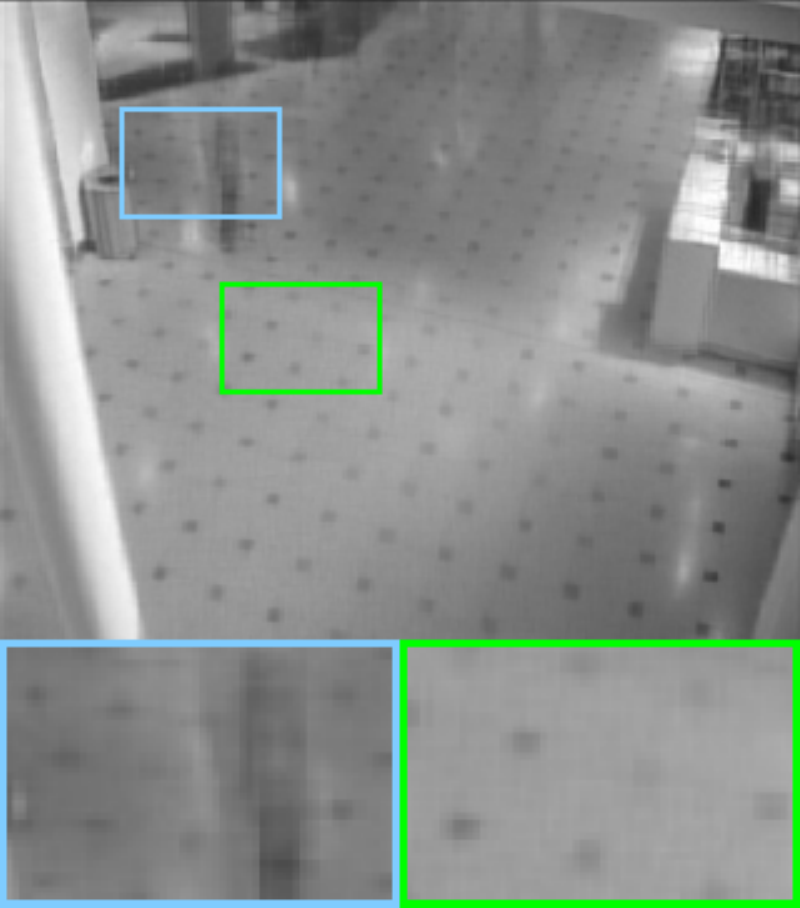}&
\includegraphics[width=0.139\textwidth]{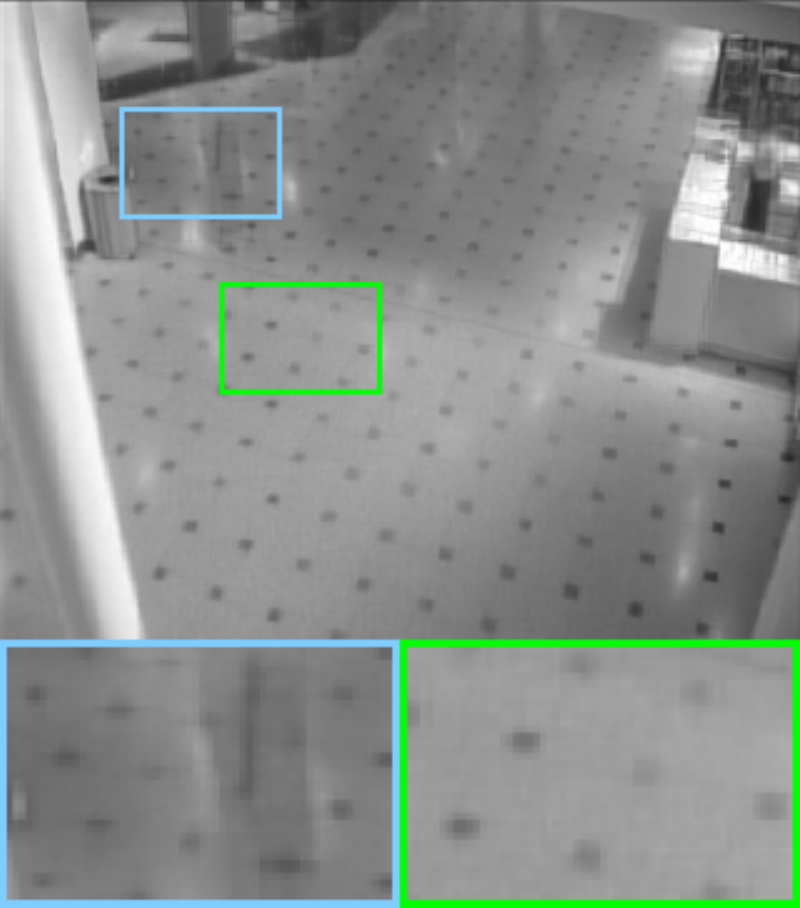}&
\includegraphics[width=0.139\textwidth]{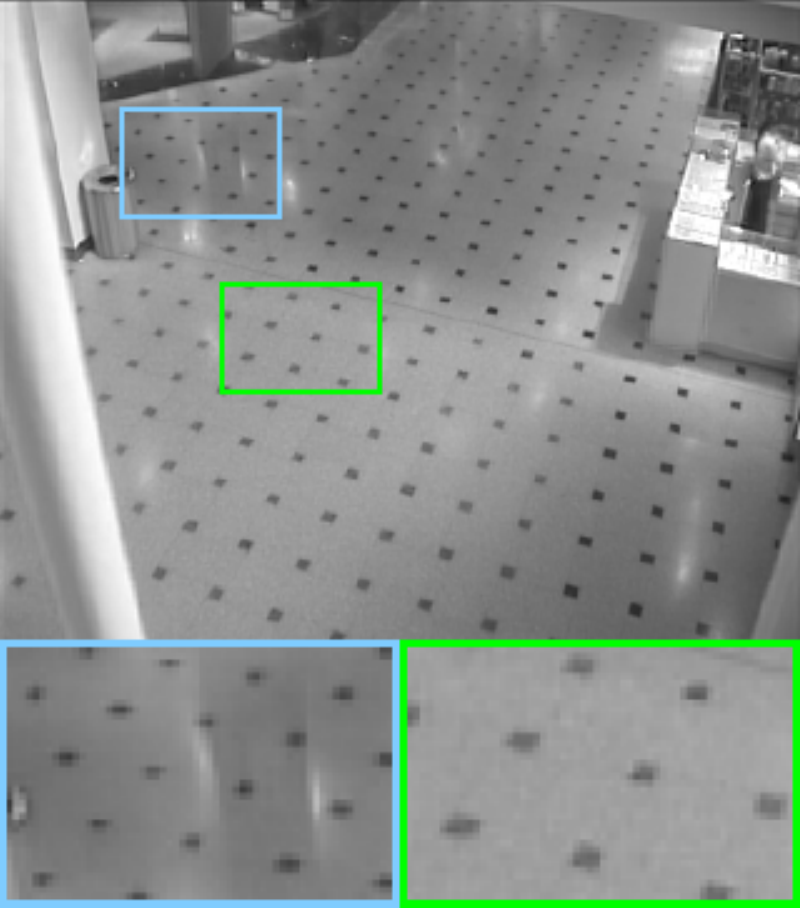}&
\includegraphics[width=0.139\textwidth]{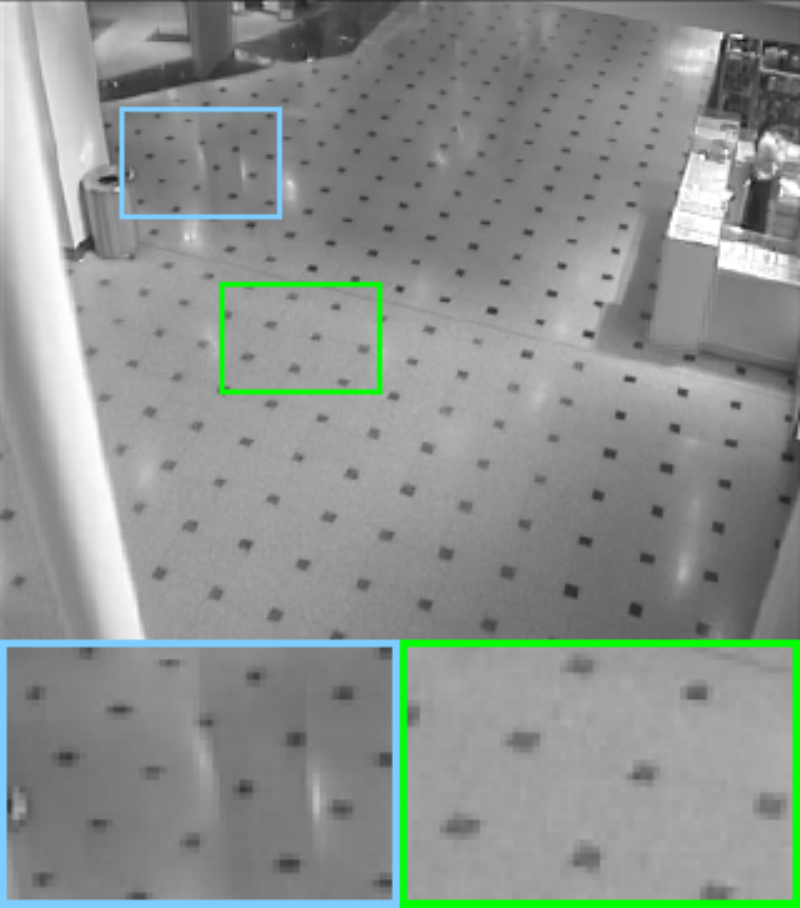}\\
\vspace{-0.3cm}
Original& FastRPCA \cite{MRPCA}&TNN \cite{TNN_TRPCA}&DCTNN \cite{DCTNN}&SSNT&SSNT-TV\\
\end{tabular}
\end{center}
\caption{The results by different methods for background subtraction on videos {\it Port}, {\it Highway}, {\it Office}, {\it PET}, and {\it Shop}.\label{TRPCA_fig}}\vspace{-0.3cm}
\end{figure*}

The results by different methods for background subtraction are shown in Fig. \ref{TRPCA_fig}. We can see that SSNT and SSNT-TV more precisely subtract the low-rank component. In addition, we can see from the zoom-in figures that SSNT and SSNT-TV more faithfully preserve the image details in the background than competing methods (e.g., the door handle in {\it Office} and the ground pattern in {\it Shop}). This is due to the nonlinear modeling ability of SSNT, which more compactly represents the low-rank tensor. 
\begin{figure*}[h]
\footnotesize
\setlength{\tabcolsep}{0.9pt}
\begin{center}
\begin{tabular}{ccccccc}
\includegraphics[width=0.139\textwidth]{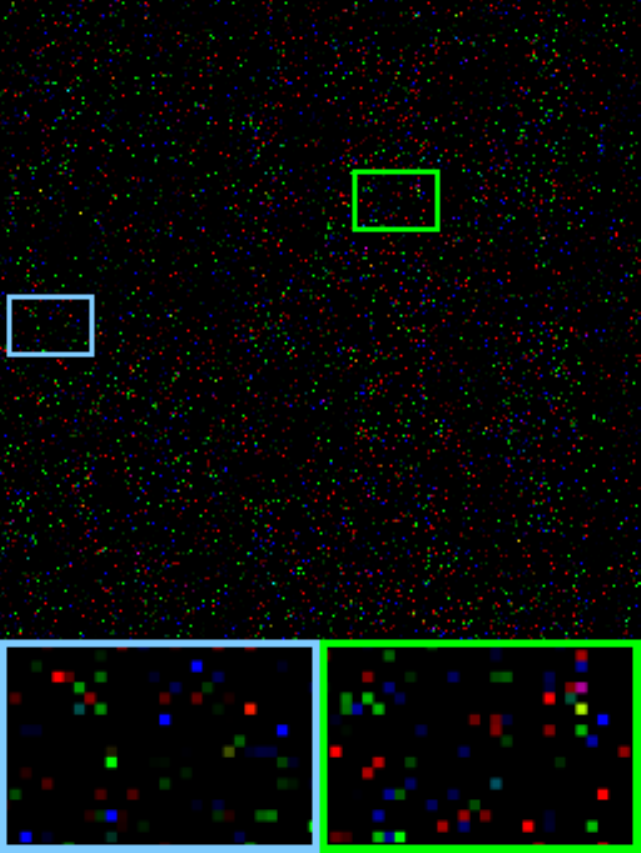}&
\includegraphics[width=0.139\textwidth]{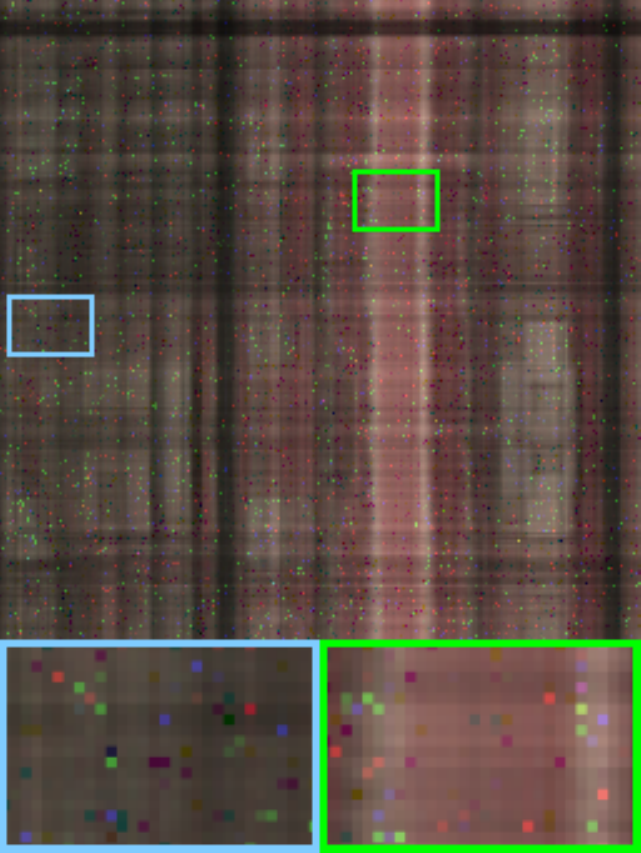}&
\includegraphics[width=0.139\textwidth]{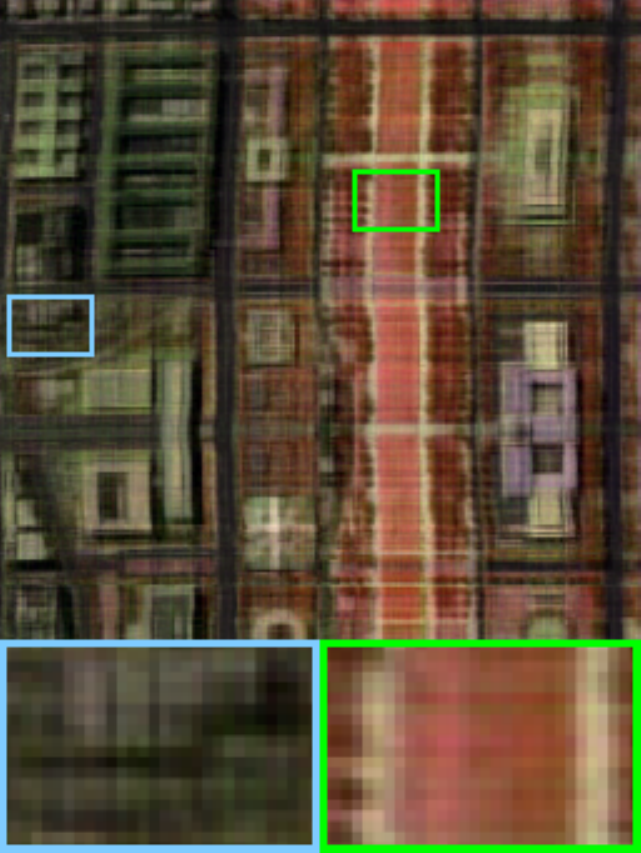}&
\includegraphics[width=0.139\textwidth]{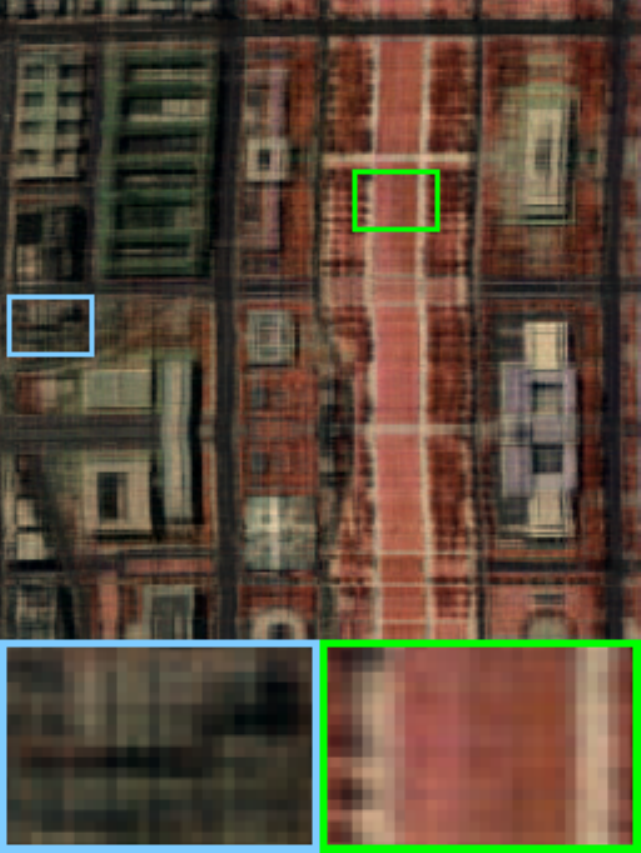}&
\includegraphics[width=0.139\textwidth]{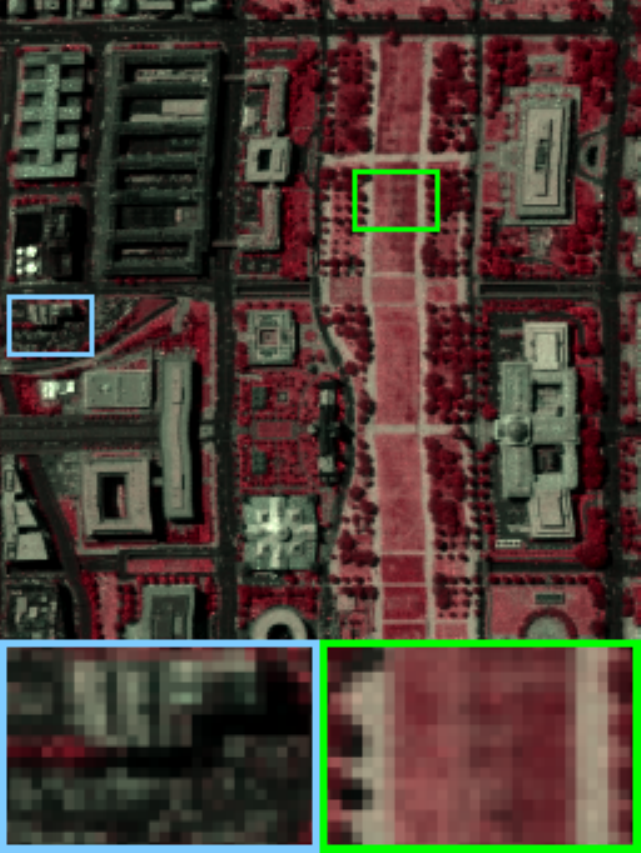}&
\includegraphics[width=0.139\textwidth]{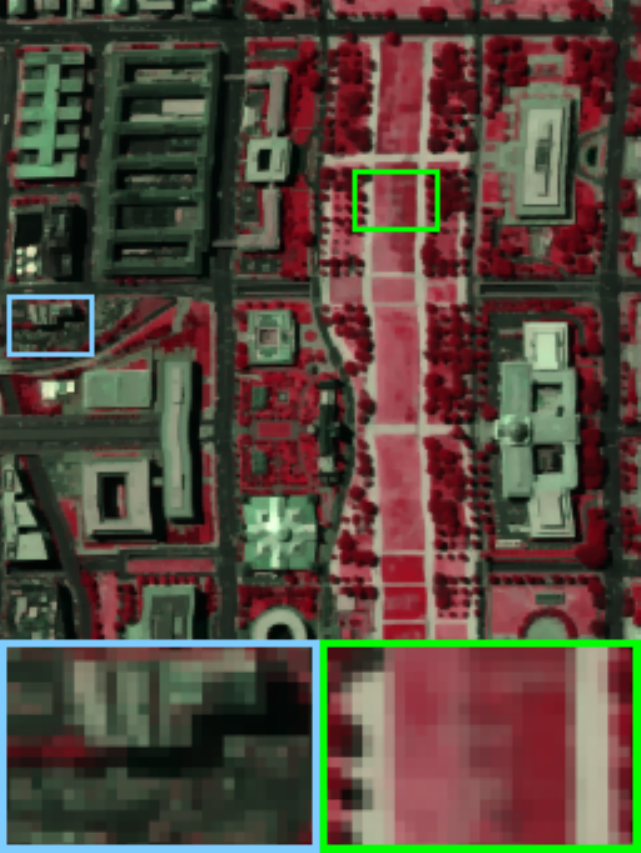}&
\includegraphics[width=0.139\textwidth]{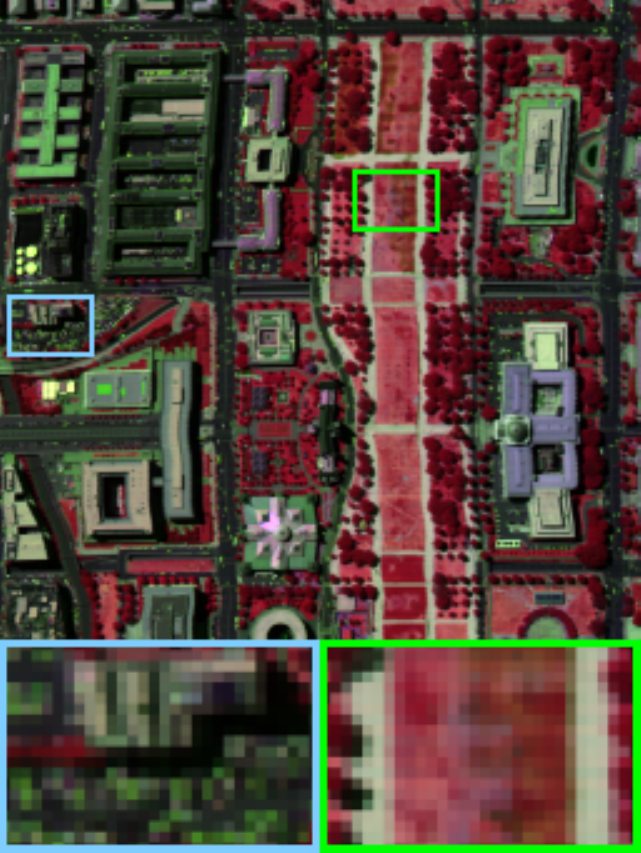}\\
\vspace{0.1cm}
PSNR 14.567 dB & PSNR 21.299 dB&PSNR 25.184 dB&PSNR 25.317 dB&PSNR 28.108 dB&PSNR 31.011 dB&PSNR Inf\\
\includegraphics[width=0.139\textwidth]{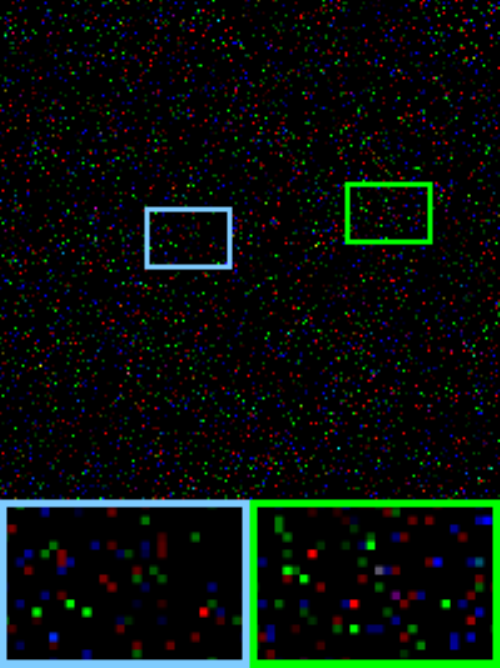}&
\includegraphics[width=0.139\textwidth]{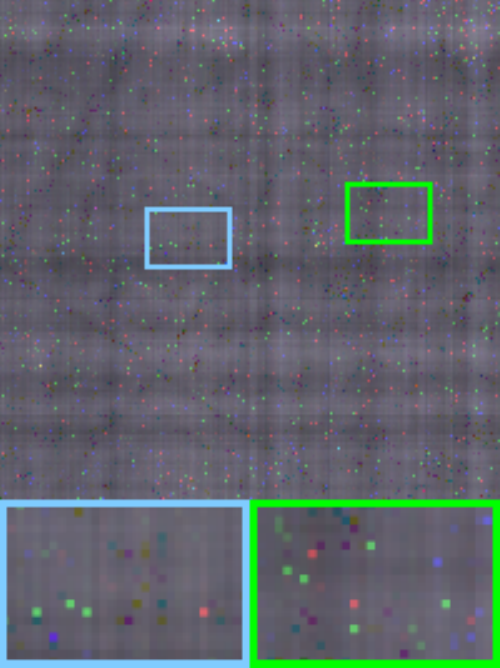}&
\includegraphics[width=0.139\textwidth]{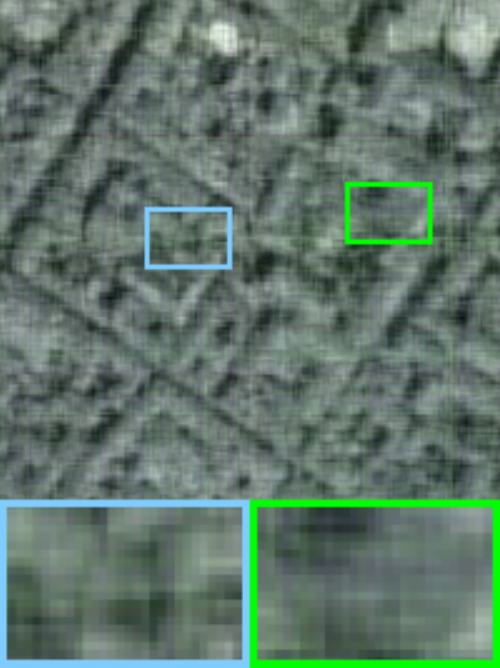}&
\includegraphics[width=0.139\textwidth]{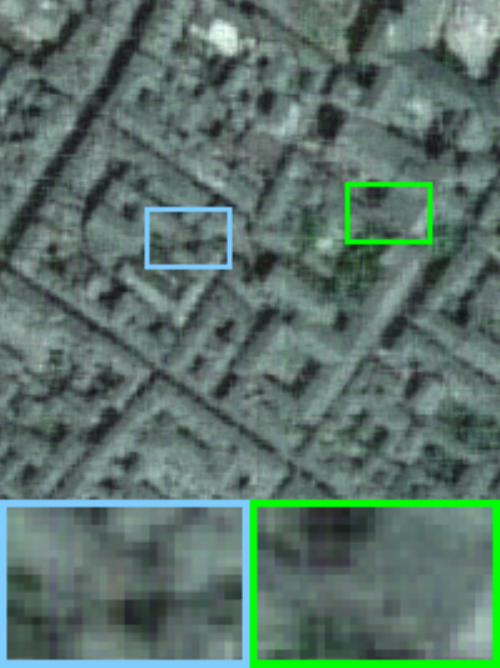}&
\includegraphics[width=0.139\textwidth]{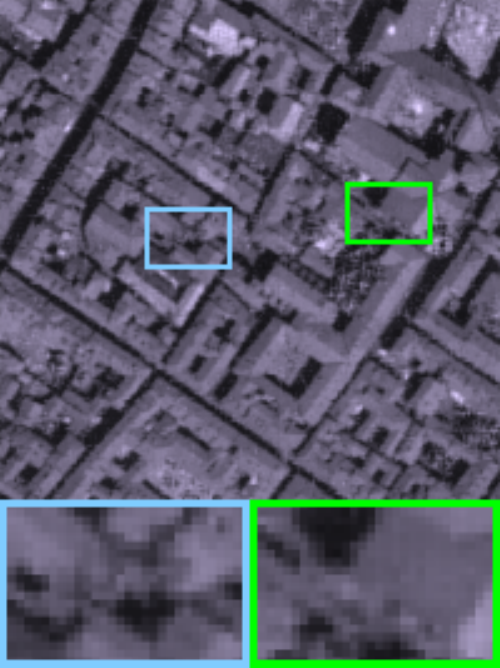}&
\includegraphics[width=0.139\textwidth]{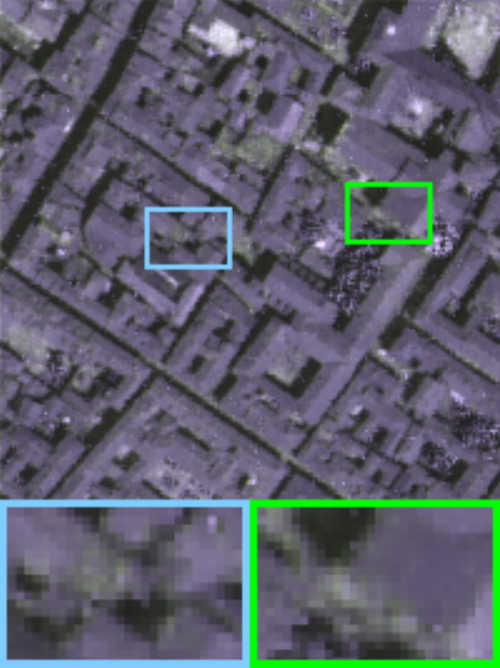}&
\includegraphics[width=0.139\textwidth]{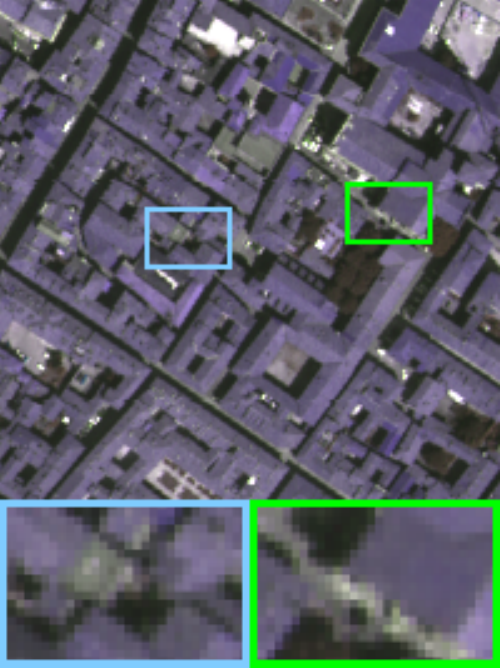}\\
\vspace{0.1cm}
PSNR 12.191 dB & PSNR 21.035 dB&PSNR 23.684 dB&PSNR 25.945 dB&PSNR 27.899 dB&PSNR 28.112 dB&PSNR Inf\\
\includegraphics[width=0.139\textwidth]{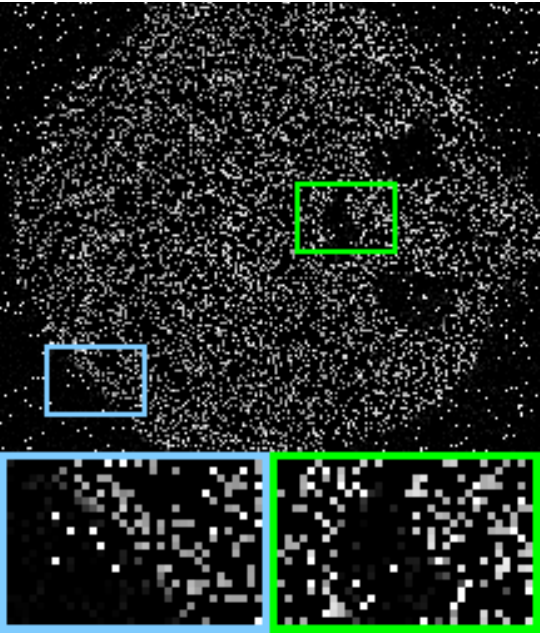}&
\includegraphics[width=0.139\textwidth]{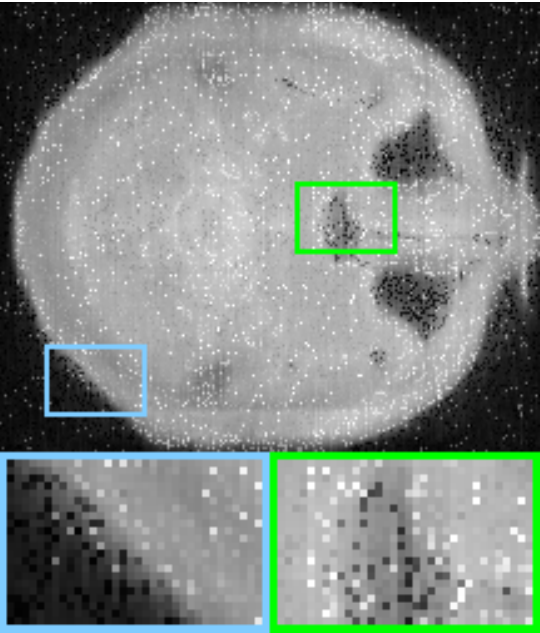}&
\includegraphics[width=0.139\textwidth]{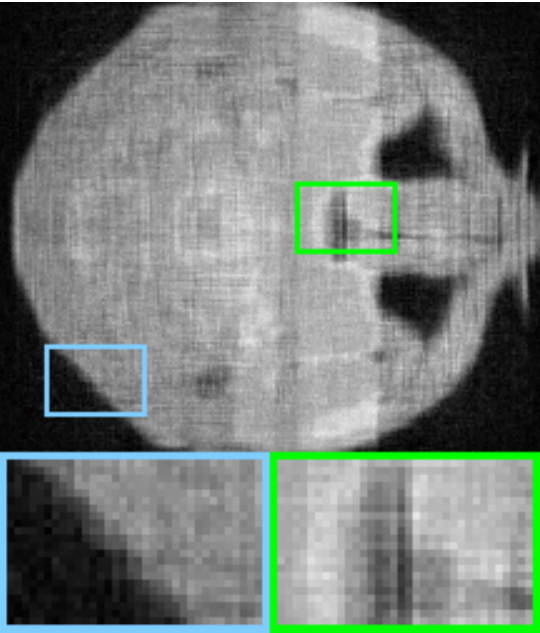}&
\includegraphics[width=0.139\textwidth]{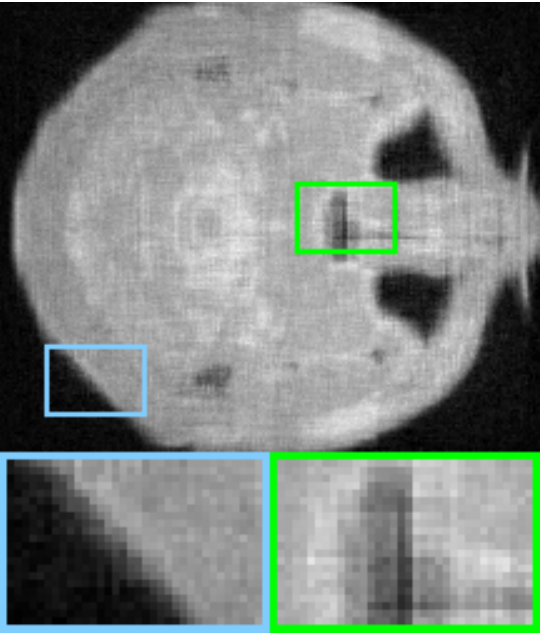}&
\includegraphics[width=0.139\textwidth]{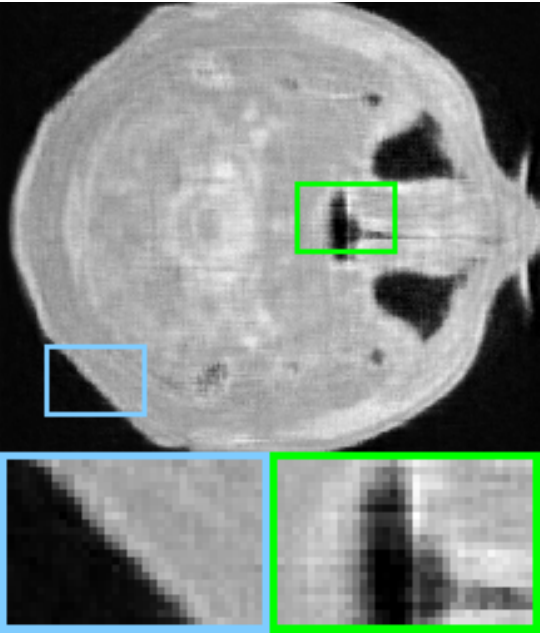}&
\includegraphics[width=0.139\textwidth]{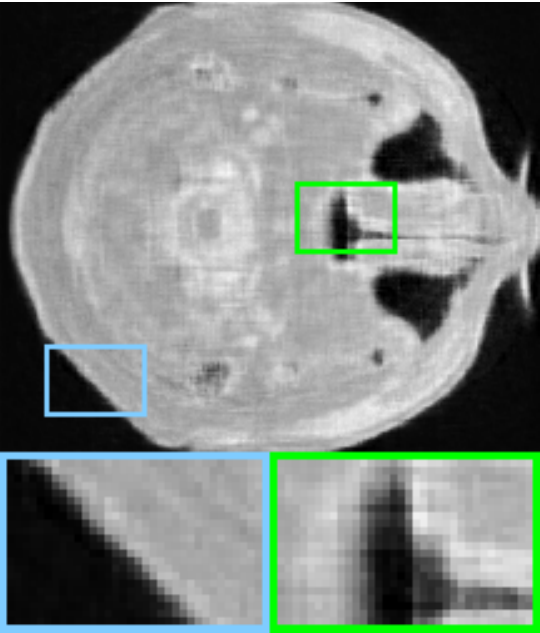}&
\includegraphics[width=0.139\textwidth]{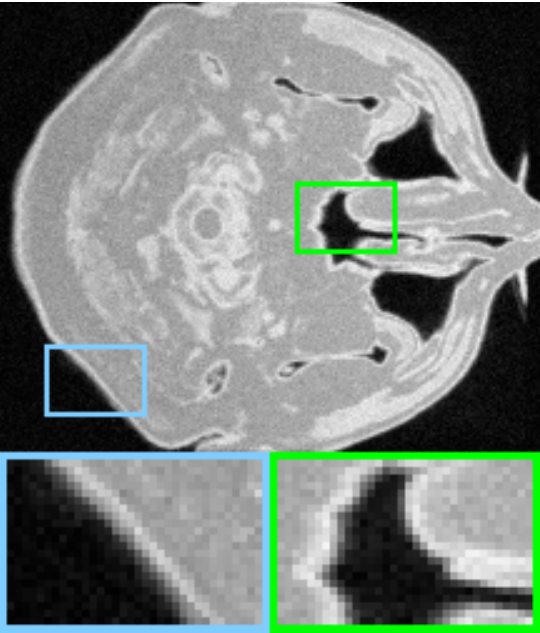}\\
PSNR 6.895 dB & PSNR 19.062 dB&PSNR 22.209 dB&PSNR 22.344 dB&PSNR 23.605 dB&PSNR 23.697 dB&PSNR Inf\\
\vspace{-0.3cm}
Observed&RTRC \cite{RTRC}&TNN \cite{TNN_TRPCA}&UTNN \cite{Haar}&SSNT&SSNT-TV&Original\\
\end{tabular}
\end{center}
\caption{The recovered results by different methods for RTC on HSI {\it WDC mall} (composed of the 50-th, 100-th, and the 150-th bands) with SR = 0.05, HSI {\it Pavia} (composed of the 1-st,10-th, and the 20-th bands) with SR = 0.05, and MRI {\it Brain} (the 10-th band) with SR = 0.25. \label{RTC_fig_1}}\vspace{-0.2cm}
\end{figure*}
\subsection{Robust Tensor Completion}
The RTC \cite{Haar} aims at recovering the low-rank tensor from the incompleted tensor with sparse error, which is formulated as
\begin{equation}
\min_{{\mathcal X},{\mathcal S}}\;\lambda\;rank({\mathcal X})+\lVert{\mathcal S}\rVert_{\ell_1},\;\;{\rm s.t.}\;\;({\mathcal X}+{\mathcal S})_\Omega = {\mathcal O}_\Omega,
\end{equation}
where $\mathcal O$ is the incompleted sparse error corrupted observation and $\mathcal X,S$ respectively stands for the low-rank and sparse component. It is easy to develop the fidelity term of our model for RTC:
\begin{equation}
L({\mathcal X},{\mathcal O}) = \lVert{\mathcal P}_\Omega({\mathcal X} - {\mathcal O})\rVert_{\ell_1}. 
\end{equation}
\par
Generally, the RTC is a more challenging task than tensor completion and background subtraction, since RTC aims at simultaneously recovering the incomplete tensor and subtracting the low-rank tensor from the observation. Thus, it requires higher capabilities of the low-rank tensor recovery method.
\subsubsection{Experimental Settings}
To illustrate the superiority of our method on RTC, we adopted the HSIs {\it Pavia} and {\it WDC mall}, the MRI {\it Brain}, and the videos {\it Highway} and {\it PET} as the experimental data. For HSIs and MRI, we firstly sample the data using different SRs and then perform sparse noise degradation with noise SR $0.1$ on the incomplete data. For videos {\it Highway} and {\it PET}, we only sample the data and regard the foreground component as the sparse error. \par
The competing methods for RTC are: The tensor ring-based method RTRC \cite{RTRC}, the linear transform-based methods TNN \cite{TNN_TRPCA} and UTNN \cite{Haar}. For our hyperparameters, we set $\lambda = N\times 10^{-7}$. We use the linear interpolation \cite{DTNN} to initialize the observed tensor for SSNT and SSNT-TV.\par
\subsubsection{Experimental Results}
The numerical results for RTC are reported in Table \ref{RTC_tab}. We can see that SSNT-TV outperforms competing methods in terms of PSNR. However, UTNN achieves better SSIM and SAM values than SSNT-TV on {\it WDC mall} with higher SR. This is due to the consideration of spatial smoothness by SSNT-TV, where the over smoothness may arise and affect the global structure of the recovered tensor.\par
Some visual results for RTC are illustrated in Fig. \ref{RTC_fig_1} and Fig. \ref{RTC_fig_2}. From Fig. \ref{RTC_fig_1}, we can see that SSNT and SSNT-TV can recover the tensor better than competing methods, where the low-rank component is well recovered and the sparse noise is accurately eliminated by our methods. Generally, SSNT-TV has smoother and cleaner results than SSNT due to the TV regularization, which results in higher PSNR values. In Fig. \ref{RTC_fig_2}, we can discover that the proposed methods have better performance for separating the low-rank component from the sparse error corrupted observation. This is due to the ability of SSNT for building a more compact low-rank representation. In addition, on the recovery of {\it Highway}, we can see that SSNT-TV performs relatively better than SSNT. This is attributed to the spatial TV constraint, which forces the low-rank component to be smooth on the spatial domain. 
\begin{table*}[!h]
\caption{The quantitative results by different methods on different data for RTC. The {\bf BEST} values are highlighted by {\bf BOLDFACE}, and the \underline{SECOND-BEST} values are highlighted by \underline{UNDERLINED}.\label{RTC_tab}}
\begin{center}
\scriptsize
\setlength{\tabcolsep}{2.9pt}
\begin{spacing}{0.95}
\begin{tabular}{clcccccccccccccccc}
\toprule
\multirow{2}*{Data}&SR&\multicolumn{3}{c}{0.05}&\multicolumn{3}{c}{0.1}&\multicolumn{3}{c}{0.15}&\multicolumn{3}{c}{0.2}&\multicolumn{3}{c}{0.25}
&\;\multirow{2}*{\tabincell{c}{Time\\(s)}}\\
\cmidrule{2-17}
~&Metric&\;\;\;PSNR&SSIM&SAM\;\;\;&PSNR&SSIM&SAM\;\;\;&PSNR&SSIM&SAM\;\;\;&PSNR&SSIM&SAM\;\;\;&PSNR&SSIM&SAM&~\\
\midrule
\multirow{6}*{\tabincell{c}{
HSI {\it WDC mall}\\{(256$\times$256$\times$191)}\\}}
~&Observed&\;\;\; 13.953&0.066&1.405 \;\;\;&13.626&0.088&1.351 \;\;\;&13.351&0.105&1.309 \;\;\;&13.109&0.117&1.271 \;\;\;&12.903&0.126&1.236&$-$\\
~&RTRC\cite{RTRC}&\;\;\; 21.299&0.577&0.268 \;\;\;&23.467&0.714&0.232 \;\;\;&25.177&0.796&0.208 \;\;\;&26.463&0.843&0.194 \;\;\;&27.660&0.878&0.181&125\\
~&TNN\cite{TNN_LRTC}&\;\;\; 25.184&0.804&0.245 \;\;\;&29.418&0.919&0.171 \;\;\;&32.307&0.955&0.137 \;\;\;&34.453&0.971&0.116 \;\;\;&36.272&0.979&0.102&184\\
~&UTNN\cite{Haar}&\;\;\; 25.317&0.818&0.218 \;\;\;&30.816&0.949&0.126 \;\;\;&\underline{34.890}&0.979&0.086 \;\;\;&37.898&\bf{0.989}&\bf{0.065} \;\;\;&40.572&\bf{0.994}&\bf{0.052}&669\\
~&SSNT&\;\;\; \underline{28.108}&\underline{0.921}&\underline{0.158} \;\;\;&\underline{33.356}&\underline{0.972}&\underline{0.106} \;\;\;&{34.641}&\underline{0.983}&\underline{0.081} \;\;\;&\underline{37.172}&\underline{0.988}&\underline{0.075} \;\;\;&\underline{38.689}&\underline{0.991}&\underline{0.062}&327\\
~&SSNT-TV&\;\;\; \bf{31.011}&\bf{0.952}&\bf{0.107} \;\;\;&\bf{35.493}&\bf{0.983}&\bf{0.076} \;\;\;&\bf{36.364}&\bf{0.986}&\bf{0.080} \;\;\;&\bf{40.565}&{0.977}&{0.088} \;\;\;&\bf{45.062}&{0.990}&{0.063}&336\\
\midrule
\multirow{6}*{\tabincell{c}{
HSI {\it Pavia}\\{(200$\times$200$\times$80)}\\}}
~&Observed&\;\;\; 11.941&0.035&1.383 \;\;\;&11.918&0.055&1.310 \;\;\;&11.894&0.069&1.259 \;\;\;&11.874&0.080&1.215 \;\;\;&11.859&0.090&1.176&$-$\\
~&RTRC\cite{RTRC}&\;\;\; 21.035&0.519&0.142 \;\;\;&22.006&0.599&0.162 \;\;\;&23.142&0.676&0.164 \;\;\;&24.151&0.737&0.160 \;\;\;&25.024&0.781&0.157&51\\
~&TNN\cite{TNN_LRTC}&\;\;\; 23.684&0.732&0.148 \;\;\;&28.133&0.902&0.118 \;\;\;&31.243&0.947&0.097 \;\;\;&33.806&0.966&0.083 \;\;\;&35.719&0.974&0.075&64\\
~&UTNN\cite{Haar}&\;\;\; 25.946&0.844&0.124 \;\;\;&30.758&0.945&0.089 \;\;\;&33.633&0.968&0.073 \;\;\;&35.662&0.977&0.065 \;\;\;&36.820&\underline{0.982}&0.060&311\\
~&SSNT&\;\;\; \underline{27.901}&\underline{0.916}&\underline{0.105} \;\;\;&\underline{32.629}&\underline{0.965}&\underline{0.049} \;\;\;&\underline{37.491}&\underline{0.989}&\underline{0.033} \;\;\;&\underline{40.318}&\underline{0.994}&\underline{0.029} \;\;\;&\underline{41.750}&\bf{0.996}&\underline{0.025}&126\\
~&SSNT-TV&\;\;\; \bf{28.112}&\bf{0.920}&\bf{0.078} \;\;\;&\bf{33.588}&\bf{0.967}&\bf{0.042} \;\;\;&\bf{39.086}&\bf{0.992}&\bf{0.029} \;\;\;&\bf{42.018}&\bf{0.996}&\bf{0.023} \;\;\;&\bf{42.406}&\bf{0.996}&\bf{0.019}&180\\
\midrule
\multirow{6}*{\tabincell{c}{
MRI {\it Brain}\\{(181$\times$217$\times$181)}\\}}
~&Observed&\;\;\; 5.793&0.091&1.386 \;\;\;&5.947&0.098&1.314 \;\;\;&6.11&0.102&1.258 \;\;\;&6.277&0.103&1.209 \;\;\;&6.451&0.104&1.164&$-$\\
~&RTRC\cite{RTRC}&\;\;\; 13.891&0.369&0.432 \;\;\;&15.766&0.465&0.372 \;\;\;&17.085&0.528&0.335 \;\;\;&18.166&0.575&0.310 \;\;\;&19.062&0.616&0.289&78\\
~&TNN\cite{TNN_LRTC}&\;\;\; 15.411&0.520&0.385 \;\;\;&17.930&0.636&0.310 \;\;\;&19.698&0.705&0.274 \;\;\;&21.058&0.750&0.251 \;\;\;&22.209&0.782&\underline{0.234}&182\\
~&UTNN\cite{Haar}&\;\;\; 15.489&0.525&0.378 \;\;\;&18.449&0.647&0.303 \;\;\;&20.092&0.708&0.273 \;\;\;&21.304&0.748&0.253 \;\;\;&22.344&0.775&0.238&294\\
~&SSNT&\;\;\; \underline{19.872}&\underline{0.695}&\underline{0.261} \;\;\;&\underline{21.616}&\underline{0.757}&\underline{0.224} \;\;\;&\underline{22.466}&\underline{0.788}&\underline{0.208} \;\;\;&\underline{23.075}&\underline{0.807}&\underline{0.196} \;\;\;&\underline{23.605}&\underline{0.824}&\bf{0.186}&191\\
~&SSNT-TV&\;\;\; \bf{20.311}&\bf{0.715}&\bf{0.249} \;\;\;&\bf{21.877}&\bf{0.769}&\bf{0.218} \;\;\;&\bf{22.642}&\bf{0.793}&\bf{0.204} \;\;\;&\bf{23.268}&\bf{0.815}&\bf{0.192} \;\;\;&\bf{23.697}&\bf{0.830}&\bf{0.186}&207\\
\bottomrule
\end{tabular}
\end{spacing}
\end{center}
\vspace{-0.4cm}
\end{table*}
\subsection{Snapshot Compressive Imaging}
The SCI is developed to capture multi-dimensional data from low-dimensional data \cite{TIT,ADMM_net}. Specifically, SCI systems can capture the abundant spectral/temporal information with low memory and computational cost by summing up the spectral/temporal signals to obtain the measurement with the help of sensing mask. One key module in SCI systems is the reconstruction of the original high-dimensional signals. Given the sensing measurement ${\bf O}$ and sensing mask $\Omega$, we formulate the low-rank tensor-based SCI reconstruction \cite{DeSCI} as follows: 
\begin{equation}
\min_{\mathcal X}\;\lambda\;rank({\mathcal X}) + \lVert \sum_k\big{(}{\mathcal P}_\Omega({\mathcal X})\big{)}^{(k)}-{\bf O}\rVert_F^2,
\end{equation}
where ${\bf O}$ is the measurement generated by summing the frontal slices of the sampled incompleted data, i.e., ${\bf O} = \sum_k\big{(}{\mathcal P}_\Omega({\mathcal X})\big{)}^{(k)}+{\bf N}$ with $\bf N$ denotes the noise. It is easy to see that the fidelity term of the proposed model for SCI is 
\begin{equation}
L({\mathcal X},{\bf O}) = \lVert \sum_k\big{(}{\mathcal P}_\Omega({\mathcal X})\big{)}^{(k)}-{\bf O}\rVert_F^2.
\end{equation}
\subsubsection{Experimental Settings}
We adopted MSIs {\it Toys} and {\it Flowers} and videos {\it Drop} and {\it Crash}\footnote{\url{https://drive.google.com/drive/folders/1d2uh9nuOL5Z7WnEQJ5HZSDMWK2VAT9sH}} as the experimental data for SCI. We firstly sampled the data using different SRs and then summing up the frontal slices to generate the sensing measurement. Gaussian noise with the standard deviation $0.1$ is performed on the sensing measurement.\par
The competing methods for SCI are: The TV-based method GAP-TV \cite{GAP-TV}, the low-rankness-based method DeSCI \cite{DeSCI}, and the sparsity-based method SeSCI \cite{SeSCI}. For our hyperparameters, we set $\lambda = N\times 10^{-5}$. We use the results of GAP-TV as the initialization of DeSCI, SSNT, and SSNT-TV.\par
\begin{figure*}[h]
\footnotesize
\setlength{\tabcolsep}{0.9pt}
\begin{center}
\begin{tabular}{ccccccc}
\vspace{0.1cm}
\includegraphics[width=0.139\textwidth]{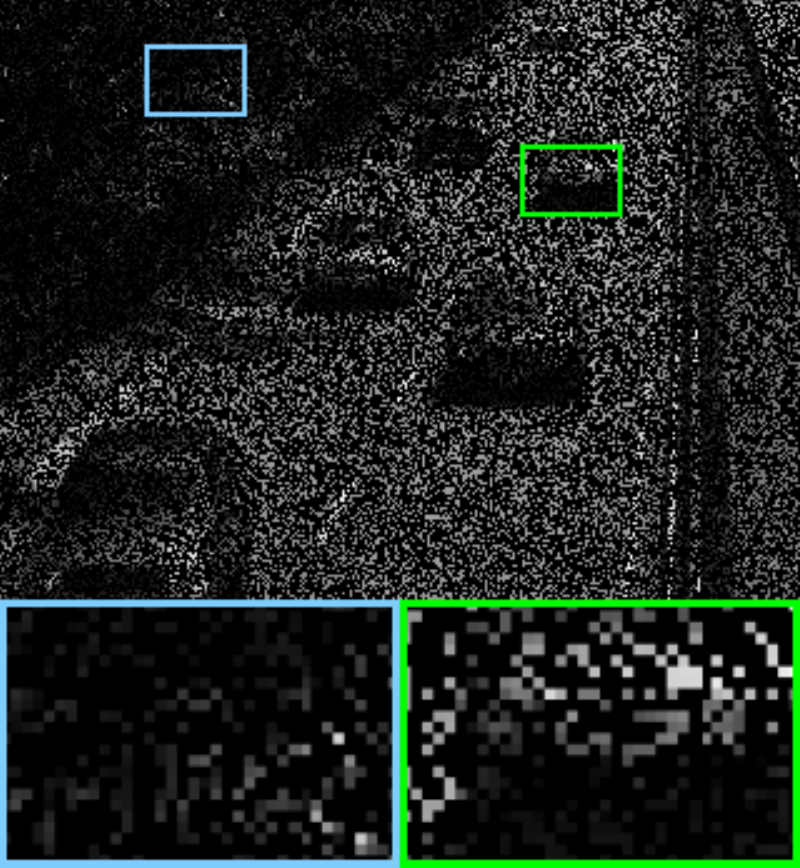}&
\includegraphics[width=0.139\textwidth]{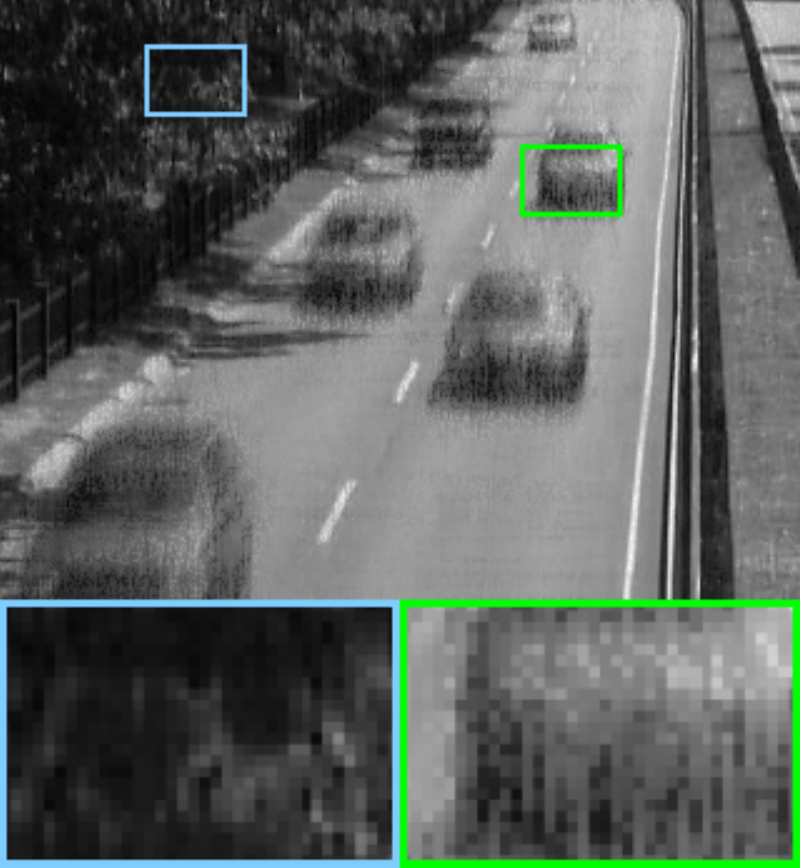}&
\includegraphics[width=0.139\textwidth]{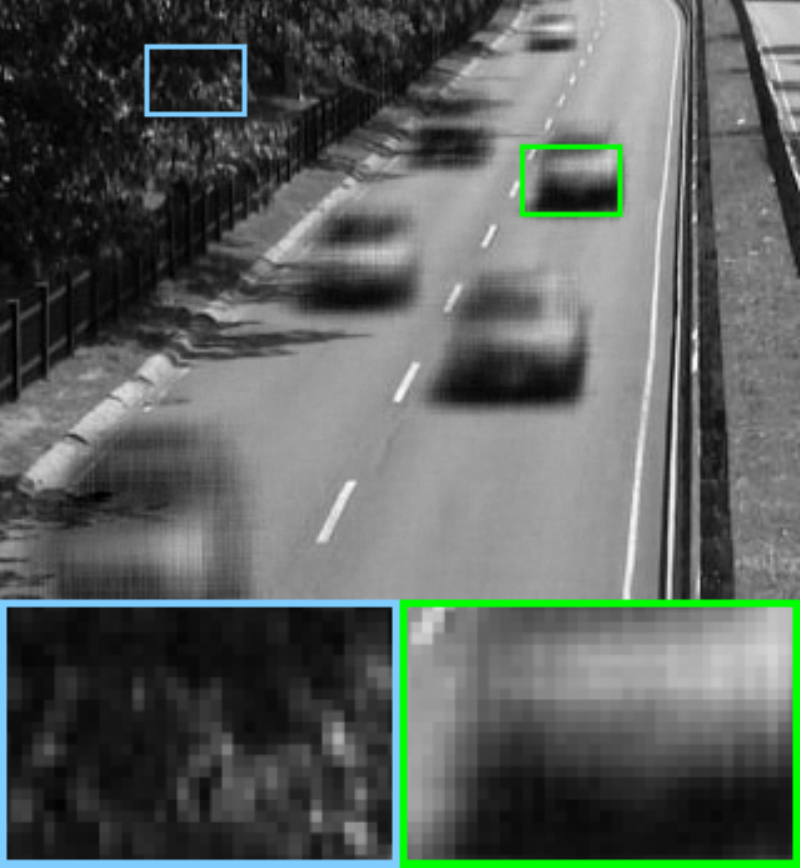}&
\includegraphics[width=0.139\textwidth]{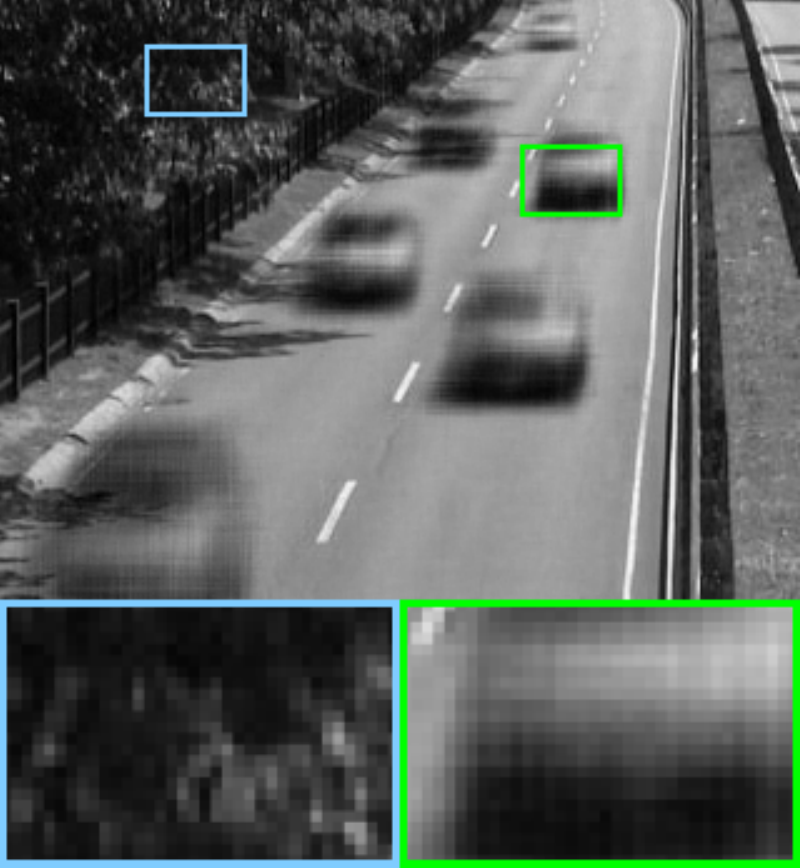}&
\includegraphics[width=0.139\textwidth]{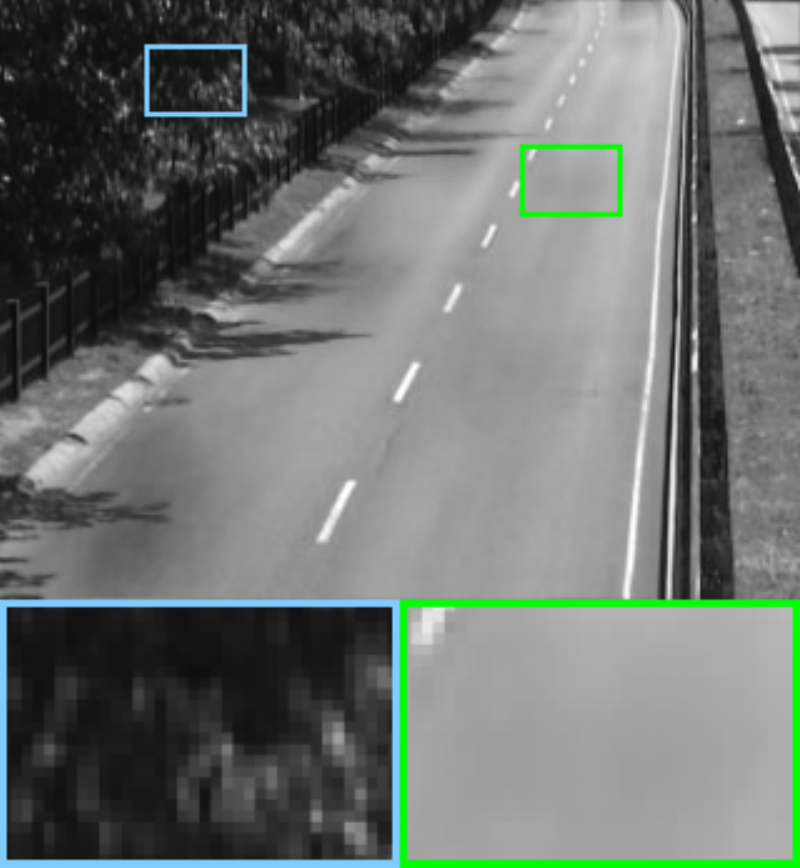}&
\includegraphics[width=0.139\textwidth]{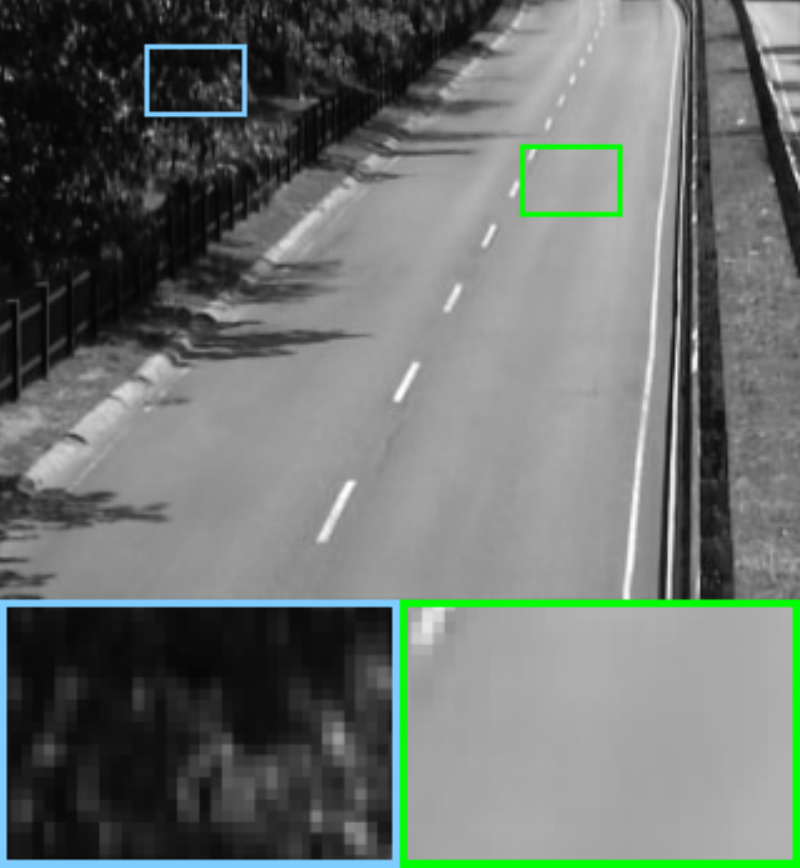}&
\includegraphics[width=0.139\textwidth]{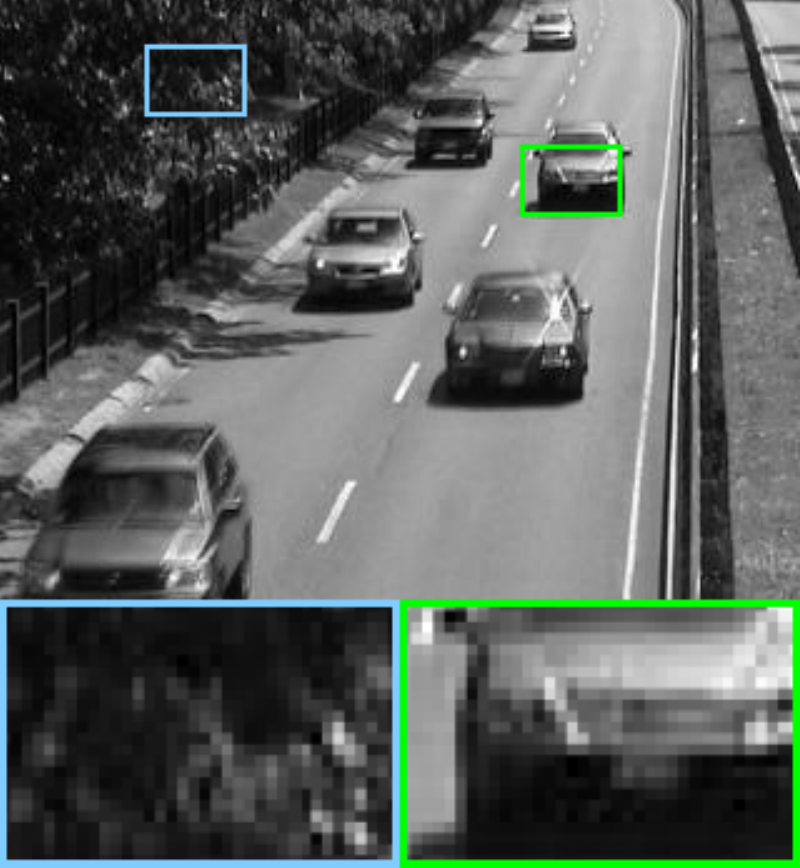}\\
\vspace{0.1cm}
\includegraphics[width=0.139\textwidth]{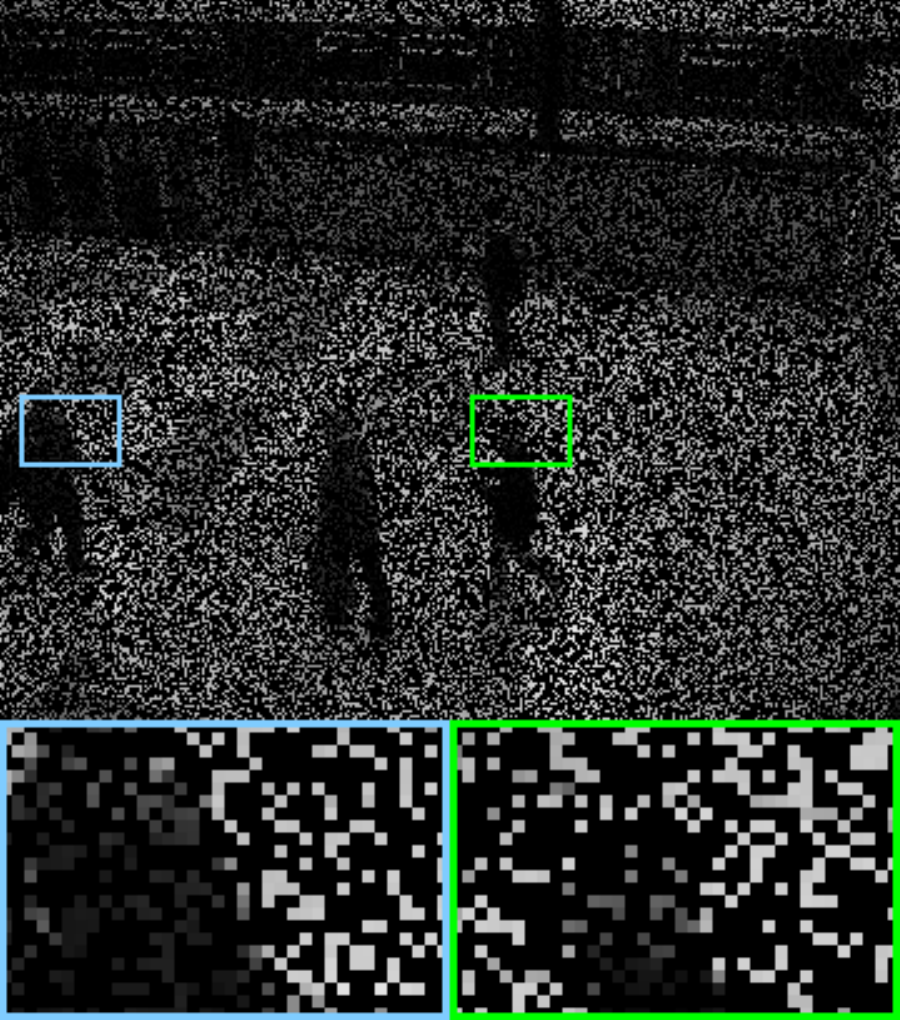}&
\includegraphics[width=0.139\textwidth]{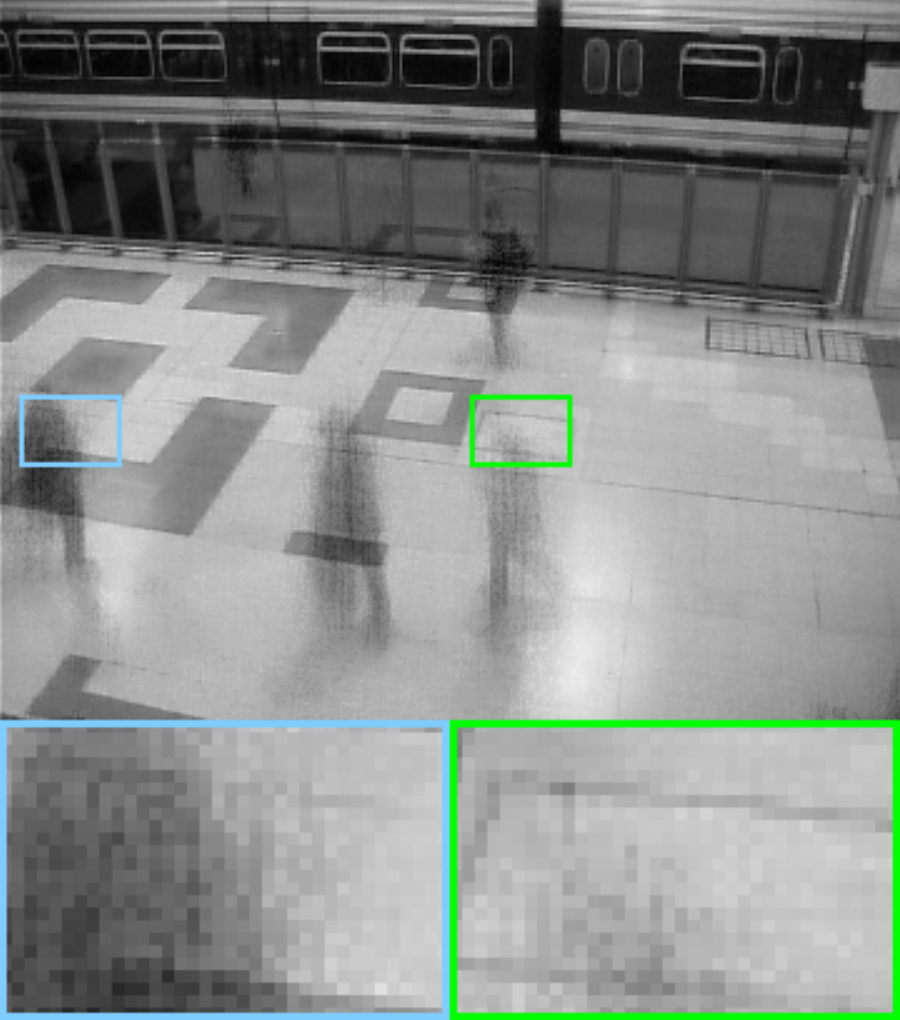}&
\includegraphics[width=0.139\textwidth]{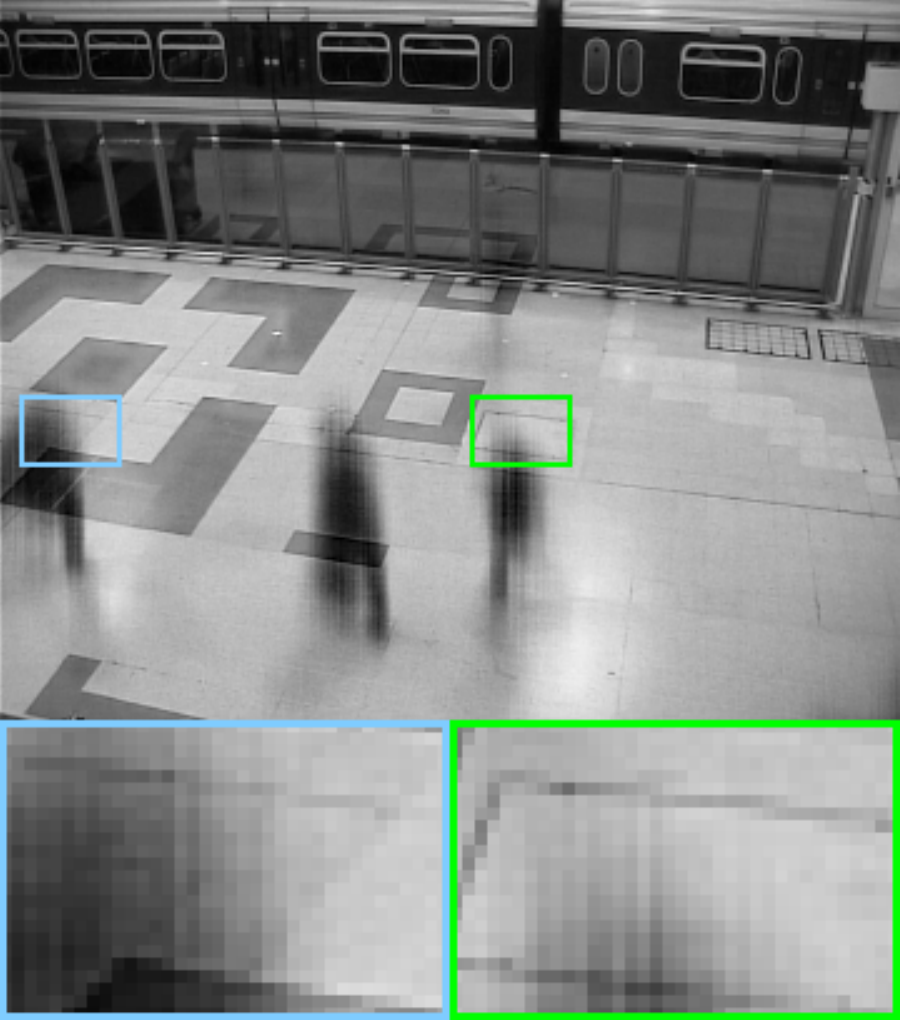}&
\includegraphics[width=0.139\textwidth]{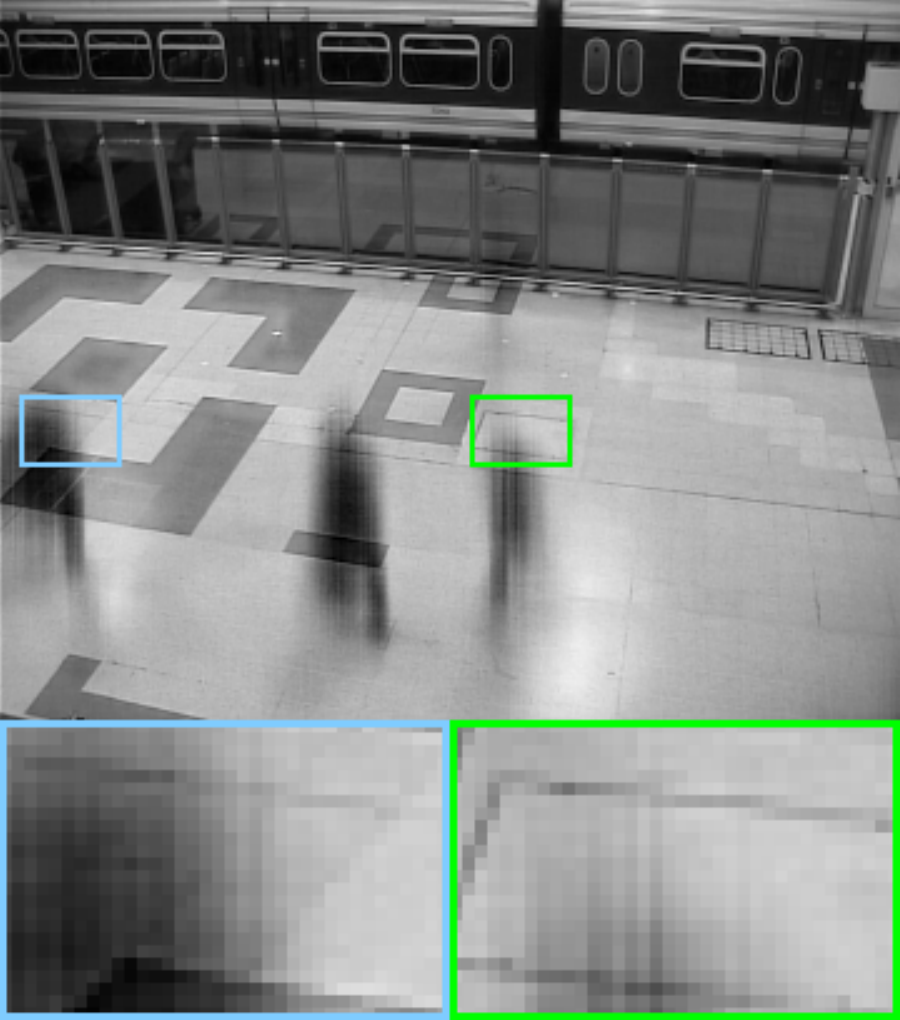}&
\includegraphics[width=0.139\textwidth]{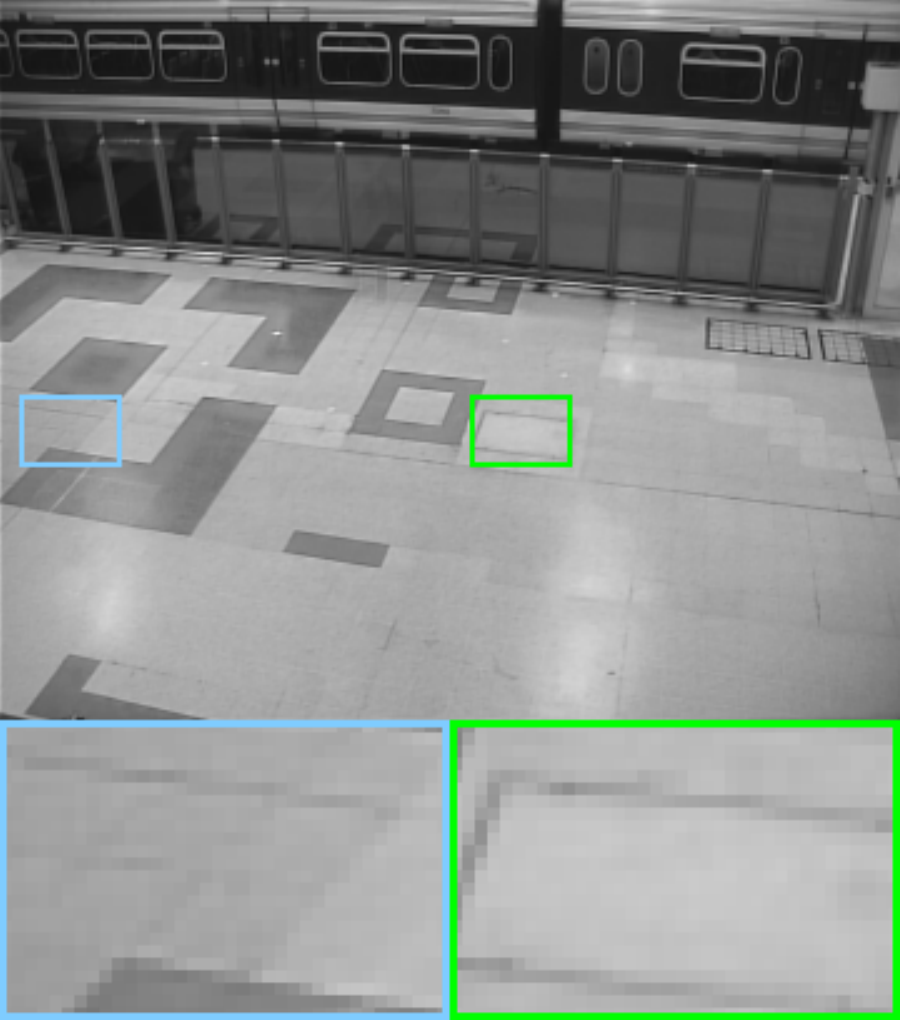}&
\includegraphics[width=0.139\textwidth]{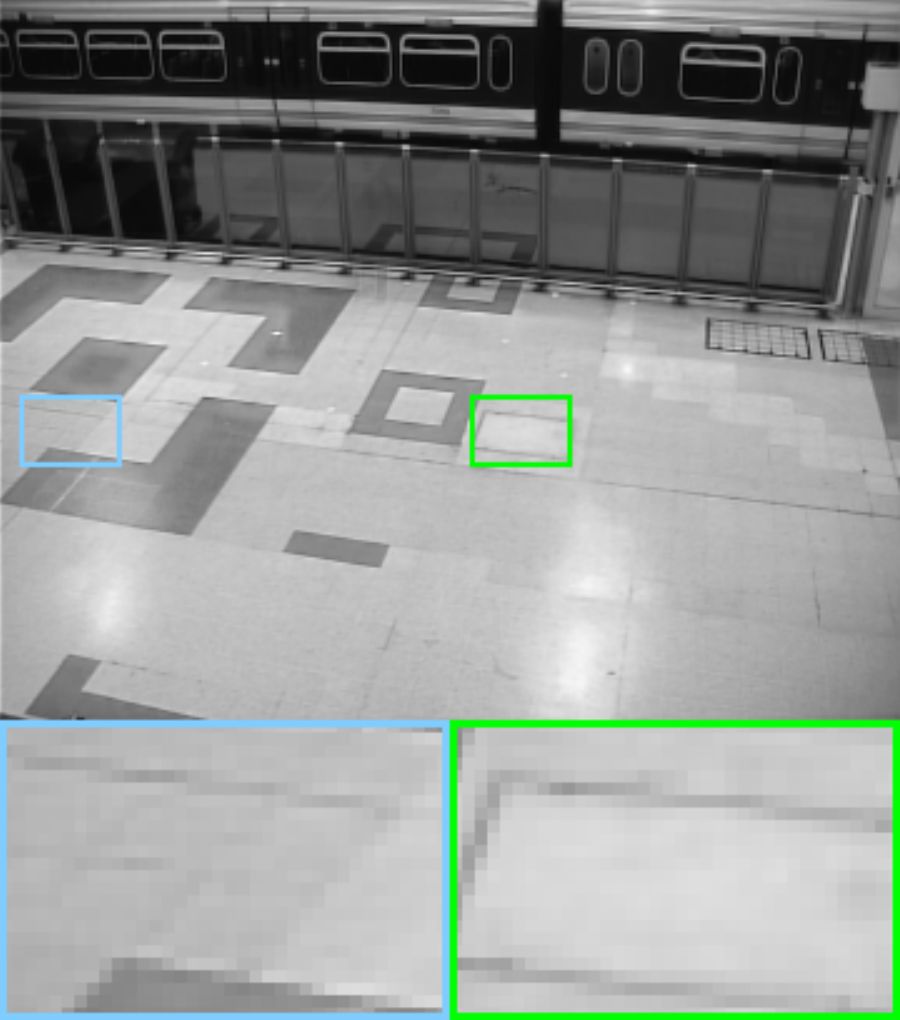}&
\includegraphics[width=0.139\textwidth]{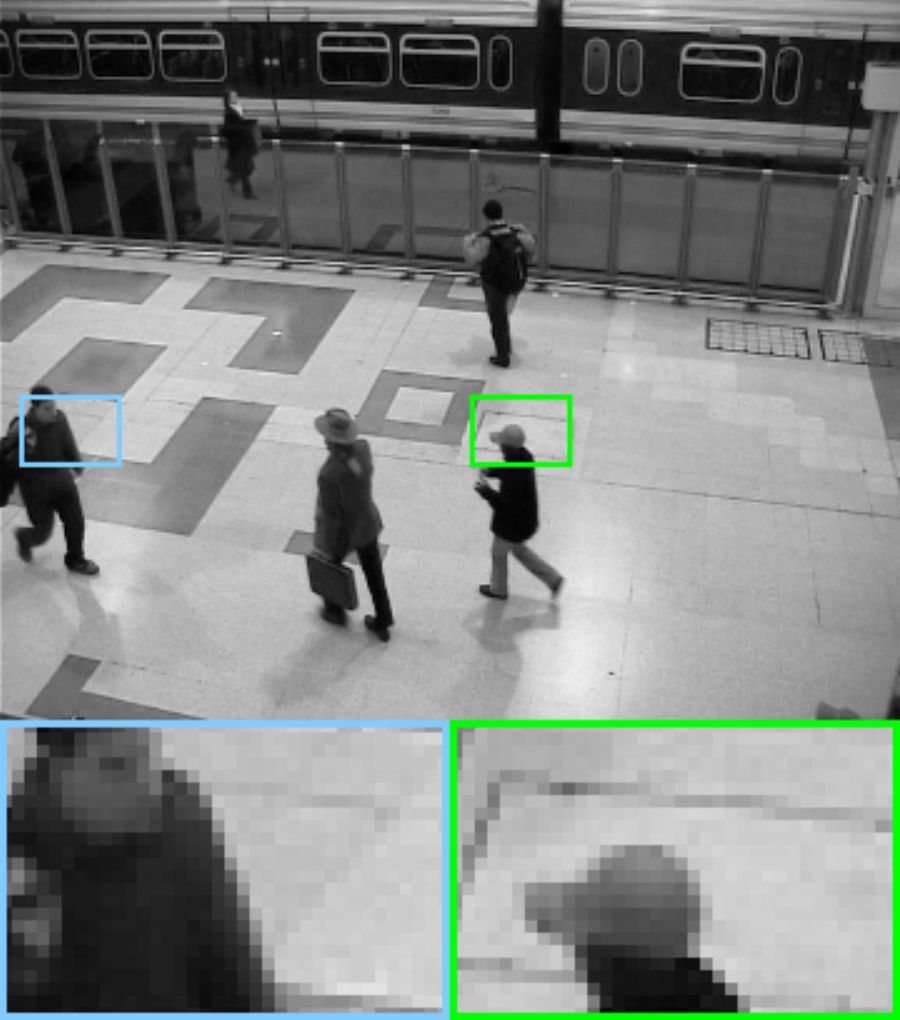}\\
\vspace{-0.3cm}
Observed&RTRC \cite{RTRC}&TNN \cite{TNN_TRPCA}&UTNN \cite{Haar}&SSNT&SSNT-TV&Original\\
\end{tabular}
\end{center}
\caption{The recovered results by different methods for RTC on videos {\it highway} and {\it PET} with SR = 0.25. \label{RTC_fig_2}}\vspace{-0.2cm}
\end{figure*}
\begin{table*}[!h]
\caption{The quantitative results by different methods on different data for SCI. The {\bf BEST} values are highlighted by {\bf BOLDFACE}, and the \underline{SECOND-BEST} values are highlighted by \underline{UNDERLINED}.\label{cs_tab}}
\begin{center}
\scriptsize
\setlength{\tabcolsep}{2.9pt}
\begin{spacing}{0.95}
\begin{tabular}{clcccccccccccccccc}
\toprule
\multirow{2}*{Data}&SR&\multicolumn{3}{c}{0.05}&\multicolumn{3}{c}{0.1}&\multicolumn{3}{c}{0.15}&\multicolumn{3}{c}{0.2}&\multicolumn{3}{c}{0.25}
&\;\multirow{2}*{\tabincell{c}{Time\\(s)}}\\
\cmidrule{2-17}
~&Metric&\;\;\;PSNR&SSIM&SAM\;\;\;&PSNR&SSIM&SAM\;\;\;&PSNR&SSIM&SAM\;\;\;&PSNR&SSIM&SAM\;\;\;&PSNR&SSIM&SAM&~\\
\midrule
\multirow{5}*{\tabincell{c}{
MSI {\it Toys}\\{(256$\times$256$\times$31)}\\}}
~&GAP-TV\cite{GAP-TV}&\;\;\; 21.488&0.642&0.774 \;\;\;&22.318&0.691&0.744 \;\;\;&22.692&0.732&0.699 \;\;\;&22.817&0.755&0.667 \;\;\;&22.766&0.772&0.648&61\\
~&SeSCI\cite{SeSCI}&\;\;\; 20.815&0.612&0.590 \;\;\;&21.574&0.689&0.602 \;\;\;&21.668&0.722&0.602 \;\;\;&21.471&0.738&0.602 \;\;\;&21.218&0.749&0.603&188\\
~&DeSCI\cite{DeSCI}&\;\;\; 19.702&0.624&0.410 \;\;\;&21.211&0.735&0.426 \;\;\;&22.148&0.785&0.437 \;\;\;&22.871&0.812&0.413 \;\;\;&23.220&0.828&0.409&2646\\
~&SSNT&\;\;\; \underline{23.876}&\underline{0.792}&\underline{0.504} \;\;\;&\underline{24.927}&\underline{0.830}&\underline{0.458} \;\;\;&\underline{25.311}&\underline{0.840}&\underline{0.508} \;\;\;&\underline{25.940}&\underline{0.862}&\underline{0.494} \;\;\;&\underline{26.464}&\underline{0.872}&\underline{0.508}&99\\
~&NoHi-TV&\;\;\; \bf{24.209}&\bf{0.803}&\bf{0.434} \;\;\;&\bf{25.424}&\bf{0.847}&\bf{0.436} \;\;\;&\bf{26.308}&\bf{0.863}&\bf{0.470} \;\;\;&\bf{26.791}&\bf{0.880}&\bf{0.450} \;\;\;&\bf{27.122}&\bf{0.885}&\bf{0.486}&174\\
\midrule
\multirow{5}*{\tabincell{c}{
MSI {\it Flowers}\\{(256$\times$256$\times$31)}\\}}
~&GAP-TV\cite{GAP-TV}&\;\;\; 22.944&0.655&0.732 \;\;\;&24.024&0.702&0.683 \;\;\;&24.585&0.741&0.633 \;\;\;&24.864&0.766&0.597 \;\;\;&25.121&0.782&0.577&66\\
~&SeSCI\cite{SeSCI}&\;\;\; 22.405&0.658&0.551 \;\;\;&23.947&0.725&0.546 \;\;\;&24.417&0.758&0.537 \;\;\;&24.578&0.777&0.531 \;\;\;&24.657&0.786&0.535&213\\
~&DeSCI\cite{DeSCI}&\;\;\; 21.150&0.633&0.465 \;\;\;&22.872&0.737&0.411 \;\;\;&23.927&0.783&0.402 \;\;\;&24.604&0.810&0.390 \;\;\;&24.693&0.826&0.382&2740\\
~&SSNT&\;\;\; \underline{26.253}&\underline{0.812}&\underline{0.479} \;\;\;&\underline{26.860}&\underline{0.852}&\underline{0.613} \;\;\;&\underline{28.505}&\bf{0.878}&\underline{0.589} \;\;\;&\underline{28.573}&\underline{0.884}&\bf{0.583} \;\;\;&\underline{29.314}&\underline{0.894}&\bf{0.564}&122\\
~&SSNT-TV&\;\;\; \bf{26.464}&\bf{0.839}&\bf{0.291} \;\;\;&\bf{27.558}&\bf{0.857}&\bf{0.605} \;\;\;&\bf{28.602}&\underline{0.877}&\bf{0.587} \;\;\;&\bf{28.955}&\bf{0.885}&\underline{0.603} \;\;\;&\bf{29.462}&\bf{0.895}&\underline{0.590}&201\\
\midrule
\multirow{5}*{\tabincell{c}{
Video {\it Drop}\\{(256$\times$256$\times$10)}\\}}
~&GAP-TV\cite{GAP-TV}&\;\;\; 23.324&0.732&0.111 \;\;\;&24.077&0.712&0.109 \;\;\;&24.495&0.714&0.106 \;\;\;&24.750&0.718&0.104 \;\;\;&25.248&0.737&0.098&21\\
~&SeSCI\cite{SeSCI}&\;\;\; 24.171&0.85&0.07 \;\;\;&26.135&0.869&0.066 \;\;\;&27.029&0.878&\underline{0.065} \;\;\;&\underline{27.430}&\underline{0.883}&\underline{0.064} \;\;\;&\underline{27.921}&0.888&0.062&125\\
~&DeSCI\cite{DeSCI}&\;\;\; 21.551&0.806&\underline{0.061} \;\;\;&22.880&0.813&0.064 \;\;\;&24.348&0.843&0.067 \;\;\;&25.169&0.860&0.066 \;\;\;&26.283&0.877&0.065&2746\\
~&SSNT&\;\;\; \underline{25.028}&\underline{0.881}&\bf{0.042} \;\;\;&\underline{26.024}&\underline{0.859}&\bf{0.042} \;\;\;&\underline{26.378}&\underline{0.869}&\bf{0.042} \;\;\;&{27.422}&\underline{0.883}&\bf{0.042} \;\;\;&{27.664}&\underline{0.889}&\bf{0.041}&122\\
~&SSNT-TV&\;\;\; \bf{26.674}&\bf{0.894}&\bf{0.042} \;\;\;&\bf{27.519}&\bf{0.898}&\underline{0.043} \;\;\;&\bf{27.396}&\bf{0.892}&\bf{0.042} \;\;\;&\bf{27.637}&\bf{0.908}&\bf{0.042} \;\;\;&\bf{27.951}&\bf{0.913}&\underline{0.042}&142\\
\midrule
\multirow{5}*{\tabincell{c}{
Video {\it Crash}\\{(256$\times$256$\times$10)}\\}}
~&GAP-TV\cite{GAP-TV}&\;\;\; 20.557&0.636&0.265 \;\;\;&21.171&0.626&0.267 \;\;\;&21.546&0.630&0.261 \;\;\;&21.806&0.642&0.258 \;\;\;&22.083&0.661&0.252&21\\
~&SeSCI\cite{SeSCI}&\;\;\; 20.016&0.698&0.203 \;\;\;&21.301&0.699&0.207 \;\;\;&21.880&0.716&0.204 \;\;\;&22.126&0.722&0.206 \;\;\;&22.345&0.734&0.205&120\\
~&DeSCI\cite{DeSCI}&\;\;\; 19.821&0.718&0.152 \;\;\;&20.378&0.708&0.177 \;\;\;&21.068&0.727&\underline{0.180} \;\;\;&21.178&0.732&0.189 \;\;\;&21.305&0.746&0.193&2466\\
~&SSNT&\;\;\; \underline{21.469}&\underline{0.787}&\underline{0.126} \;\;\;&\underline{22.117}&\underline{0.706}&\underline{0.137} \;\;\;&\underline{22.781}&\underline{0.780}&\bf{0.125} \;\;\;&\underline{22.993}&\underline{0.806}&\underline{0.137} \;\;\;&\underline{23.527}&\underline{0.829}&\underline{0.126}&124\\
~&SSNT-TV&\;\;\; \bf{21.906}&\bf{0.790}&\bf{0.125} \;\;\;&\bf{22.901}&\bf{0.796}&\bf{0.126} \;\;\;&\bf{23.201}&\bf{0.821}&\bf{0.125} \;\;\;&\bf{23.403}&\bf{0.830}&\bf{0.126} \;\;\;&\bf{23.598}&\bf{0.842}&\bf{0.125}&143\\
\bottomrule
\end{tabular}
\end{spacing}
\end{center}
\vspace{-0.6cm}
\end{table*}
\begin{figure}[!t]
\centering
\setlength{\abovecaptionskip}{0.1cm}
\includegraphics [width = 1\linewidth]{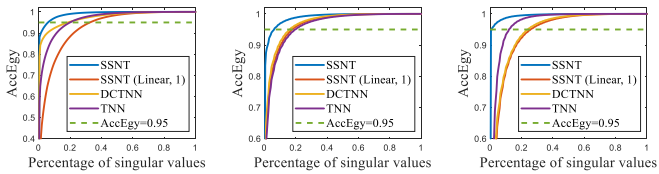}
\caption{The AccEgy (defined as $\sum_{i=1}^k\sigma_i^2/\sum_j\sigma_j^2$, where $\sigma_i$ is the $i$-th singular value) with respect to percentage of singular values of the transformed frontal slices of {\it Pavia}, {\it WDC mall}, and {\it Beads}. We can observe that SSNT obtains a better low-rank representation whose energy is concentrated in the larger singular values. Thus, SSNT could achieve more promising results.\label{en_fig}} \vspace{-0.3cm}
\end{figure}
\begin{figure*}[!t]
\footnotesize
\setlength{\tabcolsep}{0.9pt}
\begin{center}
\begin{tabular}{ccccccc}
\includegraphics[width=0.139\textwidth]{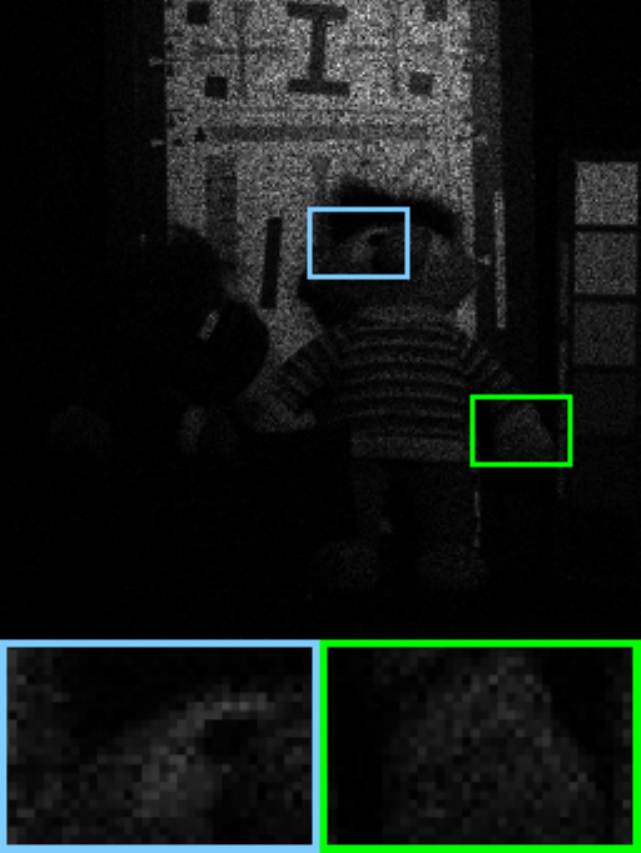}&
\includegraphics[width=0.139\textwidth]{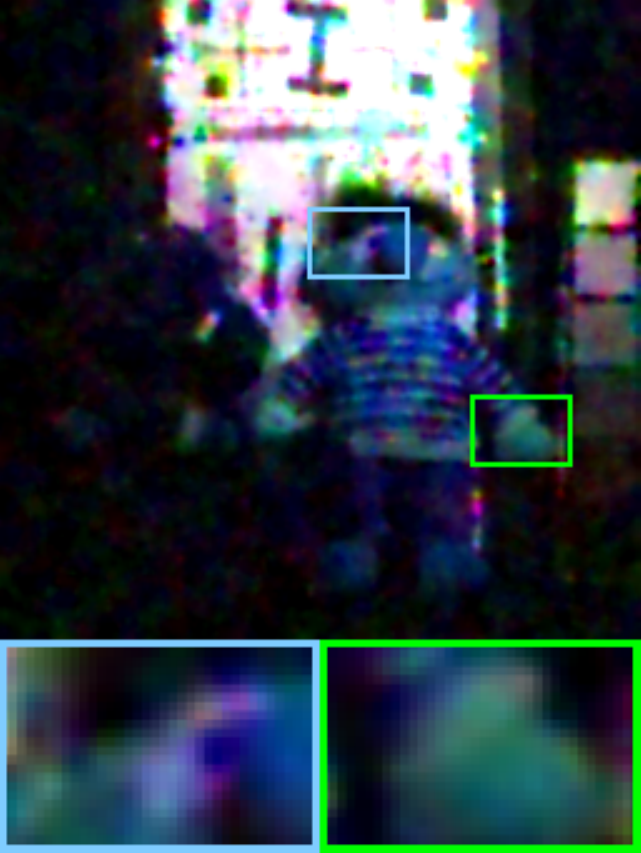}&
\includegraphics[width=0.139\textwidth]{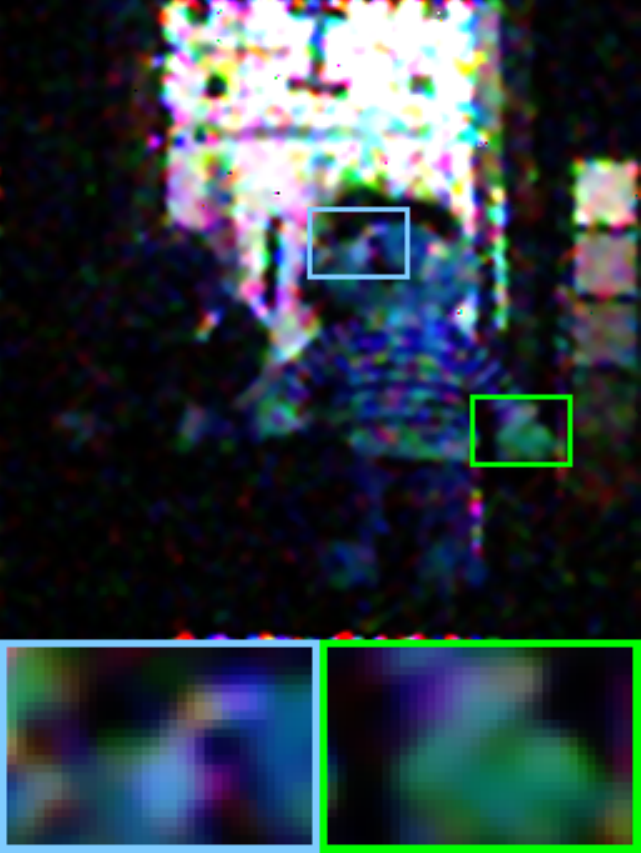}&
\includegraphics[width=0.139\textwidth]{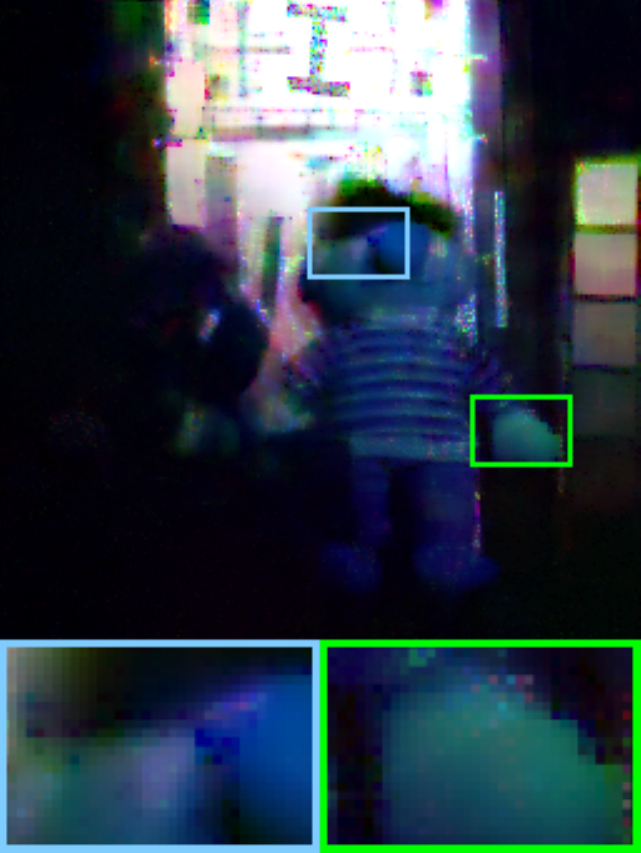}&
\includegraphics[width=0.139\textwidth]{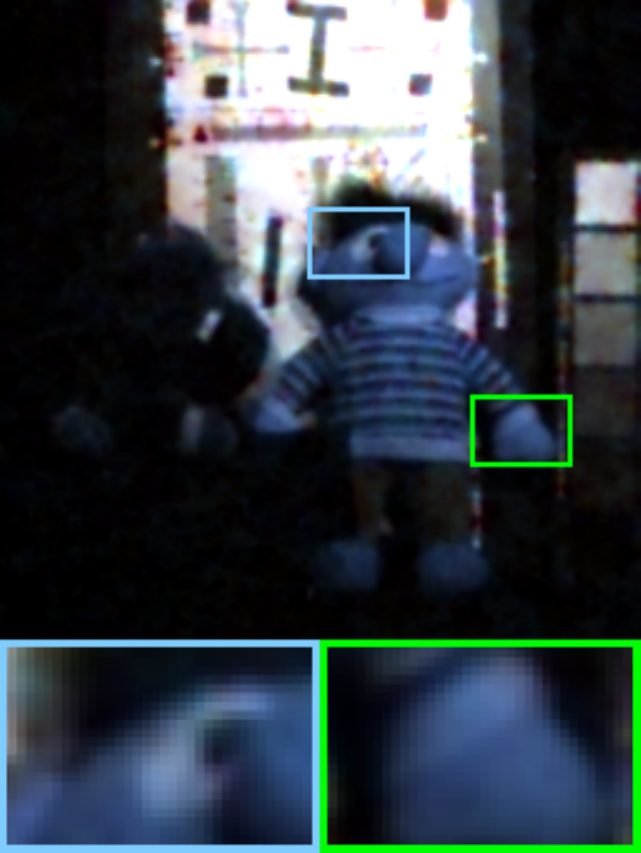}&
\includegraphics[width=0.139\textwidth]{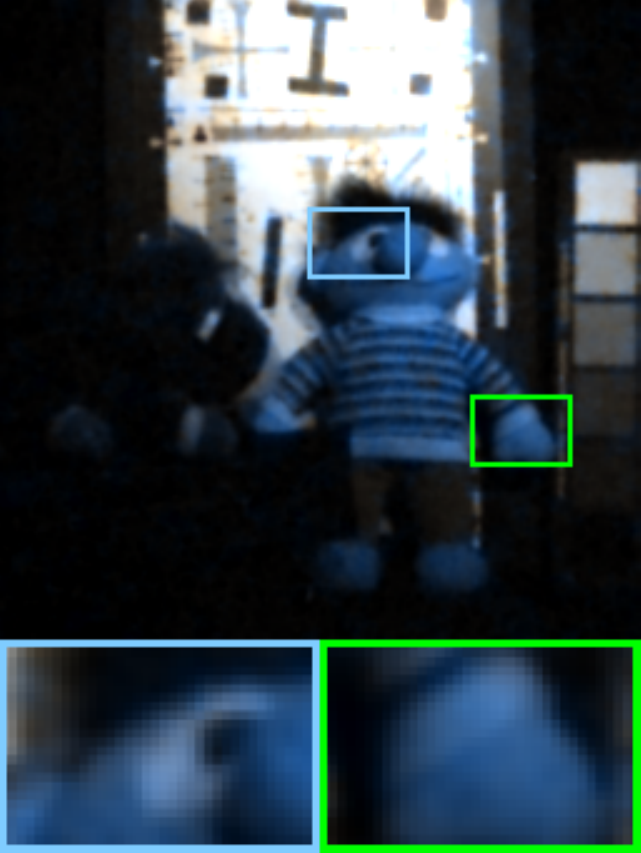}&
\includegraphics[width=0.139\textwidth]{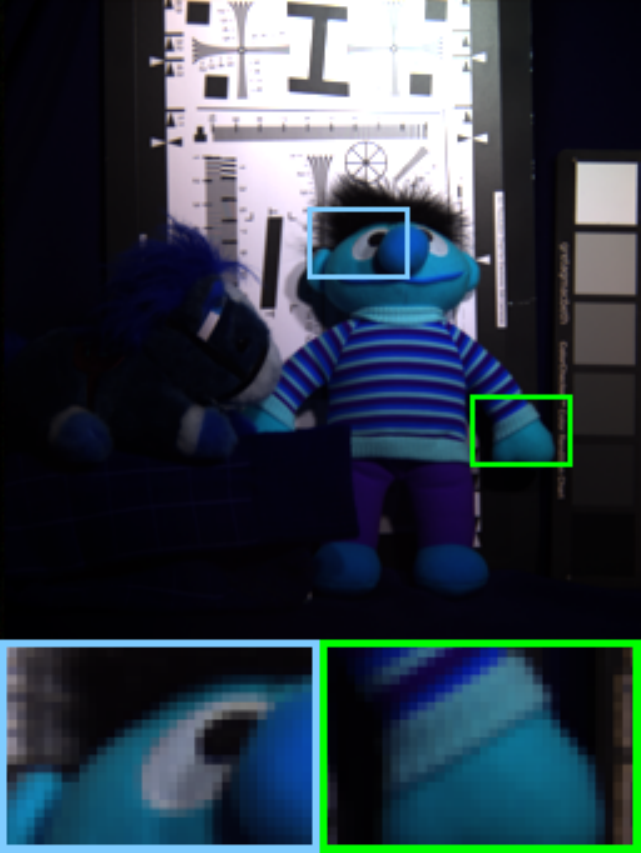}\\
\vspace{0.1cm}
~ & PSNR 22.766 dB&PSNR 21.218 dB&PSNR 23.220 dB&PSNR 26.464 dB&PSNR 27.122 dB&PSNR Inf\\

\includegraphics[width=0.139\textwidth]{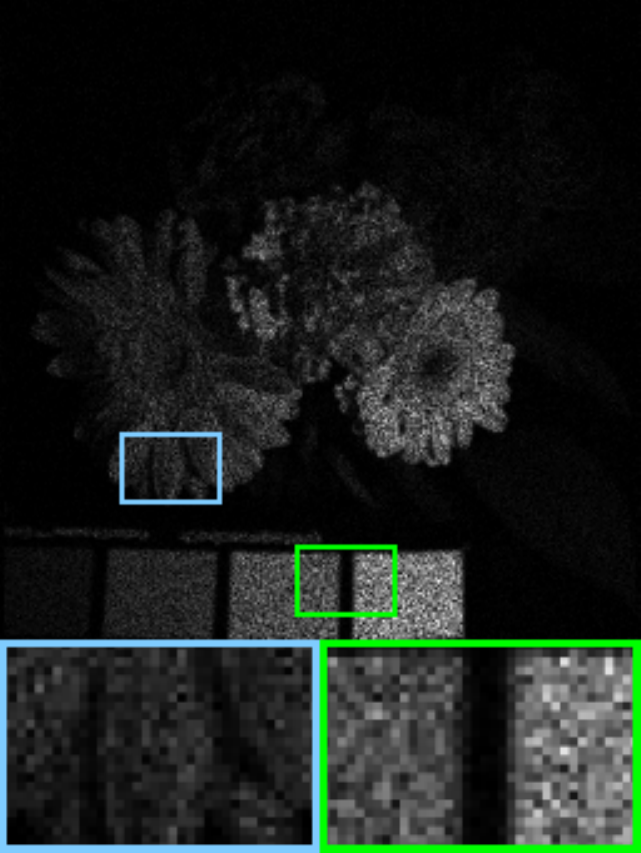}&
\includegraphics[width=0.139\textwidth]{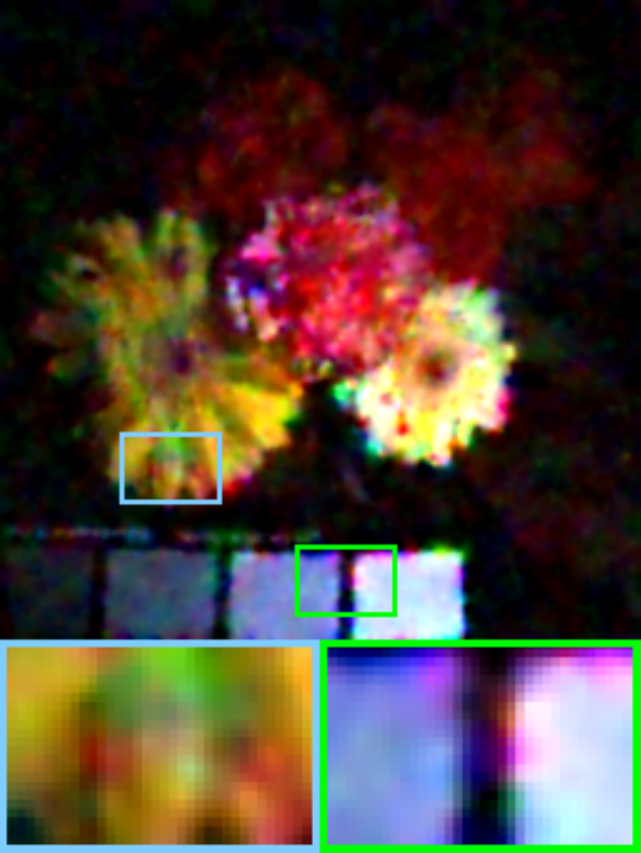}&
\includegraphics[width=0.139\textwidth]{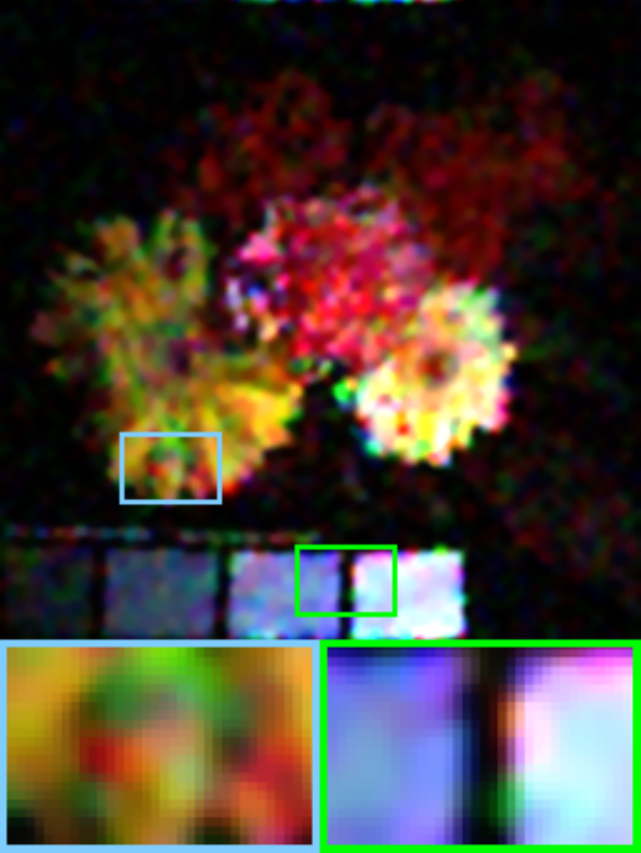}&
\includegraphics[width=0.139\textwidth]{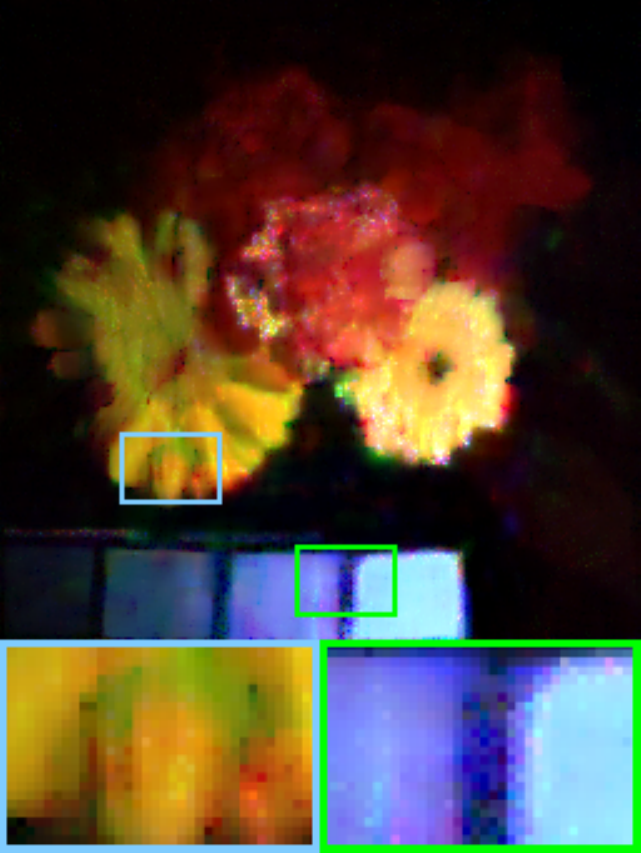}&
\includegraphics[width=0.139\textwidth]{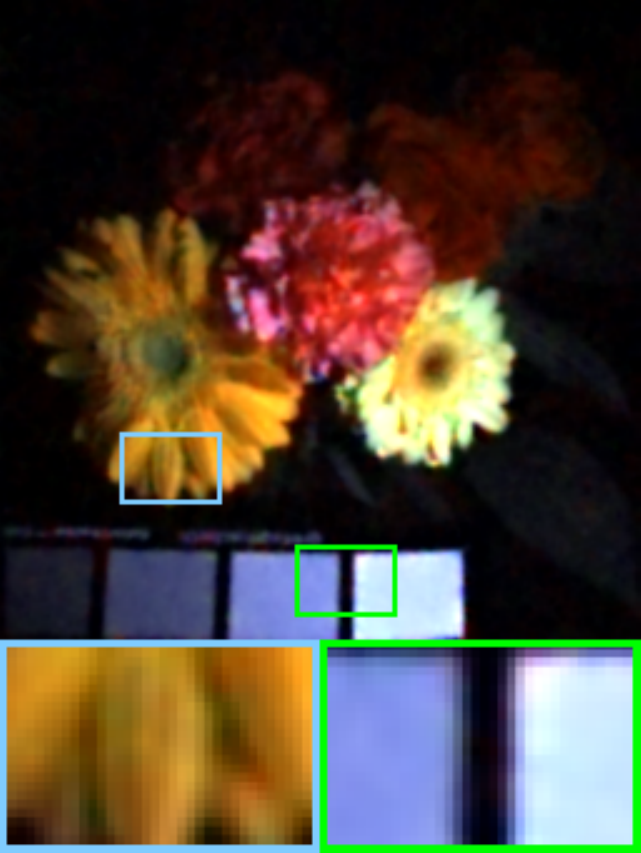}&
\includegraphics[width=0.139\textwidth]{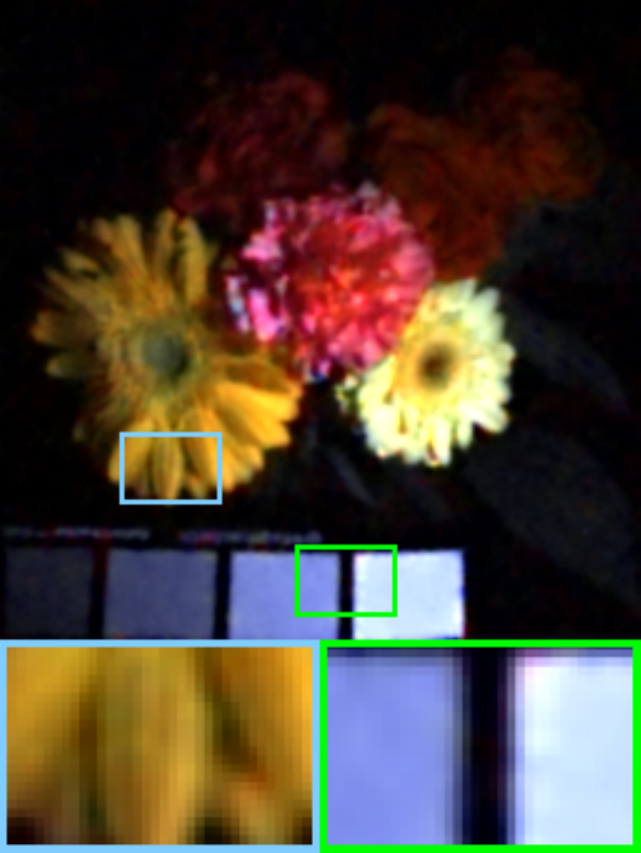}&
\includegraphics[width=0.139\textwidth]{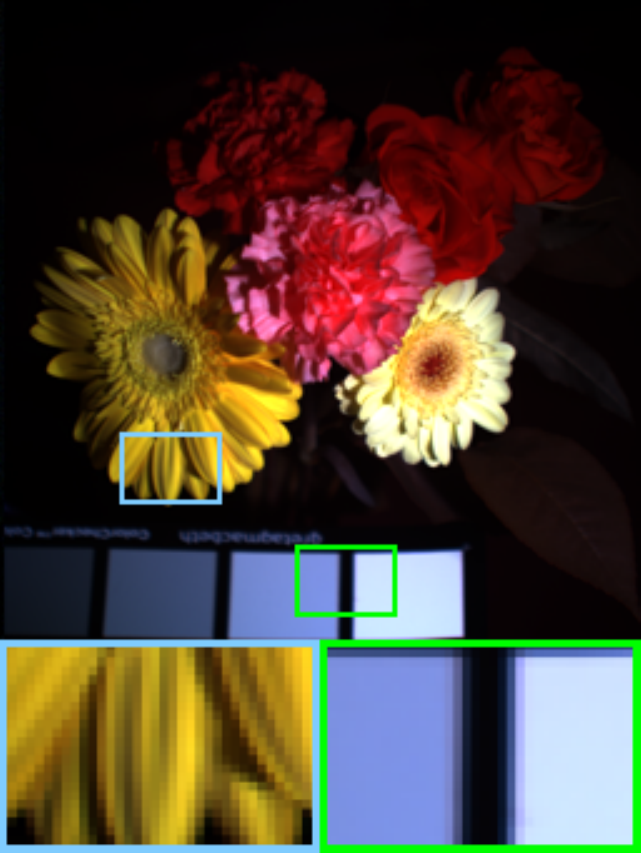}\\
\vspace{0.1cm}
~ & PSNR 25.121 dB&PSNR 24.657 dB&PSNR 24.693 dB&PSNR 29.314 dB&PSNR 29.462 dB&PSNR Inf\\
\includegraphics[width=0.139\textwidth]{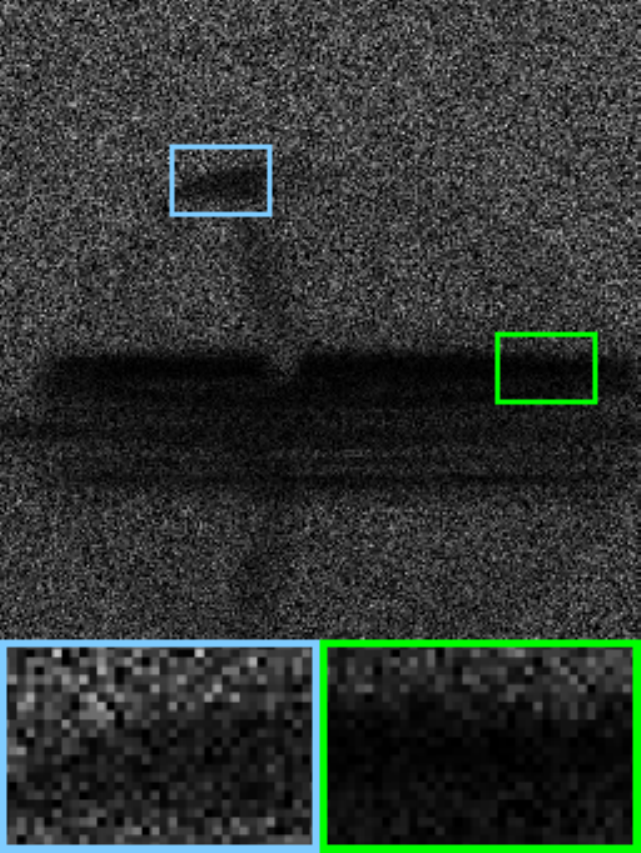}&
\includegraphics[width=0.139\textwidth]{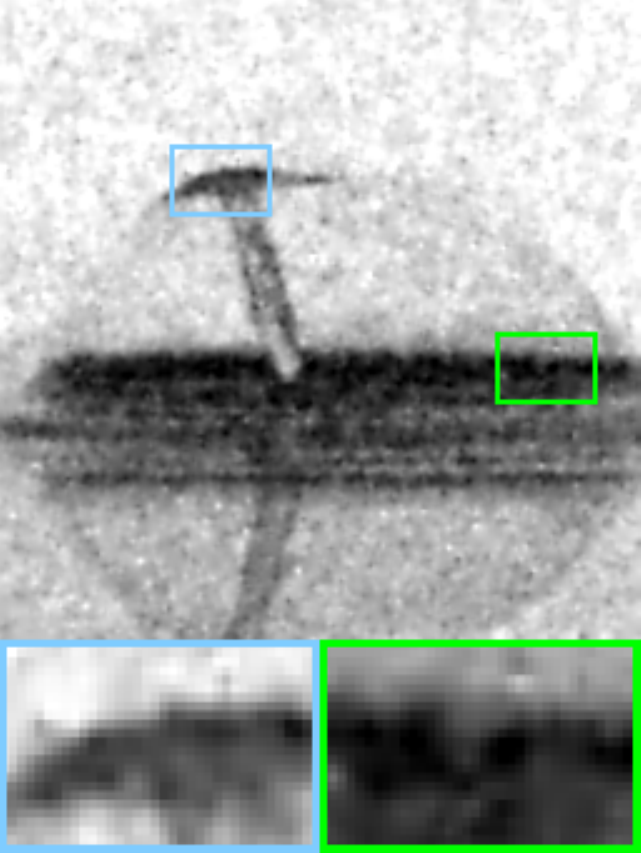}&
\includegraphics[width=0.139\textwidth]{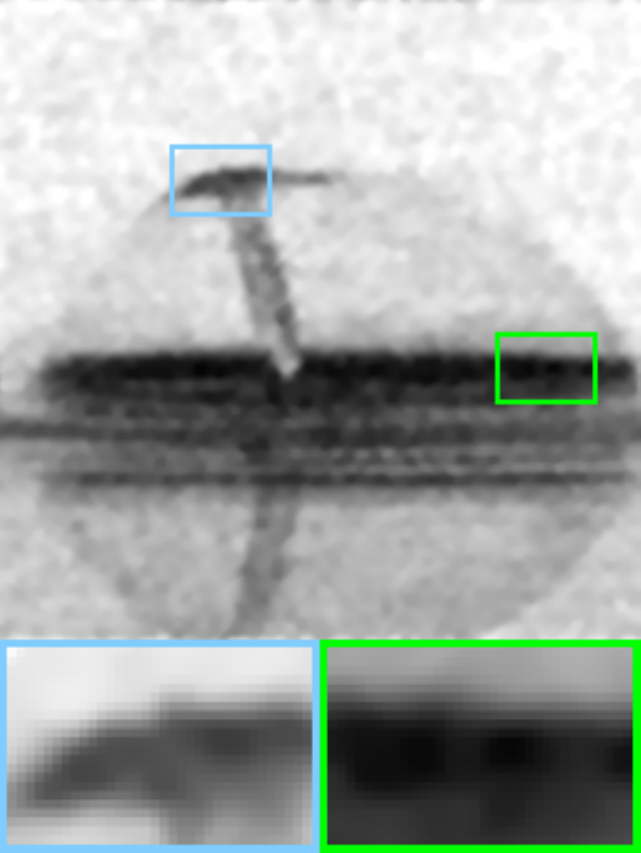}&
\includegraphics[width=0.139\textwidth]{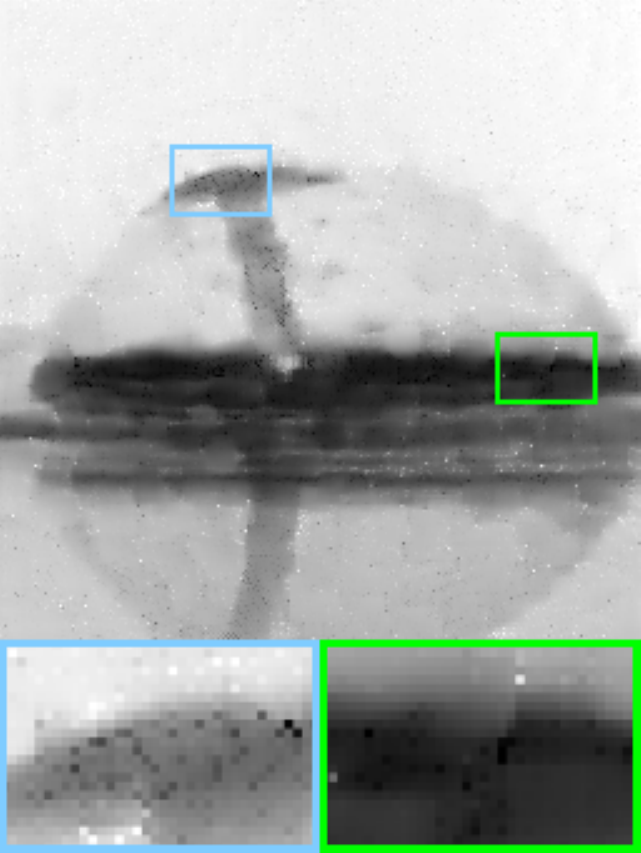}&
\includegraphics[width=0.139\textwidth]{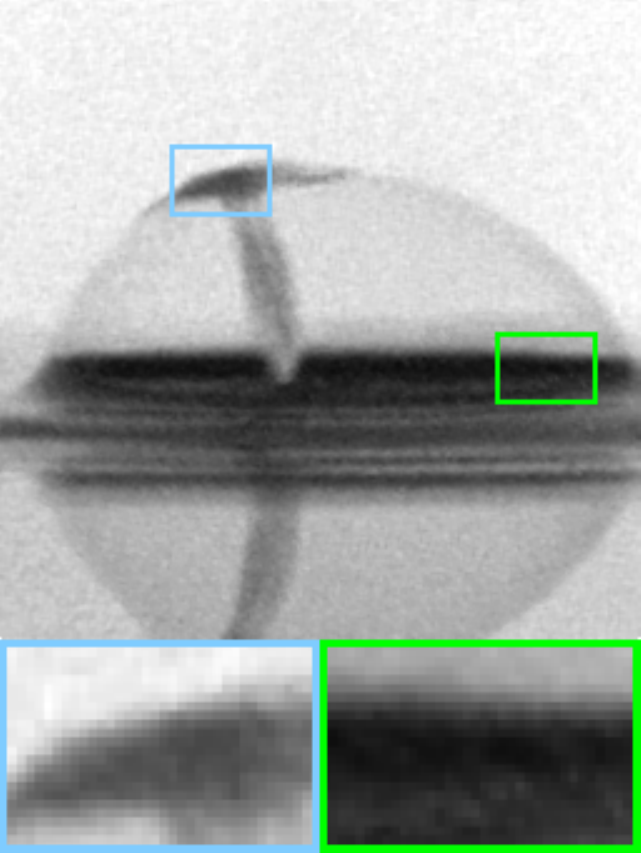}&
\includegraphics[width=0.139\textwidth]{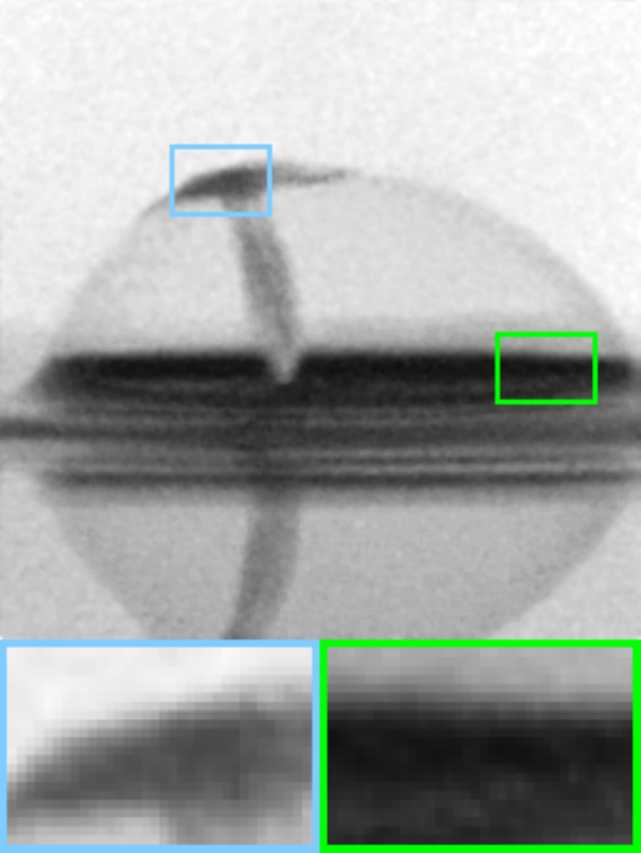}&
\includegraphics[width=0.139\textwidth]{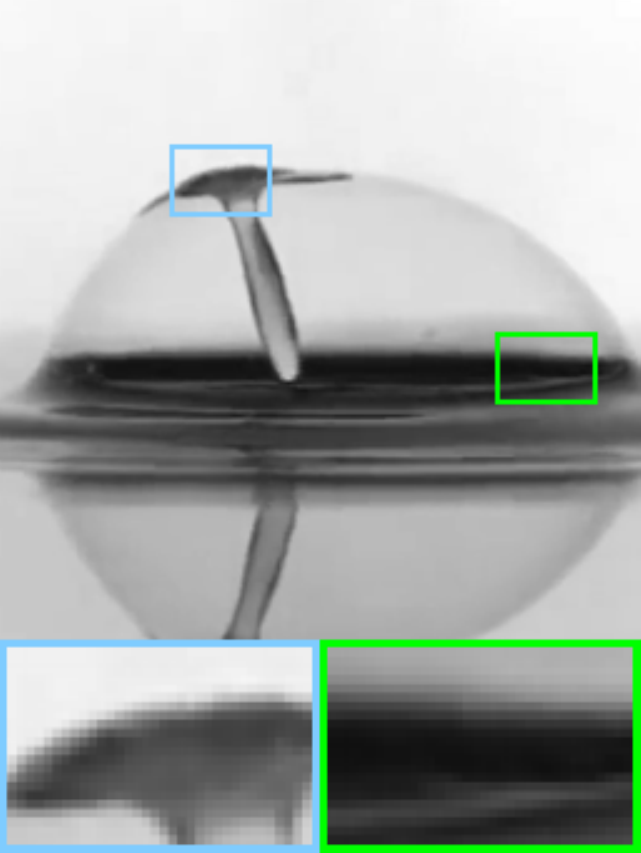}\\
\vspace{0.1cm}
~ & PSNR 25.248 dB&PSNR 27.921 dB&PSNR 26.283 dB&PSNR 27.664 dB&PSNR 27.951 dB&PSNR Inf\\
\includegraphics[width=0.139\textwidth]{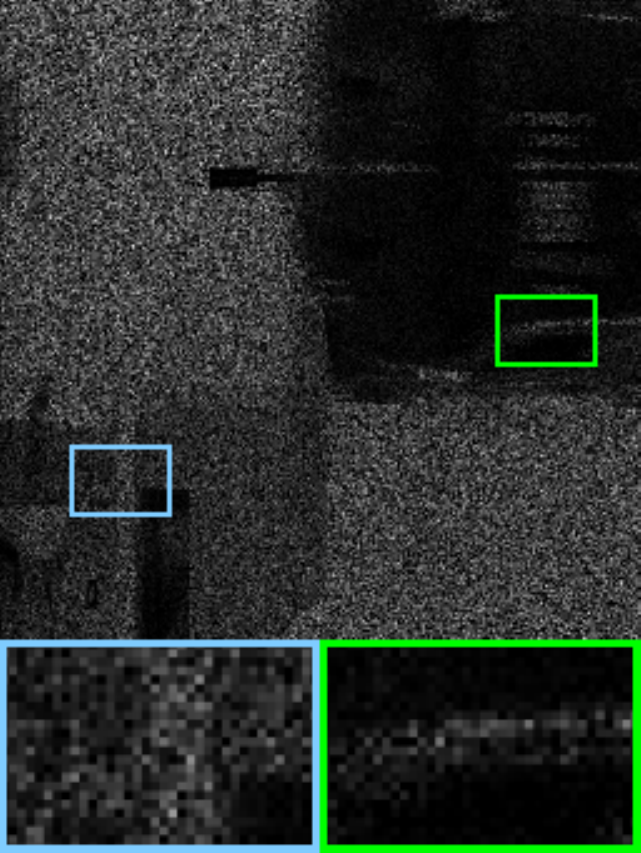}&
\includegraphics[width=0.139\textwidth]{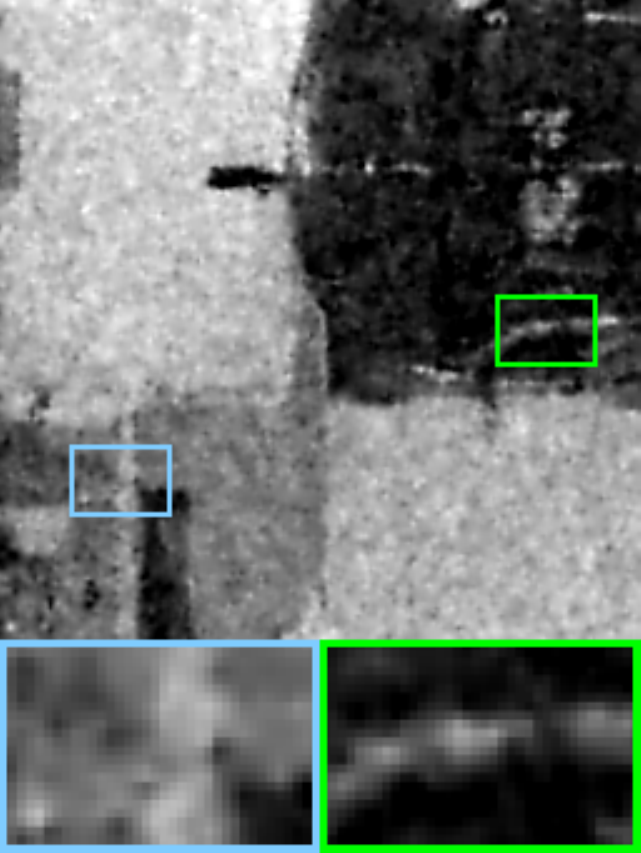}&
\includegraphics[width=0.139\textwidth]{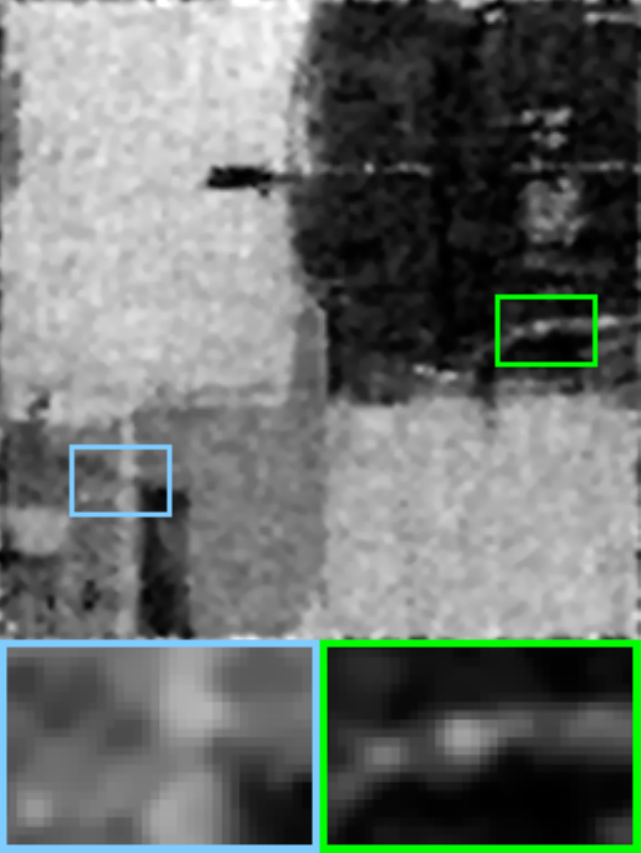}&
\includegraphics[width=0.139\textwidth]{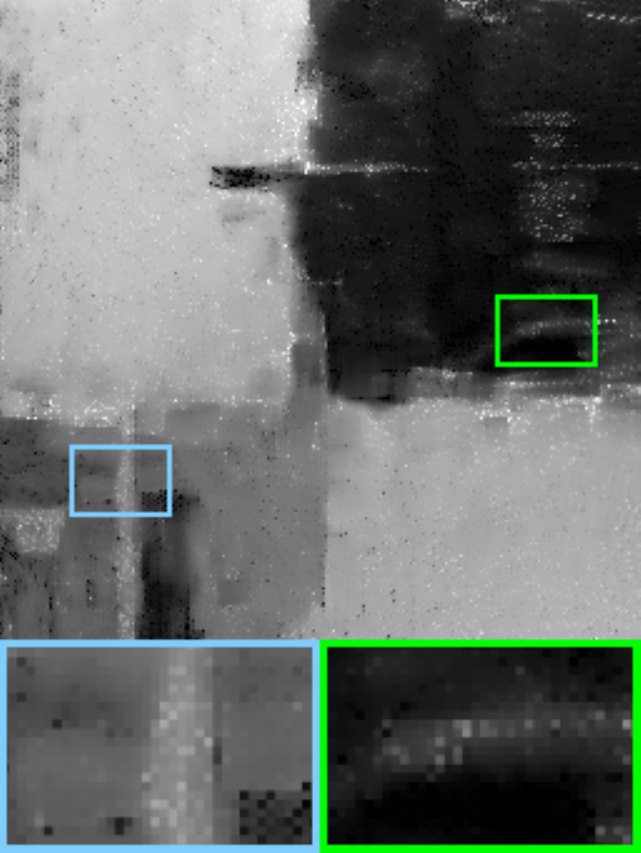}&
\includegraphics[width=0.139\textwidth]{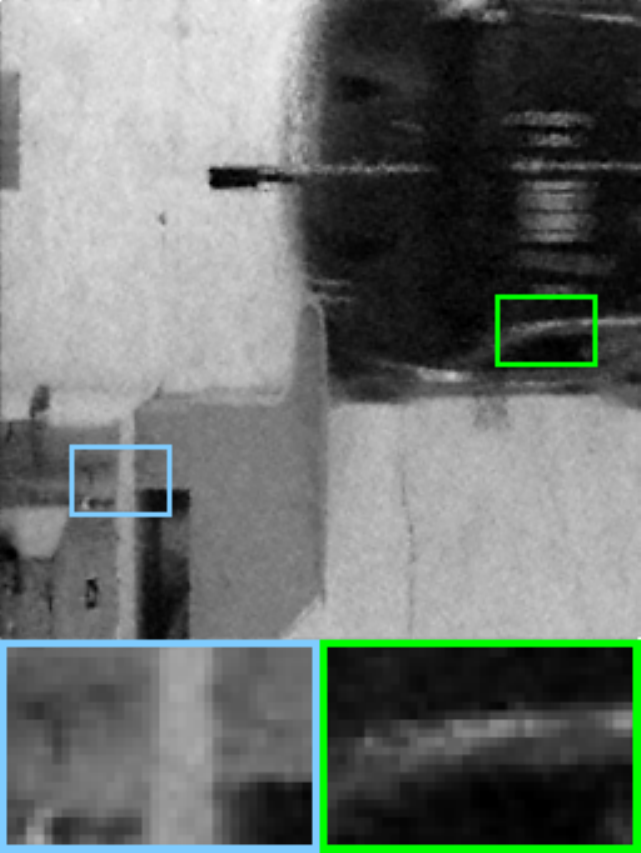}&
\includegraphics[width=0.139\textwidth]{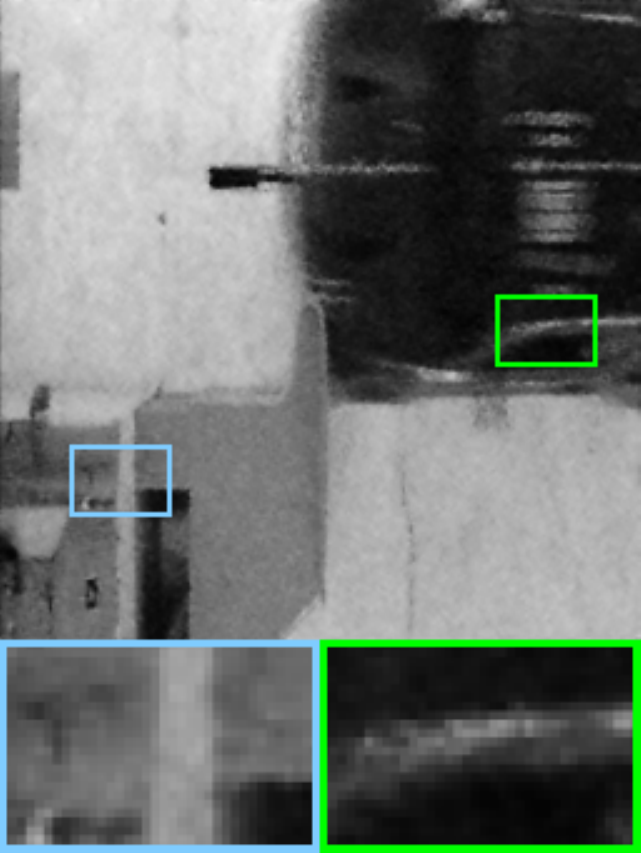}&
\includegraphics[width=0.139\textwidth]{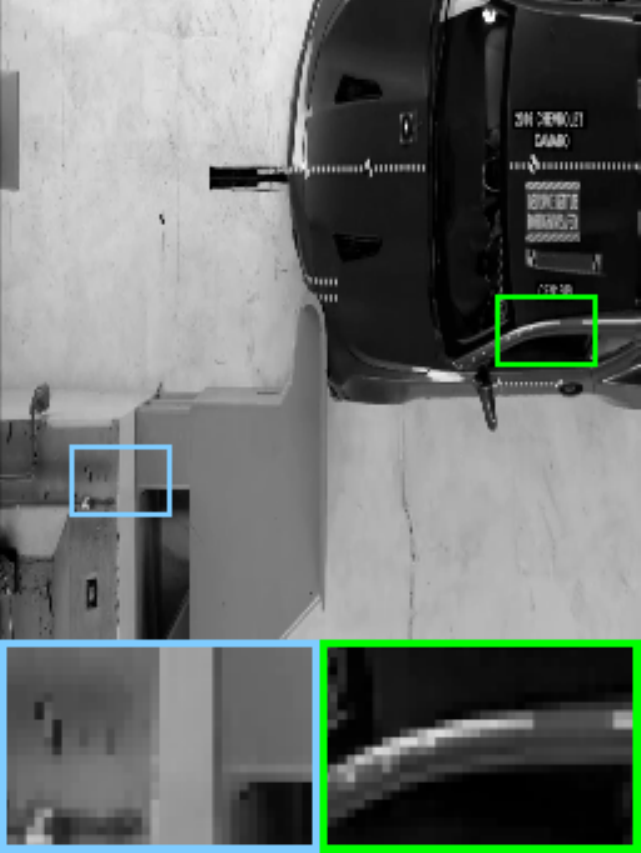}\\
~ & PSNR 22.083 dB&PSNR 22.345 dB&PSNR 21.305 dB&PSNR 23.527 dB&PSNR 23.598 dB&PSNR Inf\\
\vspace{-0.3cm}
Observed & GAP-TV \cite{GAP-TV} &SeSCI \cite{SeSCI}&DeSCI \cite{DeSCI}&SSNT&SSNT-TV&Original\\
\end{tabular}
\end{center}
\caption{The recovered results by different methods for SCI on MSIs {\it Toys} (composed of the 10-th, 20-th, and the 30-th bands) with SR = 0.25, {\it Flowers} (composed of the 10-th, 20-th, and the 30-th bands) with SR = 0.25, videos {\it Drop} with SR = 0.25, and {\it Crash} with SR = 0.25. \label{cs_fig}}\vspace{-0.1cm}
\end{figure*}
\begin{figure*}[t]
\tiny
\setlength{\tabcolsep}{0.9pt}
\begin{center}
\begin{tabular}{c}
\vspace{-0.3cm}
\includegraphics[width=0.9\textwidth]{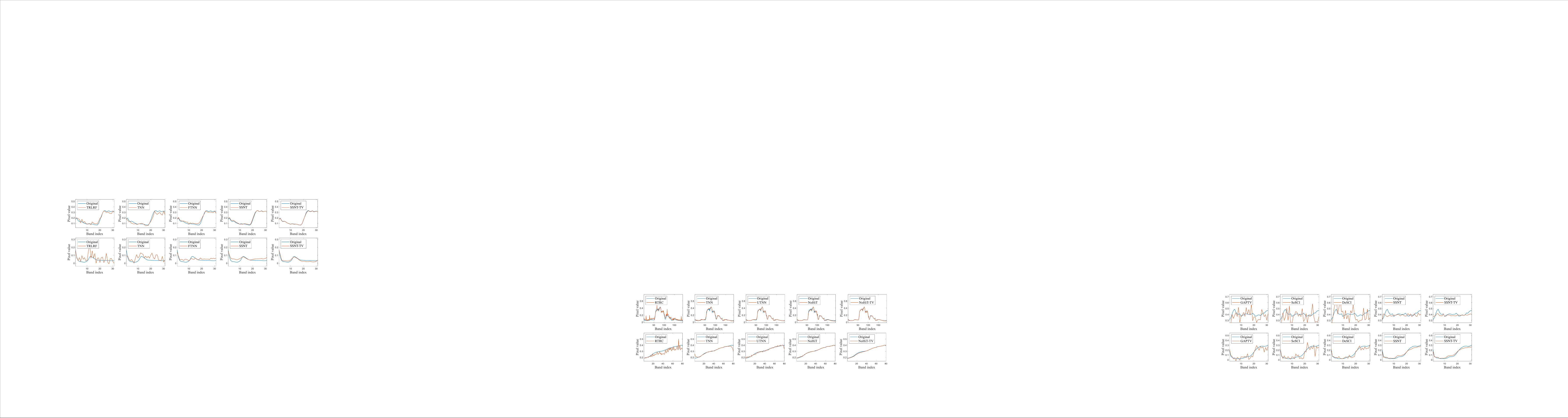}\\
\end{tabular}
\end{center}
\caption{The spectral curves of recovered results by different methods for SCI on MSIs {\it Toys} and {\it Flowers} with SR = 0.25. \label{cs_spec}}\vspace{-0.4cm}
\end{figure*}
\subsubsection{Experimental Results}
The numerical results for SCI are shown in Table \ref{cs_tab}. We can see that SSNT-TV outperforms competing methods with a considerable margin. Also, the running time comparison shows that the proposed methods are relatively efficient, which is crucial for processing large-scale sensing images.\par
The visual results for SCI are illustrated in Fig. \ref{cs_fig}. We can see that the proposed SSNT and SSNT-TV have better visual results and can recover the spatial information of the tensor more precisely and robustly. SSNT-TV has relatively smoother results than other methods due to the consideration of the spatial local smoothness. \par
In the last, we plot the spectral curves of recovered results on SCI in Fig. \ref{cs_spec}. We can see that SSNT and SSNT-TV could preserve the nonlinear spectral curves better. This is due to the nonlinear modeling capability of SSNT, where a more compact low-rank representation is obtained and the nonlinear nature of data is well preserved.
\begin{table}[t]
\caption{The quantitative results for tensor completion on MSI {\it Flowers} with SR = 0.1. SSNT (Linear) denotes the SSNT without nonlinear function. SSNT ($p$) indicates that $f$ has $p$ NoFC$_3$ layers. SSNT wo reg. denotes the SSNT without regularization.\label{l1_com_tab}}
\begin{center}
\footnotesize
\scriptsize
\setlength{\tabcolsep}{4.5pt}
\begin{spacing}{0.95}
\begin{tabular}{llcccc}
\toprule
~&Method&PSNR&SSIM&SAM&Time (s)\\
\midrule
\multirow{5}*{\bf Nonlinearity}&{SSNT (Linear)}&35.786 & 0.973 & 0.164 &291\\
~&{SSNT (ReLU)}& 36.850 & \bf0.980 & 0.122&{360} \\
~&{SSNT (LeakyReLU)}& \bf36.997 & 0.978 & 0.138& {360} \\
~&{SSNT (PReLU)} & 36.734 & 0.977 & 0.134 & 430\\
~&{SSNT (PLU)}& 36.620 & 0.979 & \bf0.106& 780\\
\midrule
\multirow{6}*{\bf Hierarchy}&{SSNT ($1$)}& 36.434 & 0.975 & 0.139& 326\\
~&{SSNT ($2$)}& 36.997 & 0.978 & 0.138 & 360\\
~&{SSNT ($3$)}& \bf37.407 & \bf0.981 & \bf 0.133& 394\\
~&{SSNT ($4$)}& 36.612 & 0.976 & 0.156&{403} \\
~&{SSNT ($5$)}& 35.863 & 0.969 & 0.203&{427} \\
~&{SSNT ($10$)}& 31.512 & 0.921 & 0.389&{467} \\
\midrule
\multirow{3}*{\bf Regularizers}&{SSNT wo reg.}& 33.397 & 0.943 & 0.306&82\\
~&{SSNT (Low-rank)}& \bf36.997 & \bf0.978 & \bf0.138& 360\\
~&{SSNT (Sparse)}& 34.179 & 0.961 & 0.284&90\\
\bottomrule
\end{tabular}
\end{spacing}
\end{center}
\vspace{-0.4cm}
\end{table}
\begin{figure}[h]
\scriptsize
\setlength{\tabcolsep}{0.9pt}
\begin{center}
\begin{tabular}{cccc}
\includegraphics[width=0.115\textwidth]{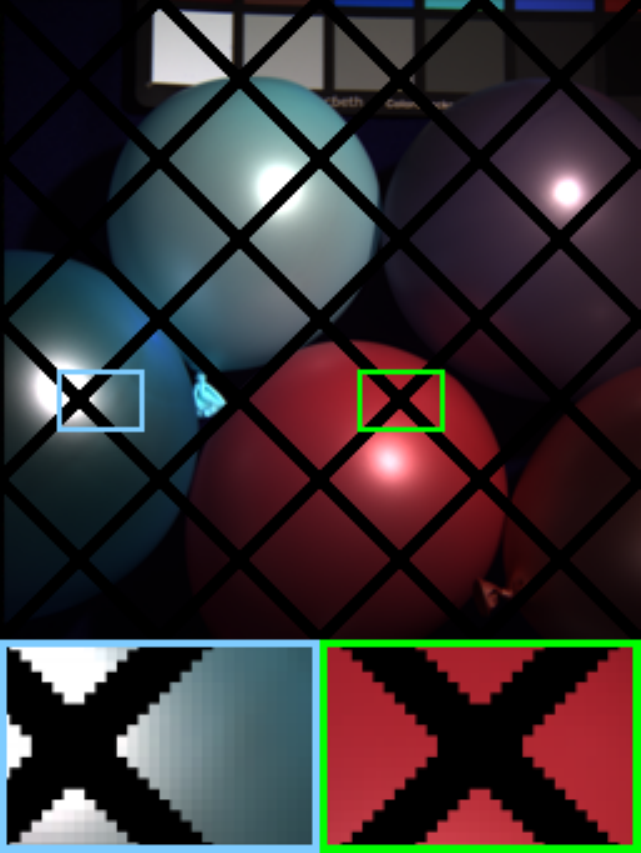}&
\includegraphics[width=0.115\textwidth]{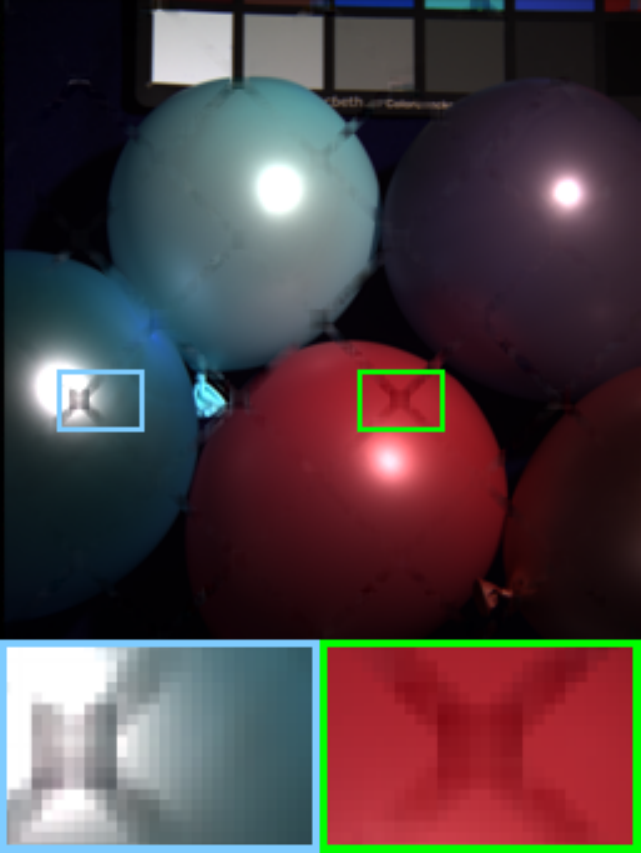}&
\includegraphics[width=0.115\textwidth]{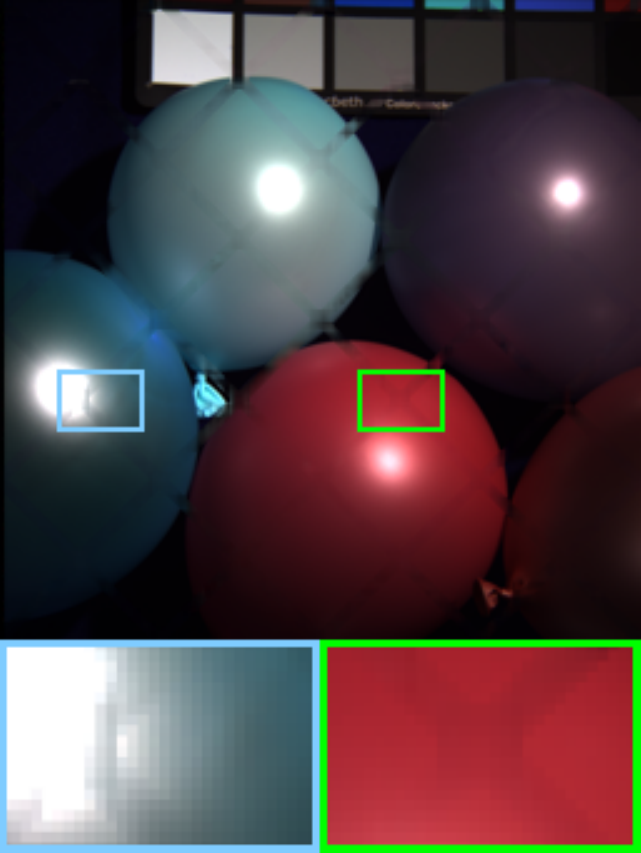}&
\includegraphics[width=0.115\textwidth]{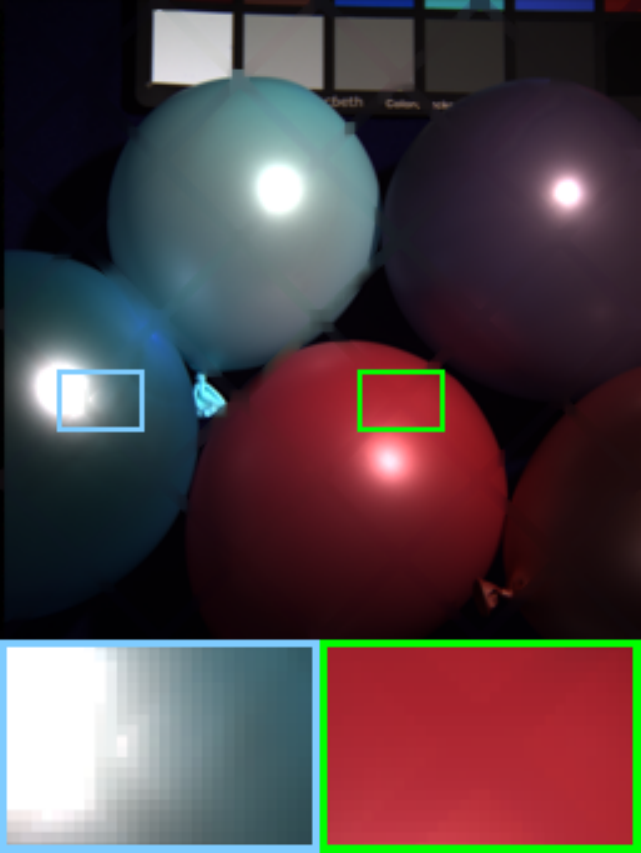}
\\
\vspace{-0.3cm}
Observed 19.8dB& FTNN\cite{FTNN} 33.9dB& SSNT 38.1dB& SSNT-TV 39.9dB\\
\end{tabular}
\end{center}
\caption{The recovered results and corresponding PSNR values by different methods for tensor completion on {\it Balloons} with structure missing. \label{str}\vspace{-0.2cm}}
\end{figure}
\vspace{-0.1cm}
\subsection{Discussions}\label{Sec:Dis}
\subsubsection{Compact Representation by SSNT}
To demonstrate that the proposed SSNT can obtain a better low-rank representation than linear transforms, we plot the accumulation energy ratio (AccEgy) with respect to the percentage of singular values of the transformed frontal slices by SSNT, SSNT (Linear, 1) (SSNT (Linear, 1) denotes that $f$ and $g$ only have one linear fully connected layer without nonlinear activation function.), DCT, and DFT in Figure \ref{en_fig}. We can observe that SSNT obtains a more compact representation with more energy concentrated in larger singular values. This could greatly benefit the recovery performance, where the data can be approximated via lower-rank representation. In contrast, SSNT (Linear, 1) obtains a less compact representation. This verifies the effectiveness of nonlinearity and the hierarchical structure of SSNT for exploring a better low-rank representation. \par
\subsubsection{Effectiveness of Nonlinearity} This section further examines the effect of nonlinearity in the proposed framework. Specifically, we compare the performance of SSNT without nonlinear layers (denoted as SSNT (Linear)) and SSNT with different nonlinear activation layers, {i.e.}, ReLU, LeakyReLU, PReLU \cite{PReLU}, and piecewise linear unit (PLU) \cite{PLU}. The results are shown in the first block of Table \ref{l1_com_tab}. We can see that the performance is considerably increased with nonlinear layers, in which the nonlinearity provides the ability to explore the nonlinear nature of the data.
\subsubsection{Effectiveness of Hierarchy} In this section, we discuss the influence of the hierarchy, i.e., the number of layers of the proposed SSNT. Specifically, we change the number of NoFC$_3$ layers in $f$ to clarify the influence. The results are displayed in the second block of Table \ref{l1_com_tab}. When the number $p$ is small, increasing $p$ could enhance the numerical performance. However, when $p$ is larger, the numerical results are not as desirable as we expected. The possible reason is that a deeper network is more likely to suffer from the vanishing gradient since mode-3 tubes can be fitted reasonably with limited NoFC$_3$ layers. 
\subsubsection{Low-Rankness vs Sparsity} The sparse modeling of the data has achieved great success \cite{spa_encoder,XieQi_PAMI}. Dose the sparsity works in our framework? To clarify this, we replace the low-rank term with the sparse term ${\mathcal L}_1 = \lambda\sum_{k=1}^{{\tilde n}_3} \lVert (f({\mathcal O}))^{(k)}\rVert_{\ell_1}$, where $\ell_1$ is the relaxation of $\ell_0$. Meanwhile, we use the SSNT without regularization ({i.e.}, ${\mathcal L}_1$ = 0) as the baseline. The results are shown in the third block of Table \ref{l1_com_tab}. We can observe that SSNT (Low-rank) considerably outperforms SSNT (Sparse), which reveals that low-rankness is more effective to represent the third-order tensor in our framework.
\subsubsection{Effectiveness of TV Regularization}
The SSNT only considers tensor low-rankness, which is limited to capture the spatial local correlation. This motivates us to perform the TV regularization on the spatial domain to faithfully explore the spatial local smoothness for better performance. To clarify this, we conduct the experiment for tensor completion where incompleted entries are structurally missed. The results are shown in Fig. \ref{str}. We can see that SSNT-TV outperforms SSNT, where the spatial information is better recovered by SSNT-TV. This verifies the effectiveness of TV regularization that the spatial local correlation is preferably exploited. 
\begin{figure}[t]
\scriptsize
\setlength{\tabcolsep}{0.9pt}
\begin{center}
\begin{tabular}{c}
\vspace{-0.2cm}
\includegraphics[width=1\linewidth]{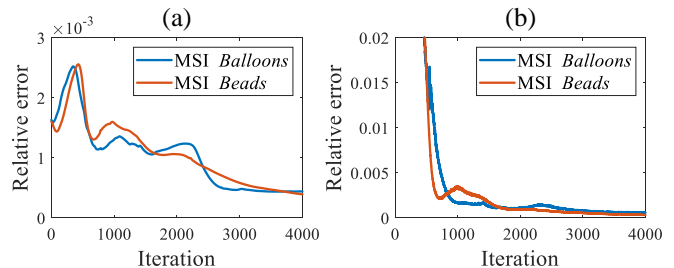}
\end{tabular}
\end{center}
\caption{The relative error with respect to iteration for tensor completion with SR = 0.25. (a) The relative error of network parameters, i.e., $\sum_{i=1}^{p+q}{\lVert {\bf W}_i^{t+1}-{\bf W}_i^{t}\rVert_F}/{\lVert {\bf W}_i^{t}\rVert_F}$. (b) The relative error of  ${\mathcal V}_p$, i.e., $\sum_{p=1,2}{\lVert {\mathcal V}_p^{t+1}-{\mathcal V}_p^{t}\rVert_F}/{\lVert {\mathcal V}_p^{t}\rVert_F}$.\label{convergence}}
\vspace{-0.3cm}
\end{figure}
\subsubsection{Convergence Analysis}
To test the convergence of the ADMM-like Algorithm \ref{alg_1}, we plot the relative error of variables with respect to the iteration number in Fig. \ref{convergence}. The downward trend of the curves verifies the convergence behavior of our method.
\section{Conclusion}\label{Sec:Con}
This paper suggests the self-supervised nonlinear transform-based TNN. Concretely, a nonlinear multilayer neural network is employed along mode-3 to represent the transform. The nuclear norm of transformed frontal slices is minimized to constrain the global low-rankness of the tensor. The proposed transform more faithfully captures the implicit low-rankness of the real-world data, so that a better low-rank representation than that of linear transforms is obtained. A spatial TV regularization term is further introduced and the ADMM-like algorithm is developed to address the proposed model. Extensive experiments on different data for tensor completion, background subtraction, RTC, and SCI demonstrate that the proposed method outperforms state-of-the-art methods of different problems.  \par
\bibliographystyle{ieeetr}
\small
\bibliography{refference}

\begin{thebibliography}{10}

\bibitem{TNN_LRTC}
Z.~Zhang, G.~Ely, S.~Aeron, N.~Hao, and M.~Kilmer, ``Novel methods for
  multilinear data completion and de-noising based on tensor-{SVD},'' in {\em
  2014 IEEE Conference on Computer Vision and Pattern Recognition},
  pp.~3842--3849, 2014.

\bibitem{FTNN}
T.~Jiang, M.~K. Ng, X.~Zhao, and T.~Huang, ``Framelet representation of tensor
  nuclear norm for third-order tensor completion,'' {\em IEEE Transactions on
  Image Processing}, vol.~29, pp.~7233--7244, 2020.

\bibitem{Haar}
G.~Song, M.~K. Ng, and X.~Zhang, ``Robust tensor completion using transformed
  tensor singular value decomposition,'' {\em Numerical Linear Algebra with
  Applications}, vol.~27, no.~3, p.~e2299, 2020.

\bibitem{GAP-TV}
X.~{Yuan}, ``Generalized alternating projection based total variation
  minimization for compressive sensing,'' in {\em 2016 IEEE International
  Conference on Image Processing (ICIP)}, pp.~2539--2543, 2016.

\bibitem{DeSCI}
Y.~{Liu}, X.~{Yuan}, J.~{Suo}, D.~J. {Brady}, and Q.~{Dai}, ``Rank minimization
  for snapshot compressive imaging,'' {\em IEEE Transactions on Pattern
  Analysis and Machine Intelligence}, vol.~41, no.~12, pp.~2990--3006, 2019.

\bibitem{MSIs_CVPR16}
Q.~{Xie}, Q.~{Zhao}, D.~{Meng}, Z.~{Xu}, S.~{Gu}, W.~{Zuo}, and L.~{Zhang},
  ``Multispectral images denoising by intrinsic tensor sparsity
  regularization,'' in {\em 2016 IEEE Conference on Computer Vision and Pattern
  Recognition (CVPR)}, pp.~1692--1700, 2016.

\bibitem{TIP_2017}
K.~H. {Jin}, M.~T. {McCann}, E.~{Froustey}, and M.~{Unser}, ``Deep
  convolutional neural network for inverse problems in imaging,'' {\em IEEE
  Transactions on Image Processing}, vol.~26, no.~9, pp.~4509--4522, 2017.

\bibitem{Tensor_app}
T.~Kolda and B.~Bader, ``Tensor decompositions and applications,'' {\em SIAM
  Review}, vol.~51, pp.~455--500, 08 2009.

\bibitem{TCI_1}
W.~{Xia}, W.~{Wu}, S.~{Niu}, F.~{Liu}, J.~{Zhou}, H.~{Yu}, G.~{Wang}, and
  Y.~{Zhang}, ``Spectral ct reconstruction—assist: Aided by self-similarity
  in image-spectral tensors,'' {\em IEEE Transactions on Computational
  Imaging}, vol.~5, no.~3, pp.~420--436, 2019.

\bibitem{TCI_MRI}
B.~Yaman, S.~Weingärtner, N.~Kargas, N.~D. Sidiropoulos, and M.~Akçakaya,
  ``Low-rank tensor models for improved multidimensional {MRI}: Application to
  dynamic cardiac $t_1$ mapping,'' {\em IEEE Transactions on Computational
  Imaging}, vol.~6, pp.~194--207, 2020.

\bibitem{tTNN}
W.~Hu, D.~Tao, W.~Zhang, Y.~Xie, and Y.~Yang, ``The twist tensor nuclear norm
  for video completion,'' {\em IEEE Transactions on Neural Networks and
  Learning Systems}, vol.~28, no.~12, pp.~2961--2973, 2017.

\bibitem{TPAMI_TC}
J.~Liu, P.~Musialski, P.~Wonka, and J.~Ye, ``Tensor completion for estimating
  missing values in visual data,'' {\em IEEE Transactions on Pattern Analysis
  and Machine Intelligence}, vol.~35, no.~1, pp.~208--220, 2013.

\bibitem{TNN_TRPCA}
C.~Lu, J.~Feng, Y.~Chen, W.~Liu, Z.~Lin, and S.~Yan, ``Tensor robust principal
  component analysis with a new tensor nuclear norm,'' {\em IEEE Transactions
  on Pattern Analysis and Machine Intelligence}, vol.~42, no.~4, pp.~925--938,
  2020.

\bibitem{exact_TC}
Z.~Zhang and S.~Aeron, ``Exact tensor completion using t-{SVD},'' {\em IEEE
  Transactions on Signal Processing}, vol.~PP, no.~6, p.~1511–1526, 2017.

\bibitem{ChaoLi}
C.~{Li}, W.~{He}, L.~{Yuan}, Z.~{Sun}, and Q.~{Zhao}, ``Guaranteed matrix
  completion under multiple linear transformations,'' in {\em 2019 IEEE/CVF
  Conference on Computer Vision and Pattern Recognition}, pp.~11128--11137,
  2019.

\bibitem{TCI_2}
C.~Y. {Lin} and J.~A. {Fessler}, ``Efficient dynamic parallel {MRI}
  reconstruction for the low-rank plus sparse model,'' {\em IEEE Transactions
  on Computational Imaging}, vol.~5, no.~1, pp.~17--26, 2019.

\bibitem{TCI_TC}
H.~{Zeng}, Y.~{Chen}, X.~{Xie}, and J.~{Ning}, ``Enhanced nonconvex low-rank
  approximation of tensor multi-modes for tensor completion,'' {\em IEEE
  Transactions on Computational Imaging}, vol.~7, pp.~164--177, 2021.

\bibitem{CP_Tucker}
T.~G. Kolda and B.~W. Bader, ``Tensor decompositions and applications,'' {\em
  SIAM Review}, vol.~51, no.~3, pp.~455--500, 2009.

\bibitem{NIPS_2013}
B.~Romera-Paredes and M.~Pontil, ``A new convex relaxation for tensor
  completion,'' in {\em Proceedings of the 26th International Conference on
  Neural Information Processing Systems - Volume 2}, p.~2967–2975, 2013.

\bibitem{TSVD}
M.~Kilmer, K.~Braman, N.~Hao, and R.~C. Hoover, ``Third-order tensors as
  operators on matrices: A theoretical and computational framework with
  applications in imaging,'' {\em SIAM Journal on Matrix Analysis and
  Applications}, vol.~34, no.~1, pp.~148--172, 2013.

\bibitem{Comp}
M.~{Kilmer}, L.~{Horesh}, H.~{Avron}, and E.~{Newman}, ``Tensor-tensor products
  for optimal representation and compression,'' {\em ArXiv, abs/2001.00046},
  2019.

\bibitem{T_prod}
E.~Kernfeld, M.~Kilmer, and S.~Aeron, ``Tensor–tensor products with
  invertible linear transforms,'' {\em Linear Algebra and its Applications},
  vol.~485, pp.~545 -- 570, 2015.

\bibitem{DCT_Zhao}
W.~Xu, X.~Zhao, and M.~K. Ng, ``A fast algorithm for cosine transform based
  tensor singular value decomposition,'' {\em ArXiv, abs/1902.03070}, 2019.

\bibitem{Mada_2018}
B.~Madathil and S.~N. George, ``{DCT} based weighted adaptive multi-linear data
  completion and denoising,'' {\em Neurocomputing}, vol.~318, pp.~120 -- 136,
  2018.

\bibitem{DCTNN}
C.~Lu, X.~Peng, and Y.~Wei, ``Low-rank tensor completion with a new tensor
  nuclear norm induced by invertible linear transforms,'' in {\em 2019 IEEE/CVF
  Conference on Computer Vision and Pattern Recognition}, pp.~5989--5997, 2019.

\bibitem{Q_rank}
H.~Kong, C.~Lu, and Z.~Lin, ``Tensor {Q}-rank: New data dependent tensor
  rank,'' {\em Machine Learning}, 2021.

\bibitem{DTNN}
T.~Jiang, X.~Zhao, H.~Zhang, and M.~K. Ng, ``Dictionary learning with low-rank
  coding coefficients for tensor completion,'' {\em ArXiv, abs/2009.12507},
  2020.

\bibitem{Patch_tube}
M.~K. Ng, X.~Zhang, and X.~Zhao, ``Patched-tube unitary transform for robust
  tensor completion,'' {\em Pattern Recognition}, vol.~100, p.~107181, 2020.

\bibitem{Universial}
A.~Kratsios, ``The universal approximation property: Characterization,
  construction, representation, and existence,'' {\em Annals of Mathematics and
  Artificial Intelligence}, 01 2021.

\bibitem{TV_1}
T.~F. {Chan} and {Chiu-Kwong Wong}, ``Total variation blind deconvolution,''
  {\em IEEE Transactions on Image Processing}, vol.~7, no.~3, pp.~370--375,
  1998.

\bibitem{PReLU}
K.~He, X.~Zhang, S.~Ren, and J.~Sun, ``Delving deep into rectifiers: Surpassing
  human-level performance on {I}mage{N}et classification,'' in {\em 2015 IEEE
  International Conference on Computer Vision}, pp.~1026--1034, 2015.

\bibitem{nuc_grad}
G.~A. Watson, ``Characterization of the subdifferential of some matrix norms,''
  {\em Linear Algebra and its Applications}, vol.~170, pp.~33 -- 45, 1992.

\bibitem{ADAM}
D.~Kingma and J.~Ba, ``Adam: A method for stochastic optimization,'' {\em
  International Conference on Learning Representations}, 12 2014.

\bibitem{admmdiptv}
P.~Cascarano, A.~Sebastiani, and M.~C. Comes, ``{ADMM-DIPTV}: combining total
  variation and deep image prior for image restoration,'' {\em ArXiv,
  abs/2009.11380}, 2020.

\bibitem{TPAMI_pos}
X.~Zhang and M.~K.-P. Ng, ``Low rank tensor completion with poisson
  observations,'' {\em IEEE Transactions on Pattern Analysis and Machine
  Intelligence}, 2021.

\bibitem{TRLRF}
L.~Yuan, C.~Li, D.~Mandic, J.~Cao, and Q.~Zhao, ``Tensor ring decomposition
  with rank minimization on latent space: An efficient approach for tensor
  completion,'' in {\em Proceedings of the AAAI Conference on Artificial
  Intelligence}, vol.~33, pp.~9151--9158, Jul. 2019.

\bibitem{CAVE}
F.~Yasuma, T.~Mitsunaga, D.~Iso, and S.~K. Nayar, ``Generalized assorted pixel
  camera: Postcapture control of resolution, dynamic range, and spectrum,''
  {\em IEEE Transactions on Image Processing}, vol.~19, no.~9, pp.~2241--2253,
  2010.

\bibitem{sam}
B.~R. {Shivakumar} and S.~V. {Rajashekararadhya}, ``Performance évaluation of
  spectral angle mapper and spectral correlation mapper classifiers over
  multiple remote sensor data,'' in {\em 2017 Second International Conference
  on Electrical, Computer and Communication Technologies}, pp.~1--6, 2017.

\bibitem{MRPCA}
A.~Aravkin, S.~Becker, V.~Cevher, and P.~Olsen, ``A variational approach to
  stable principal component pursuit,'' in {\em Conference on Uncertainty in
  Artificial Intelligence (UAI)}, July 2014.

\bibitem{RTRC}
H.~{Huang}, Y.~{Liu}, Z.~{Long}, and C.~{Zhu}, ``Robust low-rank tensor ring
  completion,'' {\em IEEE Transactions on Computational Imaging}, vol.~6,
  pp.~1117--1126, 2020.

\bibitem{TIT}
D.~L. {Donoho}, ``Compressed sensing,'' {\em IEEE Transactions on Information
  Theory}, vol.~52, no.~4, pp.~1289--1306, 2006.

\bibitem{ADMM_net}
J.~{Ma}, X.~{Liu}, Z.~{Shou}, and X.~{Yuan}, ``Deep tensor admm-net for
  snapshot compressive imaging,'' in {\em 2019 IEEE/CVF International
  Conference on Computer Vision}, pp.~10222--10231, 2019.

\bibitem{SeSCI}
P.~{Yang}, L.~{Kong}, X.~Y. {Liu}, X.~{Yuan}, and G.~{Chen}, ``Shearlet
  enhanced snapshot compressive imaging,'' {\em IEEE Transactions on Image
  Processing}, vol.~29, pp.~6466--6481, 2020.

\bibitem{PLU}
A.~Nicolae, ``{PLU:} the piecewise linear unit activation function,'' {\em
  ArXiv, abs/1809.09534}, 2018.

\bibitem{spa_encoder}
J.~Deng, Z.~Zhang, E.~Marchi, and B.~Schuller, ``Sparse autoencoder-based
  feature transfer learning for speech emotion recognition,'' in {\em 2013
  Humaine Association Conference on Affective Computing and Intelligent
  Interaction}, pp.~511--516, 2013.

\bibitem{XieQi_PAMI}
Q.~Xie, Q.~Zhao, and D.~Meng, ``Kronecker-basis-representation based tensor
  sparsity and its applications to tensor recovery,'' {\em IEEE Transactions on
  Pattern Analysis and Machine Intelligence}, vol.~40, no.~8, pp.~1888--1902,
  2018.

\end{thebibliography}
\end{document}